\documentclass[12pt,english]{article}
\usepackage{times}
\usepackage[T1]{fontenc}
\usepackage{geometry}
\usepackage{graphicx}
\usepackage{setspace}
\usepackage{amssymb}

\makeatletter

\providecommand{\tabularnewline}{\\}

\usepackage{bm}

\usepackage{babel}
\makeatother
\begin{document}

\section*{\noindent Multi-Objective System-by-Design for \emph{mm}-Wave Automotive
Radar Antennas}

\noindent ~

\noindent \vfill

\noindent P. Rosatti,$^{(1)}$ M. Salucci,$^{(1)}$ \emph{Senior Member,
IEEE}, L. Poli,$^{(1)}$ \emph{Member, IEEE}, and A. Massa,$^{(2)(1)(3)}$
\emph{Fellow, IEEE}

\noindent \vfill

\noindent ~

\noindent {\footnotesize $^{(1)}$} \emph{\footnotesize ELEDIA Research
Center} {\footnotesize (}\emph{\footnotesize ELEDIA}{\footnotesize @}\emph{\footnotesize UniTN}
{\footnotesize - University of Trento)}{\footnotesize \par}

\noindent {\footnotesize DICAM - Department of Civil, Environmental,
and Mechanical Engineering}{\footnotesize \par}

\noindent {\footnotesize Via Mesiano 77, 38123 Trento - Italy}{\footnotesize \par}

\noindent \textit{\emph{\footnotesize E-mail:}} {\footnotesize \{}\emph{\footnotesize pietro.rosatti}{\footnotesize ,}
\emph{\footnotesize marco.salucci}{\footnotesize ,} \emph{\footnotesize lorenzo.poli}{\footnotesize ,}
\emph{\footnotesize andrea.massa}{\footnotesize \}@}\emph{\footnotesize unitn.it}{\footnotesize \par}

\noindent {\footnotesize Website:} \emph{\footnotesize www.eledia.org/eledia-unitn}{\footnotesize \par}

\noindent {\footnotesize ~}{\footnotesize \par}

\noindent {\footnotesize $^{(2)}$} \emph{\footnotesize ELEDIA Research
Center} {\footnotesize (}\emph{\footnotesize ELEDIA}{\footnotesize @}\emph{\footnotesize UESTC}
{\footnotesize - UESTC)}{\footnotesize \par}

\noindent {\footnotesize School of Electronic Engineering, Chengdu
611731 - China}{\footnotesize \par}

\noindent \textit{\emph{\footnotesize E-mail:}} \emph{\footnotesize andrea.massa@uestc.edu.cn}{\footnotesize \par}

\noindent {\footnotesize Website:} \emph{\footnotesize www.eledia.org/eledia}{\footnotesize -}\emph{\footnotesize uestc}{\footnotesize \par}

\noindent ~

\noindent {\footnotesize $^{(3)}$} \emph{\footnotesize ELEDIA Research
Center} {\footnotesize (}\emph{\footnotesize ELEDIA@TSINGHUA} {\footnotesize -
Tsinghua University)}{\footnotesize \par}

\noindent {\footnotesize 30 Shuangqing Rd, 100084 Haidian, Beijing
- China}{\footnotesize \par}

\noindent {\footnotesize E-mail:} \emph{\footnotesize andrea.massa@tsinghua.edu.cn}{\footnotesize \par}

\noindent {\footnotesize Website:} \emph{\footnotesize www.eledia.org/eledia-tsinghua}{\footnotesize \par}

\noindent \vfill

\noindent \textbf{\emph{This work has been submitted to the IEEE for
possible publication. Copyright may be transferred without notice,
after which this version may no longer be accessible.}}

\noindent \vfill

\newpage
\section*{Multi-Objective System-by-Design for \emph{mm}-Wave Automotive Radar
Antennas}

~

\noindent ~

\noindent ~

\begin{flushleft}P. Rosatti, M. Salucci, L. Poli, and A. Massa\end{flushleft}

\vfill

\begin{abstract}
\noindent The computationally-efficient solution of multi-objective
optimization problems (\emph{MOP}s) arising in the design of modern
electromagnetic (\emph{EM}) microwave devices is addressed. Towards
this end, a novel System-by-Design (\emph{SbD}) method is developed
to effectively explore the solution space and to provide the decision
maker with a set of optimal trade-off solutions minimizing multiple
and (generally) contrasting objectives. The proposed \emph{MO-SbD}
method proves a high computational efficiency, with a remarkable time
saving with respect to a competitive state-of-the-art \emph{MOP} solution
strategy, thanks to the {}``smart'' integration of evolutionary-inspired
concepts and operators with artificial intelligence (\emph{AI}) and
machine learning (\emph{ML}) techniques. Representative numerical
results are reported to provide the interested users with useful insights
and guidelines on the proposed optimization method as well as to assess
its effectiveness in designing \emph{mm}-wave automotive radar antennas. 

\noindent \vfill
\end{abstract}
\noindent \textbf{Key words}: Microwave Design, Radar Antennas, Evolutionary
Algorithms (\emph{EA}s), Artificial Intelligence (\emph{AI}), Machine
Learning (\emph{ML}), Multi-Objective Optimization Problem (\emph{MOP}),
System-by-Design (\emph{SbD}).

\newpage
\section{Introduction and Motivation}

\noindent The design of modern high-performance microwave devices
such as filters \cite{Zhang 2021}, directional couplers \cite{Wang 2015},
and antennas \cite{Watanabe 2020}, generally involves a number of
challenging and (sometimes) contrasting requirements on their electromagnetic
(\emph{EM}) behavior along with tight geometrical, weight, and cost
constraints \cite{Goudos 2021}. Such a synthesis problem is often
formulated as a single-objective optimization problem (\emph{SOP})
whose solution is the global minimum of a cost function that quantifies
the mismatch between the \emph{EM}/structural features of the synthesized
device and the user-defined objectives. Several and effective solution
strategies for solving \emph{SOP}s have been presented in the scientific
literature based on deterministic \cite{Amari 2000} as well as stochastic
\cite{Goudos 2021}\cite{Rocca 2009}-\cite{Goudos 2016} optimization
algorithms. However, since conflicting requirements are typically
at hand, the arising \emph{SOP} solution turns out to be only a particular
trade-off among all the optimization targets. Otherwise, a set of
optimal compromise solutions belonging to the so-called Pareto Front
(\emph{PF}) can be obtained by natively addressing the design process
as a multi-objective optimization problem (\emph{MOP}) \cite{Nagar 2017}-\cite{Nagar 2017b}.
This approach allows the designer to select the most suitable design
that fulfils a specific application \cite{Rocca 2022} or that features
other profitable properties. 

\noindent Regardless of the single- or multi-objective formulation,
full-wave (\emph{FW}) \emph{EM} software are usually used to accurately
predict the \emph{EM} behavior of each trial design/solution \cite{Goudos 2021}.
Indeed, more and more reliable and precise computational tools are
nowadays available \cite{Harrington 1993}-\cite{Chen 2022} for the
analysis of complex \emph{EM} devices. Because of the reliability
in modeling non-linear and coupling \emph{EM} phenomena, which are
generally neglected or approximated by simplified analytical approaches
\cite{Balanis 2016}\cite{Jackson 1991}, they play a key-role in
the study of modern microwave devices. 

\noindent However, the integration of \emph{FW} solvers in iterative
optimization loops (e.g., evolutionary-algorithm (\emph{EA}) optimizations
\cite{Goudos 2021}\cite{Rocca 2009}) is not straightforward due
to the high computational burden. To counteract such an issue, significant
efforts have been devoted and the proposed approaches can be classified
according to the following countermeasures: (\emph{i}) the exploitation
of the \emph{a-priori} knowledge on the solution through model-based
approaches \cite{Massa 2015}, (\emph{ii}) the reduction of the dimensionality
of the solution space by looking for {}``smart'' definitions of
the unknowns/degrees-of-freedom (\emph{DoF}s) \cite{Salucci 2022},
(\emph{iii}) a suitable initialization of the optimization process
\cite{Oliveri 2011}, and (\emph{iv}) the reduction of the \emph{CPU}
time for the evaluation of the {}``fitness'' of a trial solution
to the design objectives by means of space mapping strategies \cite{Koziel 2010}\cite{Zhang 2018}
or artificial intelligence (\emph{AI}) and machine learning (\emph{ML})
methods \cite{Koziel 2021.a}-\cite{Long 2021}.

\noindent As for this latter framework, \emph{ML} techniques based
on the Learning-by-Examples (\emph{LBE}) \cite{Massa 2018} theory
that also exploit complex Deep Learning (\emph{DL}) \cite{Massa 2019}
architectures are recently gaining an increasing attention from both
the scientific community and the industrial one. As a matter of fact,
computationally-efficient surrogate models (\emph{SM}s), which are
able to accurately predict complex \emph{EM} dynamics \cite{Wu 2021}-\cite{An 2018},
can be derived starting from a training set of input/output (\emph{I/O})
examples computed with the \emph{FW} solver.

\noindent Based on the {}``interactive collaboration'' between \emph{EA}s
and \emph{LBE} methods, effective and efficient techniques for the
synthesis of \emph{EM} devices have been recently proposed according
to the so-called System-by-Design (\emph{SbD}) paradigm, which is
defined as the \char`\"{}\emph{task-oriented design, definition, and
integration of system components to yield EM devices with user-desired
performance having the minimum costs, the maximum scalability, and
suitable reconfigurability properties}\char`\"{} \cite{Massa 2022}.
The potential and the flexibility of such an approach have been demonstrated
in the design of a plethora of devices including wide-angle impedance
layers (\emph{WAIM}s) \cite{Oliveri 2022}, metamaterial lenses \cite{Salucci 2019},
holographic \emph{EM} skins \cite{Oliveri 2021b}, polarizers \cite{Arnieri 2022},
nano-structures \cite{Nagar 2017b}, reflectarrays \cite{Oliveri 2020},
airborne radomes \cite{Salucci 2022}, \emph{5G} arrays \cite{Massa 2022},
time modulated arrays \cite{Massa 2022}, and automotive radar antennas
\cite{Massa 2022}. However, to the best of the authors' knowledge,
\emph{SO-SbD} methods have been proposed so far mainly without exploring
the pros \& cons of \emph{MO} implementations.

\noindent Towards this end, this paper proposes a new \emph{SbD}-based
technique for the \emph{AI}-driven solution of complex \emph{EM} syntheses
formulated as \emph{MOP}s. From a methodological point of view, the
main novelties of this work over the existing literature can be summarized
as follows: (\emph{i}) the integration of evolutionary-inspired concepts
and operators within a new \emph{MO-SbD} solution scheme that adaptively
exploits a \emph{ML}-based \emph{SM} to accurately predict the {}``fitness''
of each guess solution to the design problem at hand, (\emph{ii})
the derivation of a new set of {}``\emph{SbD}-dominance'' criteria
that allows a fruitful exploitation of the available information on
the reliability index of the \emph{SM} predictions, (\emph{iii}) the
implementation of a reinforced learning (\emph{RL}) strategy to improve
the accuracy of the \emph{SM} during the \emph{SbD} optimization loop,
and (\emph{iv}) the customization of the \emph{MO-SbD} scheme to the
synthesis of \emph{mm}-wave automotive radar antennas \cite{Hash 2012}.

\noindent The paper is organized as follows. Section \ref{sec:2 - Mathematical-Formulation}
gives the mathematical formulation of the \emph{SbD-MOP}. The description
of the proposed \emph{MO-SbD} is detailed in Sect. \ref{sub:SbD-MOP-Solution-Methodology}
along with its customization to the synthesis problem of designing
\emph{mm}-wave antennas for automotive radars (Sect. \ref{sub:Application-to-mm-Wave}).
The assessment of the \emph{MO-SbD} method is carried out by first
considering standard \emph{MOP} benchmark functions (Sect. \ref{sub:4.1 - Sensitivity-Analysis})
to derive some general insights on its performance as well as useful
guidelines for its application to specific applicative problems. Successively,
Section \ref{sub:4.2 - Design-of-77GHz-Automotive} is devoted to
assess the effectiveness and the computational efficiency of the \emph{MO-SbD}
in synthesizing \emph{mm}-wave antennas for automotive radars. Finally,
some conclusions and final remarks are drawn (Sect. \ref{sec:Conclusions-and-Final}).

\section{\noindent Mathematical Formulation \label{sec:2 - Mathematical-Formulation}}

\noindent Let $K$ be the dimensionality of the solution space, $\mathbb{R}^{K}$,
of a \emph{MO} design problem, $\underline{\Omega}$ being a generic
element of $\mathbb{R}^{K}$ ($\underline{\Omega}\triangleq\left\{ \Omega_{k};\, k=1,...,K\right\} $),
aimed at determining the set $\mathbf{A}$ of $A$ ($A>1$) Pareto-optimal
trade-off solutions, $\mathbf{A}\triangleq\left\{ \underline{\Omega}^{\left(a\right)};\, a=1,...,A\right\} $,
that fulfils $Q$ ($Q>1$) user-defined objectives coded into the
cost function vector $\underline{\Phi}$ ($\underline{\Phi}\left(\underline{\Omega}\right)\triangleq\left[\Phi_{q}\left(\underline{\Omega}\right);\, q=1,...,Q\right]$,
$\underline{\Phi}\left(\underline{\Omega}\right)\in\mathbb{R}^{Q}$).

\noindent According to the reference \emph{MOP} literature \cite{Deb 2001}\cite{Deb 2003}\cite{Laumanns 2002},
$\mathbf{A}$ is composed by the Pareto Front (\emph{PF}) of all \emph{non-dominated}
solutions spanning the $K$-dimensional search space $\mathbb{R}^{K}$,
the dominance of one solution, $\underline{\Omega}^{\left(1\right)}$,
over another one, $\underline{\Omega}^{\left(2\right)}\neq\underline{\Omega}^{\left(1\right)}$
(i.e., $\underline{\Omega}^{\left(1\right)}\prec\underline{\Omega}^{\left(2\right)}$)
being mathematically stated as follows (Fig. 1)\begin{equation}
\left\{ \begin{array}{ll}
\Phi_{p}\left(\underline{\Omega}^{\left(1\right)}\right)<\Phi_{p}\left(\underline{\Omega}^{\left(2\right)}\right) & \,\,\,\,\, p\in\left\{ 1,...,Q\right\} \\
\Phi_{q}\left(\underline{\Omega}^{\left(1\right)}\right)\leq\Phi_{q}\left(\underline{\Omega}^{\left(2\right)}\right) & \,\,\,\,\, q=1,...,Q;\, q\ne p.\end{array}\right.\label{eq:standard-dominance}\end{equation}
However, because of the continuous/real nature of the $Q$ components
of $\underline{\Phi}$, the number of entries of $\mathbf{A}$ is
infinite ($A\rightarrow\infty$) and the resulting optimization task
turns out to be unfeasible \cite{Deb 2003}\cite{Laumanns 2002}.
In practical scenarios, it is thus convenient to look for an alternative/approximated
definition of $\mathbf{A}$ that complies with the following assumptions:
(\emph{a}) the number of entries of the \emph{PF}, $A$, is finite
(i.e., $\mathbf{A}$ is a bounded set) and (\emph{b}) these $A$ solutions
are uniformly spread in the solution space $\mathbb{R}^{K}$ so that
they are representative of all possible non-dominated solutions that
comply with (\ref{eq:standard-dominance}). Towards this end, the
concept of \emph{$\varepsilon$}-\emph{dominance} is exploited. Accordingly,
$\underline{\Omega}^{\left(1\right)}$ is said to $\varepsilon$\emph{-dominate}
$\underline{\Omega}^{\left(2\right)}$ (i.e., $\underline{\Omega}^{\left(1\right)}\prec_{\varepsilon}\underline{\Omega}^{\left(2\right)}$)
if one of the following conditions holds true (Fig. 1) \cite{Deb 2003}\cite{Laumanns 2002}\begin{equation}
\left\{ \begin{array}{ll}
\left\lfloor \frac{\Phi_{p}\left(\underline{\Omega}^{\left(1\right)}\right)}{\varepsilon_{p}}\right\rfloor <\left\lfloor \frac{\Phi_{p}\left(\underline{\Omega}^{\left(2\right)}\right)}{\varepsilon_{p}}\right\rfloor  & \,\,\,\,\, p\in\left\{ 1,...,Q\right\} \\
\left\lfloor \frac{\Phi_{q}\left(\underline{\Omega}^{\left(1\right)}\right)}{\varepsilon_{q}}\right\rfloor \leq\left\lfloor \frac{\Phi_{q}\left(\underline{\Omega}^{\left(2\right)}\right)}{\varepsilon_{q}}\right\rfloor  & \,\,\,\,\, q=1,...,Q;\, q\ne p\end{array}\right.\label{eq:1 - epsilon-dominance}\end{equation}
or\begin{equation}
\left\{ \begin{array}{ll}
\left\lfloor \frac{\Phi_{q}\left(\underline{\Omega}^{\left(1\right)}\right)}{\varepsilon_{q}}\right\rfloor =\left\lfloor \frac{\Phi_{q}\left(\underline{\Omega}^{\left(2\right)}\right)}{\varepsilon_{q}}\right\rfloor  & \,\,\,\,\, q=1,...,Q\\
d_{\varepsilon}\left(\underline{\Omega}^{\left(1\right)}\right)<d_{\varepsilon}\left(\underline{\Omega}^{\left(2\right)}\right).\end{array}\right.\label{eq:2 - epsilon-dominance}\end{equation}

\noindent In (\ref{eq:1 - epsilon-dominance}), which is the {}``quantized''
counterpart of (\ref{eq:standard-dominance}), and (\ref{eq:2 - epsilon-dominance}),
$\left\lfloor \,.\,\right\rfloor $ is the integer floor function,
while $d_{\varepsilon}\left(\underline{\Omega}\right)$ is the Euclidean
distance between $\underline{\Phi}\left(\underline{\Omega}\right)$
and its {}``$\varepsilon$-quantized'' version, $\widehat{\underline{\Phi}}\left(\underline{\Omega}\right)$
{[}$\widehat{\underline{\Phi}}\left(\underline{\Omega}\right)\triangleq\left[\widehat{\Phi}_{q}\left(\underline{\Omega}\right);\, q=1,...,Q\right]$
being $\widehat{\Phi}_{q}\left(\underline{\Omega}\right)\triangleq\left\lfloor \frac{\Phi_{q}\left(\underline{\Omega}\right)}{\varepsilon_{q}}\right\rfloor \times\varepsilon_{q}${]}
(Fig. 1)\begin{equation}
d_{\varepsilon}\left(\underline{\Omega}\right)=\sqrt{\sum_{q=1}^{Q}\left[\Phi_{q}\left(\underline{\Omega}\right)-\widehat{\Phi}_{q}\left(\underline{\Omega}\right)\right]^{2}}.\label{eq:}\end{equation}
Moreover, $\underline{\varepsilon}$ is the set of $Q$ user-defined
real values ($\underline{\varepsilon}\triangleq\left\{ \varepsilon_{q}>0;\, q=1,...,Q\right\} $)
that allows the designer to \emph{a-priori} control the accuracy of
the approximation of the ideal \emph{PF}. A large value of its $q$-th
($q=1,...,Q$) entry, $\varepsilon_{q}$, results in a coarser representation
of the \emph{PF} along the $q$-th objective, and vice-versa. 

\noindent An illustrative example of both dominance conditions, (\ref{eq:standard-dominance})
and (\ref{eq:1 - epsilon-dominance})(\ref{eq:2 - epsilon-dominance}),
is shown in Fig. 1. According to (\ref{eq:standard-dominance}), it
turns out that $\underline{\Omega}^{\left(1\right)}\prec\underline{\Omega}^{\left(3\right)}$
and $\underline{\Omega}^{\left(2\right)}\prec\underline{\Omega}^{\left(3\right)}$,
but $\underline{\Omega}^{\left(1\right)}\nprec\underline{\Omega}^{\left(2\right)}$
and $\underline{\Omega}^{\left(2\right)}\nprec\underline{\Omega}^{\left(1\right)}$
(i.e., $\underline{\Omega}^{\left(1\right)}$ and $\underline{\Omega}^{\left(2\right)}$
are two non-dominated solutions). Otherwise, the solution $\underline{\Omega}^{\left(1\right)}$
$\varepsilon$-dominates both $\underline{\Omega}^{\left(2\right)}$
and $\underline{\Omega}^{\left(3\right)}$ (i.e., $\underline{\Omega}^{\left(1\right)}\prec_{\varepsilon}\underline{\Omega}^{\left(2\right)}$
and $\underline{\Omega}^{\left(1\right)}\prec_{\varepsilon}\underline{\Omega}^{\left(3\right)}$)
according to (\ref{eq:2 - epsilon-dominance}), while $\underline{\Omega}^{\left(2\right)}\prec_{\varepsilon}\underline{\Omega}^{\left(3\right)}$.

\noindent Within the \emph{$\varepsilon$}-dominance framework, the
statement of the \emph{MO} problem at hand can be then phrased as
follows:

\begin{quotation}
\noindent \textbf{\emph{$\varepsilon$-MOP}} - Given a set of $Q$
objectives, $\underline{\Phi}$, and defined the accuracy vector $\underline{\varepsilon}$,
find the {}``optimal'' \emph{PF} of $A^{opt}$ ($A^{opt}>1$) non-$\varepsilon$-dominated
solutions, $\mathbf{A}^{opt}$,

\noindent \begin{equation}
\mathbf{A}^{opt}=\left\{ \left[\underline{\Omega}^{\left(a\right)}:\,\nexists\, b\neq a\rightarrow\underline{\Omega}^{\left(b\right)}\prec_{\varepsilon}\underline{\Omega}^{\left(a\right)}\right];\, a,b=1,...,A^{opt}\right\} ,\label{eq:}\end{equation}
that includes the global optimum $\underline{\Omega}^{opt}$ ($\underline{\Omega}^{opt}\in\mathbf{A}^{opt}$),
$\underline{\Omega}^{opt}=\arg\left\{ \min_{\underline{\Omega}}\left[\Phi_{q}\left(\underline{\Omega}\right);\, q=1,...,Q\right]\right\} $.

\end{quotation}

\section{\emph{MO-SbD} Solution Method \label{sub:SbD-MOP-Solution-Methodology}}

\noindent In order to solve the $\varepsilon$-\emph{MOP} with high
computational efficiency, the problem at hand is \emph{}recast within
the \emph{SbD} framework as follows

\begin{quotation}
\noindent $SbD$\textbf{\emph{-MOP}} - Iteratively ($i$ being the
iteration index, $i=1,...,I$) generate a sequence of \emph{PF}s \{$\mathbf{A}_{i}$;
$i=0,...,I$\}, that converges to the optimal one $\mathbf{A}^{opt}$
($\mathbf{A}^{SbD}\to\mathbf{A}^{opt}$, $\mathbf{A}^{SbD}=\left.\mathbf{A}_{i}\right\rfloor _{i=I_{SbD}}$,
$I_{SbD}$ being the convergence iteration) in an overall computation
time $\Delta t^{SbD}$ fulfilling the condition\begin{equation}
\Delta t^{SbD}\ll\Delta t^{StD},\label{eq:SbD-Time}\end{equation}
$\Delta t^{StD}$ being the time needed by a standard optimization
method to reach the same \emph{SbD} convergence condition, which is
defined as\begin{equation}
\Phi_{q}\left(\underline{\Omega}^{SbD}\right)\leq\zeta_{q}\,\,\,\,\,\,\,\, q=1,...,Q\label{eq:}\end{equation}
where $\underline{\Omega}^{SbD}\in\mathbf{A}^{SbD}$ and $\zeta_{q}$
is the convergence threshold for the $q$-th ($q=1,...,Q$) objective.
\end{quotation}
This latter formulation is then addressed with an innovative synthesis
strategy leveraging on the {}``interactive collaboration'' between
evolutionary-inspired operators drawn from the $\varepsilon$-\emph{MOEA}
algorithm \cite{Deb 2003} (referred to as the {}``\emph{StD}''
method in the following) and \emph{AI} techniques.

\noindent In the following, the discussion on the \emph{SbD} implementation
will focus on three main items: (\emph{1}) the \emph{AI}-based mechanisms
for dealing with a faithful, but also computationally efficient, prediction
of the \emph{EM} behavior of each guess design and, in turn, the estimation
of its fitness to the problem at hand {[}i.e., $\underline{\Phi}\left(\underline{\Omega}\right)${]};
(\emph{2}) the multiple-agent strategy for sampling the solution space;
(\emph{3}) the work-flow of the algorithmic implementation of the
\emph{MO-SbD}.

\noindent As for (\emph{1}), let us observe that the overall computational
cost of the synthesis based on a standard iterative optimization is
equal to\begin{equation}
\Delta t^{StD}=C_{FW}^{StD}\times\Delta t_{FW}\label{eq:Std-Time}\end{equation}
where $\Delta t_{FW}$ is the average time for evaluating/computing
the $Q$ objectives/cost-function-terms associated to any trial guess
solution $\underline{\Omega}$, \{$\Phi_{q}\left(\underline{\Omega}\right)$;
$q=1,...,Q$\}, and $C_{FW}^{StD}=P+I$ is the total number of \emph{FW}
calls during the iterative multi-agent process, $P$ being the population
of agents/trial solutions. Thus, the computational goal of a \emph{SbD}-based
design (\ref{eq:SbD-Time}) can be yielded by alternatively or (better)
simultaneously acting on both factors in (\ref{eq:Std-Time}).

\noindent For instance, a recent trend is that of reducing the number
of computationally-expensive \emph{FW} calls (i.e., $C_{FW}^{SbD}\ll C_{FW}^{StD}$)
by exploiting a fast surrogate model (\emph{SM}) to predict the fitness
value, $\underline{\Phi}\left(\underline{\Omega}\right)$, of the
remaining ($C_{FW}^{StD}-C_{FW}^{SbD}$) solutions (i.e., $\Delta t_{SM}\ll\Delta t_{FW}$).
According to this guideline, a \emph{SM} based on the Ordinary Kriging
(\emph{OK}) \emph{LBE} technique \cite{Jones 1998}\cite{Sacks 1989}\cite{Salucci 2018}
is built at each $i$-th ($i=1,...,I$) \emph{SbD} iteration, $\mathcal{F}_{i}\left\{ \underline{\Omega}\right\} $,
starting from the information learned from a training set $\mathbf{T}_{i}$
of $T_{i}$ \emph{I/O} pairs\begin{equation}
\mathbf{T}_{i}=\left\{ \left[\underline{\Omega}_{i}^{\left(t\right)},\,\underline{\Phi}\left(\underline{\Omega}_{i}^{\left(t\right)}\right)\right];\, t=1,...,T_{i}\right\} .\label{eq:}\end{equation}
Such a methodological choice is motivated by the solid mathematical
background of the \emph{OK} theory that allows the arising \emph{SM}
to output deterministic predictions, $\underline{\widetilde{\Phi}}\left(\underline{\Omega}\right)$,
which interpolate the (\emph{FW}-computed) \emph{I/O} samples of the
corresponding training set, $\mathbf{T}$ {[}i.e., $\underline{\widetilde{\Phi}}\left(\underline{\Omega}\right)\triangleq\underline{\widetilde{\Phi}}\left(\left.\underline{\Omega}\right|\mathbf{T}\right)${]}.
Moreover, the by-product of using an \emph{OK}-based predictor is
the availability of a {}``reliability index'' for every fitness
prediction \cite{Jones 1998}\cite{Sacks 1989}\cite{Salucci 2018},
$\underline{\delta}\left(\underline{\Omega}\right)$ {[}$\underline{\delta}\left(\underline{\Omega}\right)\triangleq\underline{\delta}\left(\left.\underline{\Omega}\right|\mathbf{T}\right)${]},
whose $q$-th ($q=1,...,Q$) entry, $\delta_{q}\left(\underline{\Omega}\right)$
{[}$\delta_{q}\left(\underline{\Omega}\right)\triangleq\delta_{q}\left(\left.\underline{\Omega}\right|\mathbf{T}\right)${]},
is given by $\delta_{q}\left(\underline{\Omega}\right)=\sqrt{\eta_{q}\left(\underline{\Omega}\right)}$,
$\eta_{q}\left(\underline{\Omega}\right)$ being the estimated mean
squared error (\emph{MSE}) of the \emph{OK} prediction, $\widetilde{\Phi}_{q}\left(\underline{\Omega}\right)$,
in approximating the corresponding actual value, $\Phi_{q}\left(\underline{\Omega}\right)$
\cite{Jones 1998}\cite{Sacks 1989}. Quantitatively, a smaller value
of $\delta_{q}\left(\underline{\Omega}\right)$ means a higher level
of reliability of the prediction {[}i.e., $\widetilde{\Phi}_{q}\left(\underline{\Omega}\right)\approx\Phi_{q}\left(\underline{\Omega}\right)${]},
and vice-versa %
\footnote{\noindent The \emph{OK} prediction is faithful {[}i.e., $\widetilde{\Phi}_{q}\left(\underline{\Omega}\right)=\Phi_{q}\left(\underline{\Omega}\right)$,
$q=1,...,Q${]} with null uncertainty {[}i.e., $\delta_{q}\left(\underline{\Omega}\right)=0$,
$q=1,...,Q${]} if $\underline{\Omega}$ coincides with one training
sample \cite{Jones 1998} (i.e., $\underline{\Omega}\in\mathbf{T}$).%
} \cite{Jones 1998}\cite{Sacks 1989}\cite{Salucci 2018}.

\noindent By exploiting the information on the {}``reliability''
of the \emph{OK} predictions, the {}``confidence hyper-volume''
of a prediction $\underline{\widetilde{\Phi}}\left(\underline{\Omega}\right)$,
$\mathcal{V}_{\widetilde{\Phi}}$, is defined as the $Q$-dimensional
region with the highest probability of containing the actual value
$\underline{\Phi}\left(\underline{\Omega}\right)$ (Fig. 2)\begin{equation}
\mathcal{V}_{\widetilde{\Phi}}=\left\{ \mathcal{L}_{q}\left(\underline{\Omega}\right)\leq\widetilde{\Phi}_{q}\left(\underline{\Omega}\right)\leq\mathcal{U}_{q}\left(\underline{\Omega}\right);\, q=1,...,Q\right\} .\label{eq:confidence-volume}\end{equation}
where $\mathcal{L}_{q}\left(\underline{\Omega}\right)$ and $\mathcal{U}_{q}\left(\underline{\Omega}\right)$
are the lower and the upper confidence bounds of the \emph{OK} prediction,
$\widetilde{\Phi}_{q}\left(\underline{\Omega}\right)$ ($q=1,...,Q$):
$\mathcal{L}_{q}\left(\underline{\Omega}\right)=\widetilde{\Phi}_{q}\left(\underline{\Omega}\right)-\delta_{q}\left(\underline{\Omega}\right)$
and $\mathcal{U}_{q}\left(\underline{\Omega}\right)=\widetilde{\Phi}_{q}\left(\underline{\Omega}\right)+\delta_{q}\left(\underline{\Omega}\right)$.

\noindent Concerning the sampling of the solution space (\emph{2})
and similarly to the \emph{StD} method \cite{Deb 2003}, the proposed
\emph{SbD} -based strategy generates ($i=0$) and evolves ($i=1,...,I$)
both a variable-size \emph{PF}, $\mathbf{A}_{i}$, of $A_{i}$ solutions
($\mathbf{A}_{i}=\left\{ \underline{\Omega}_{i}^{\left(a\right)};\, a=1,...,A_{i}\right\} $)
and a fixed-size population, $\mathbf{P}_{i}$, of $P$ individuals
($\mathbf{P}_{i}=\left\{ \underline{\Omega}_{i}^{\left(p\right)};\, p=1,...,P\right\} $).
While the \emph{PF} is updated exclusively with \emph{FW}-evaluated
individuals to guarantee the unambiguous knowledge of the best trade-off
solutions, a more sophisticated evolution mechanism, which involves
both \emph{FW}-simulated and \emph{SM}-predicted individuals, is used
for evolving the population.

\noindent More specifically, the concept of \emph{SbD}-\emph{dominance}
(i.e., $\underline{\Omega}^{\left(1\right)}\prec_{SbD}\underline{\Omega}^{\left(2\right)}$)
between any pair of solutions $\underline{\Omega}^{\left(1\right)}$
and $\underline{\Omega}^{\left(2\right)}$ evaluated either with the
\emph{FW} solver or with the \emph{SM} is introduced. It is based
on the following rules: (\emph{i}) an individual whose objectives
have been simulated is preferred to a predicted one because of its
higher trustworthiness; (\emph{ii}) a simulated individual is \emph{SbD}-dominated
by a predicted one only if taking into account the corresponding confidence
hyper-volume. Mathematically, such guidelines are coded into the relations
in Tab. I that are also pictorially illustrated in Fig. 3 ($Q=2$).
As a representative example, let us consider the case reported in
Fig. 3(\emph{c}) where $\underline{\Omega}^{\left(1\right)}$ is predicted
and $\underline{\Omega}^{\left(2\right)}$ is simulated. It turns
out that $\underline{\Omega}^{\left(1\right)}\prec_{SbD}\underline{\Omega}^{\left(2\right)}$
because $\underline{\Omega}^{\left(1\right)}$ would dominate $\underline{\Omega}^{\left(2\right)}$
even in the {}``worst case'' when $\underline{\Phi}\left(\underline{\Omega}^{\left(1\right)}\right)=\underline{\mathcal{U}}\left(\underline{\Omega}^{\left(1\right)}\right)$.

\noindent To adaptively increase the accuracy of the predictor {[}$\underline{\widetilde{\Phi}}_{i}\left(\underline{\Omega}\right)\to\underline{\Phi}\left(\underline{\Omega}\right)$,
$i=1,...,I${]}, while the \emph{SbD} minimization process converges
towards the optimal \emph{PF} ($i\to I_{SbD}$), a \emph{RL} strategy
is applied to progressively update the \emph{SM} each time a new individual
is elected for being evaluated with the \emph{FW} solver. It is worth
pointing out that the designer has the full control of the total time
of the \emph{SbD} optimization, $\Delta t^{SbD}$, since the number
of \emph{RL FW} calls is \emph{a-priori} limited to $T_{RL}\leq T_{RL}^{\max}$,
$T_{RL}^{\max}$ being a user-defined maximum value. By integrating
the \emph{AI}-based mechanisms for predicting the \emph{EM} behavior
of a guess design (\emph{1}) and the strategy for sampling the solution
space with evolutionary-inspired operators \cite{Deb 2003} (\emph{2}),
the work-flow of the \emph{MO-SbD} in Fig. 4 consists of two phases
performed in the following chronological order: (\emph{1}.) \emph{Initialization}
($i=0$) and (\emph{2.}) \emph{Iterative Loop} ($i=1,...,I$). More
in detail,

\noindent \textbf{(}\textbf{\emph{1.}}\textbf{) Initialization} ($i=0$) 

\begin{itemize}
\item \noindent Set $T_{RL}=0$. Build an initial training set $\mathbf{T}_{0}$
of $T_{0}$ \emph{I/O} pairs ($\mathbf{T}_{0}$ $=$ \{$\left[\underline{\Omega}_{0}^{\left(t\right)},\,\underline{\Phi}\left(\underline{\Omega}_{0}^{\left(t\right)}\right)\right]$;
$t=1,...,T_{0}$\}) where the set of solutions $\left\{ \underline{\Omega}_{0}^{\left(t\right)};\, t=1,...,T_{0}\right\} $
is generated by means of the Latin Hypercube Sampling (\emph{LHS})
method \cite{Massa 2022};
\item Train the initial \emph{SM}, $\mathcal{F}_{0}\left\{ \underline{\Omega}\right\} $,
with $\mathbf{T}_{0}$;
\item Initialize the \emph{PF} $\mathbf{A}_{0}$ ($\mathbf{A}_{0}=\left\{ \underline{\Omega}_{0}^{\left(a\right)};\, a=1,...,A_{0}\right\} $,
$\mathbf{A}_{0}\subseteq\mathbf{T}_{0}$, $A_{0}\leq T_{0}$) by applying
the operator $\mathcal{N}_{\varepsilon}\left\{ \mathbf{G}\right\} $,
which extracts all $G'\leq G$ non-$\varepsilon$-dominated solutions
within a generic set $\mathbf{G}$ of $G$ solutions ($\mathbf{G}=\left\{ \underline{\Omega}^{\left(g\right)};\, g=1,...,G\right\} $)
\begin{equation}
\mathcal{N}_{\varepsilon}\left\{ \mathbf{G}\right\} =\left\{ \left[\underline{\Omega}^{\left(h\right)}\in\mathbf{G}\,:\,\nexists g\in\left(g=1,...,G\right)\rightarrow\underline{\Omega}^{\left(g\right)}\prec_{\varepsilon}\underline{\Omega}^{\left(h\right)}\right];\, h=1,...,G'\right\} ,\label{eq:non-eps-dominated-subset}\end{equation}
to $\mathbf{T}_{0}$ ($\mathbf{A}_{0}=\mathcal{N}_{\varepsilon}\left\{ \underline{\Omega}_{0}^{\left(t\right)};\, t=1,...,T_{0}\right\} $);
\item Choose $P$ individuals of the initial population, $\mathbf{P}_{0}=\left\{ \underline{\Omega}_{0}^{\left(p\right)};\, p=1,...,P\right\} $,
according to the following rule\begin{equation}
\underline{\Omega}_{0}^{\left(p\right)}=\left\{ \begin{array}{ll}
\left.\underline{\Omega}_{0}^{\left(a\right)}\right\rfloor _{a=p} & \mathrm{if}\,\, p\le\min\left(P,\, A_{0}\right)\\
\mathcal{R}\left\{ \underline{\Omega}_{0}^{\left(t\right)};\, t=1,...,T_{0}\right\}  & \mathrm{if}\,\, P>A_{0}\end{array}\right.\label{eq:}\end{equation}
($p=1,...,P$), $\mathcal{R}\left\{ \mathbf{G}\right\} $ being the
operator that randomly extracts one entry of $\mathbf{G}$ {[}i.e.,
$\mathcal{R}\left\{ \mathbf{G}\right\} =\underline{\Omega}^{\left(r\right)}\in\mathbf{G}$,
$r=\mathrm{rand}\left(1,...,G\right)${]};
\end{itemize}
\textbf{(}\textbf{\emph{2.}}\textbf{) Iterative Loop} ($i=1,...,I$)

\begin{itemize}
\item Generate a new solution $\underline{\Omega}_{i}^{\left(\mathcal{M}\right)}$
by applying the following sequence of evolutionary-based operations

\begin{itemize}
\item \emph{Selection} - Randomly pick one solution from the $i$-th \emph{PF},
$\underline{\Omega}_{i}^{\left(\mathcal{R}\right)}=\mathcal{R}\left\{ \mathbf{A}_{i}\right\} $,
and one non-dominated solution from the current population, $\underline{\Omega}_{i}^{\left(\mathcal{N}\right)}=\mathcal{R}\left\{ \mathcal{N}\left\{ \mathbf{P}_{i}\right\} \right\} $,
$\mathcal{N}\left\{ \mathbf{G}\right\} $ being, analogously to (\ref{eq:non-eps-dominated-subset}),
the operator that extracts all $G'\leq G$ (standard) non-dominated
(\ref{eq:standard-dominance}) solutions within a generic set $\mathbf{G}$
of $G$ solutions ($\mathbf{G}=\left\{ \underline{\Omega}^{\left(g\right)};\, g=1,...,G\right\} $);
\item \emph{Crossover} - Apply the simulated binary crossover (\emph{SBX})
recombination operator $\mathcal{X}\left\{ \,.\,\right\} $ \cite{Deb 1995}
to generate a new individual: $\underline{\Omega}_{i}^{\left(\mathcal{X}\right)}=\mathcal{X}\left\{ \underline{\Omega}_{i}^{\left(\mathcal{R}\right)},\,\underline{\Omega}_{i}^{\left(\mathcal{N}\right)}\right\} $;
\item \emph{Mutation} - Determine the final offspring $\underline{\Omega}_{i}^{\left(\mathcal{M}\right)}$
by applying the polynomial mutation operator $\mathcal{M}\left\{ \,.\,\right\} $
\cite{Deb 1996} to $\underline{\Omega}_{i}^{\left(\mathcal{X}\right)}$:
$\underline{\Omega}_{i}^{\left(\mathcal{M}\right)}=\mathcal{M}\left\{ \underline{\Omega}_{i}^{\left(\mathcal{X}\right)}\right\} $;
\end{itemize}
\item Input $\underline{\Omega}_{i}^{\left(\mathcal{M}\right)}$ into the
$\left(i-1\right)$-th \emph{SM}, \emph{}$\mathcal{F}_{i-1}\left\{ \,.\,\right\} $,
to let it predict the objectives vector, $\underline{\widetilde{\Phi}}_{i-1}\left(\underline{\Omega}_{i}^{\left(\mathcal{M}\right)}\right)$,
and the corresponding confidence levels, $\underline{\delta}_{i-1}\left(\underline{\Omega}_{i}^{\left(\mathcal{M}\right)}\right)$;
\item Verify whether $\underline{\Omega}_{i}^{\left(\mathcal{M}\right)}$
is $\varepsilon$-dominated by any of the $A_{i-1}$ solutions belonging
to the \emph{PF} $\mathbf{A}_{i-1}$ at the $\left(i-1\right)$-th
iteration. To reduce the computational cost, each $a$-th ($a=1,...,A_{i-1}$)
$\varepsilon$-dominance check (\ref{eq:1 - epsilon-dominance})(\ref{eq:2 - epsilon-dominance})
is performed by replacing the unknown actual objectives of $\underline{\Omega}_{i}^{\left(\mathcal{M}\right)}$
{[}i.e., $\underline{\Phi}\left(\underline{\Omega}_{i}^{\left(\mathcal{M}\right)}\right)${]}
with their predictions, $\underline{\widetilde{\Phi}}_{i-1}\left(\underline{\Omega}_{i}^{\left(\mathcal{M}\right)}\right)$.
Finally, if $\underline{\Omega}_{i}^{\left(\mathcal{M}\right)}$ is
non-$\varepsilon$-dominated (i.e., $\nexists a\in\left\{ 1,...,A_{i-1}\right\} $
such that $\underline{\Omega}_{i-1}^{\left(a\right)}\prec_{\varepsilon}\underline{\Omega}_{i}^{\left(\mathcal{M}\right)}$),
then compute $\underline{\Phi}\left(\underline{\Omega}_{i}^{\left(\mathcal{M}\right)}\right)$
with the \emph{FW} solver, else let $\mathcal{F}_{i}\left\{ \,.\,\right\} \leftarrow\mathcal{F}_{i-1}\left\{ \,.\,\right\} $,
$T_{i}\leftarrow T_{i-1}$, $\mathbf{T}_{i}\leftarrow\mathbf{T}_{i-1}$,
and jump (avoid) the following step concerned with the \emph{RL};
\item Reinforce the training set $\mathbf{T}_{i-1}$ by adding the new I/O
training sample $\left\{ \underline{\Omega}_{i}^{\left(\mathcal{M}\right)},\,\underline{\Phi}\left(\underline{\Omega}_{i}^{\left(\mathcal{M}\right)}\right)\right\} $
(i.e., $\mathbf{T}_{i}=\mathbf{T}_{i-1}\bigcup\left\{ \underline{\Omega}_{i}^{\left(\mathcal{M}\right)},\,\underline{\Phi}\left(\underline{\Omega}_{i}^{\left(\mathcal{M}\right)}\right)\right\} $)
and increase the indexes $T_{RL}$ and $T_{i}$ {[}i.e., $T_{RL}\leftarrow\left(T_{RL}+1\right)$
and $T_{i}\leftarrow\left(T_{i-1}+1\right)${]}. Update the \emph{SM},
$\mathcal{F}_{i}\left\{ \underline{\Omega}\right\} $, with $\mathbf{T}_{i}$;
\item Update the \emph{PF} ($\mathbf{A}_{i}\leftarrow\mathbf{A}_{i-1}$)
by applying the operator $\mathcal{N}_{\varepsilon}$ to the set $\left\{ \underline{\Omega}_{i}^{\left(\mathcal{M}\right)},\,\mathbf{A}_{i-1}\right\} $
(i.e., $\mathbf{A}_{i}=\mathcal{N}_{\varepsilon}\left\{ \underline{\Omega}_{i}^{\left(\mathcal{M}\right)},\,\mathbf{A}_{i-1}\right\} $);
\item Derive from $\mathbf{P}_{i-1}$ the two complementary subsets $\mathbf{P}_{i-1}^{pred}$
and $\mathbf{P}_{i-1}^{sim}$ ($\mathbf{P}_{i-1}=\mathbf{P}_{i-1}^{pred}\bigcup\mathbf{P}_{i-1}^{sim}$)
that contain all the individuals with predicted or simulated objectives,
respectively. By exploiting the \emph{SbD} updating rules (Tab. II),
decide whether the new individual $\underline{\Omega}_{i}^{\left(\mathcal{M}\right)}$
should be included in the new population $\mathbf{P}_{i}$, by substituting
one properly-chosen individual of $\mathbf{P}_{i-1}$, or discarded
(i.e., $\mathbf{P}_{i}\leftarrow\mathbf{P}_{i-1}$ - Tab. II) according
to the \emph{SbD}-dominance rules in Tab. I;
\item Stop the iterative process and set $i=I_{SbD}$ if $\mathbf{A}_{i}$
fulfils the stationarity condition\begin{equation}
\sqrt{\frac{1}{W}\sum_{w=0}^{W-1}\left[\Gamma\left(\mathbf{A}_{i-w}\right)-\overline{\Gamma}_{i}\right]^{2}}\leq\gamma\label{eq:PF-Stationarity}\end{equation}
over a window of $W$ iterations, $\gamma$, $\Gamma\left(\mathbf{A}_{i-w}\right)$,
and $\overline{\Gamma}_{i}$ being a user-defined threshold, the maximum
crowding distance at the $\left(i-w\right)$-th iteration, and its
average over the iteration window, respectively (see Appendix I) \cite{Roudenko 2004}
or if the maximum number of allowed \emph{FW} evaluations has been
reached (i.e., $T_{RL}=T_{RL}^{\max}$). Otherwise, let $i\leftarrow\left(i+1\right)$
and repeat the iterative loop;
\item Output the \emph{SbD}-optimal \emph{PF} by setting $\mathbf{A}^{SbD}=\left.\mathbf{A}_{i}\right\rfloor _{i=I_{SbD}}$.
\end{itemize}
Since $C_{FW}^{SbD}=\left(T_{0}+T_{RL}\right)$ and being $T_{RL}\leq T_{RL}^{\max}\ll I$,
the time saving of the \emph{SbD} over the \emph{StD} is given by%
\footnote{\noindent Let us assume that in practical scenarios the time required
to train, $\Delta t_{train}$, and test, $\Delta t_{test}$, the \emph{SM}
is negligible since $\left(\Delta t_{train},\Delta t_{test}\right)\ll\Delta t_{FW}$
\cite{Massa 2022}.%
}\begin{equation}
\Delta t=\frac{\Delta t^{StD}-\Delta t^{SbD}}{\Delta t^{StD}}\approx\frac{\left(P+I\right)-\left(T_{0}+T_{RL}\right)}{\left(P+I\right)}.\label{eq:time-saving}\end{equation}
Such a result indicates that the \emph{SbD} is advantageous over the
\emph{StD} when the condition $\left(T_{0}+T_{RL}\right)\ll\left(P+I\right)$
holds true.

\subsection{\emph{MO-SbD} as Applied to \emph{mm}-Wave Automotive Radar Antenna
Design \label{sub:Application-to-mm-Wave}}

\noindent The \emph{MO-SbD} method is then applied to the synthesis
of \emph{mm}-wave automotive radar antennas. More in detail, the \emph{MOP}
at hand is that of designing series-fed radiators lying on the $\left(x,\, y\right)$-plane
(Fig. 5) and complying with the following (conflicting) goals: (\emph{i})
a proper input impedance matching over the frequency bandwidth $\left[f_{\min},\, f_{\max}\right]$,
(\emph{ii}) a good suppression of the sidelobes on the elevation ($\phi=90$
{[}deg{]} - Fig. 5) plane to avoid interferences from/towards the
sky (i.e., $\theta>0$ {[}deg{]}) and the road (i.e., $\theta<0$
{[}deg{]}), and (\emph{iii}) a good beam pointing stability towards
the broadside direction ($\theta_{0}=0$ {[}deg{]}) within the operative
frequency band \cite{Hash 2012}. Mathematically, these requirements
are coded into $Q=3$ objectives {[}i.e., $\Phi_{1}\left(\underline{\Omega}\right)=\Phi_{S_{11}}\left(\underline{\Omega}\right)$,
$\Phi_{2}\left(\underline{\Omega}\right)=\Phi_{SLL}\left(\underline{\Omega}\right)$,
and $\Phi_{3}\left(\underline{\Omega}\right)=\Phi_{BDD}\left(\underline{\Omega}\right)${]}.

\noindent As for the resonance objective (\emph{i}), let us define
the following ($q=1$)-th cost function\begin{equation}
\Phi_{S_{11}}\left(\underline{\Omega}\right)\triangleq\frac{1}{B}\sum_{b=1}^{B}\frac{\left|S_{11}\left(\left.f_{b}\right|\underline{\Omega}\right)\right|-S_{11}^{th}}{\left|S_{11}^{th}\right|}\times\mathcal{H}\left\{ \left|S_{11}\left(\left.f_{b}\right|\underline{\Omega}\right)\right|-S_{11}^{th}\right\} \label{eq:PHI_S11}\end{equation}
where $f_{b}=f_{\min}+\left(b-1\right)\times\frac{\left(f_{\max}-f_{\min}\right)}{\left(B-1\right)}$
is the $b$-th ($b=1,...,B$) frequency sample ($f_{b}\in\left[f_{\min},\, f_{\max}\right]$),
$S_{11}$ is the input reflection coefficient, $S_{11}^{th}$ is the
user-defined target threshold, and $\mathcal{H}\left\{ \,.\,\right\} $
is the Heaviside function ($\mathcal{H}\left\{ \alpha\right\} =1$
if $\alpha>0$, $\mathcal{H}\left\{ \alpha\right\} =0$ otherwise). 

\noindent Moreover, the two requirements on beam pattern shaping (\emph{ii})
and (\emph{iii}) are formulated as\begin{equation}
\Phi_{SLL}\left(\underline{\Omega}\right)=\frac{1}{B}\sum_{b=1}^{B}\frac{SLL\left(\left.f_{b}\right|\underline{\Omega}\right)-SLL^{th}}{\left|SLL^{th}\right|}\times\mathcal{H}\left\{ SLL\left(\left.f_{b}\right|\underline{\Omega}\right)-SLL^{th}\right\} \label{eq:PHI_SLL}\end{equation}
and\begin{equation}
\Phi_{BDD}\left(\underline{\Omega}\right)=\frac{1}{B}\sum_{b=1}^{B}\frac{\left|BDD\left(\left.f_{b}\right|\underline{\Omega}\right)\right|-BDD^{th}}{BDD^{th}}\times\mathcal{H}\left\{ \left|BDD\left(\left.f_{b}\right|\underline{\Omega}\right)\right|-BDD^{th}\right\} ,\label{eq:PHI_BDD}\end{equation}
respectively. In (\ref{eq:PHI_SLL}), $SLL$ and $SLL^{th}$ are the
simulated and the target sidelobe levels, while the beam direction
deviation (\emph{BDD}) cost term (\ref{eq:PHI_BDD}) is a function
of the pointing direction, $BDD\left(\left.f_{b}\right|\underline{\Omega}\right)\triangleq\arg\left\{ \max_{\theta}\mathcal{P}\left(\theta,\,\phi=90\right)\right\} $,
$\mathcal{P}\left(\theta,\phi\right)$ being the power pattern, and
the maximum allowed deviation $BDD^{th}$.

\section{Numerical Results\label{sec:4 - Numerical-Results}}

The aim of this Section is two-fold. On the one hand, it deals with
a preliminary assessment of the \emph{MO-SbD} method against \emph{MOP}
benchmark functions with analytically-known \emph{PF}s (Sect. \ref{sub:4.1 - Sensitivity-Analysis})
to also discuss the sensitivity of the optimization performance on
its key calibration parameters by providing useful guidelines for
the optimal setup. On the other hand, the scope of this section is
that of assessing the suitability of such a \emph{SbD} implementation
to a real and challenging \emph{MO} synthesis task. Towards this end,
the design of automotive radar antennas operating in the 77 GHz band
\cite{Hash 2012} has been chosen (Sect. \ref{sub:4.2 - Design-of-77GHz-Automotive})
as a representative test case.

\subsection{Assessment against \emph{MOP} Benchmark Functions\label{sub:4.1 - Sensitivity-Analysis}}

\noindent In order to give the interested reader some insights on
the \emph{MO-SbD} working, effectiveness, and computational efficiency,
as well as to assess its dependence on the control parameters, a set
of numerical experiments, drawn from an extensive validation carried
out on \emph{MOP} benchmark functions, is presented hereinafter.

\noindent The first test case is concerned with the optimization of
the \emph{DTLZ1} function with $K=3$ \emph{DoF}s and $Q=2$ objectives
\cite{Deb 2001b}. The control parameters of both the \emph{StD}/\emph{SbD}
methods have been set according to the literature guidelines \cite{Deb 2003}\cite{Roudenko 2004}
as follows: $I=1.5\times10^{4}$, $P=15$, $W=40$, $\gamma=2\times10^{-2}$,
and $\varepsilon_{q}=2\times10^{-2}$ ($q=1,...,Q$). To provide a
quantitative measure of the {}``quality'' of a \emph{PF} $\mathbf{A}$,
the error index, defined in \cite{Deb 2003} as\begin{equation}
\xi\left(\mathbf{A}\right)=\frac{1}{A}\sum_{a=1}^{A}\min_{a'\in\left\{ 1,...,A^{act}\right\} }\left\Vert \underline{\Phi}\left(\underline{\Omega}^{\left(a\right)}\right)-\underline{\Phi}\left(\underline{\Omega}^{\left(a'\right)}\right)\right\Vert _{2}\label{eq:}\end{equation}
where $\left\Vert \,.\,\right\Vert $ is the $\ell_{2}$-norm and
$\mathbf{A}^{act}=\left\{ \underline{\Omega}^{\left(a'\right)};\, a'=1,...,A^{act}\gg A\right\} $
is the actual (analytically-computed and densely sampled) \emph{PF},
has been evaluated.

\noindent The first analysis investigates on the dependence of the
optimization performance on the number of \emph{FW} calls %
\footnote{\noindent Dealing with the benchmark functions, a \emph{FW} call means
an exact computation of the objectives associated to a guess solution.%
} for training and updating the \emph{SM} while evolving the \emph{PF}
towards $\mathbf{A}^{SbD}$. Accordingly, the sensitivity analysis
has considered a variation of the number of initial training samples,
$T_{0}$, and of the maximum number of allowed \emph{RL} calls, $T_{RL}^{\max}$
(Sect. \ref{sub:SbD-MOP-Solution-Methodology}). It turns out that
the setup $T_{0}=\left(10\times K\right)$ (i.e., $T_{0}=30$) and
$T_{RL}^{\max}=\left(0.2\times I\right)$ (i.e., $T_{RL}^{\max}=3.0\times10^{3}$)
is a suitable compromise between a faithful approximation of the \emph{PF}
(i.e., $\left.\xi\left(\mathbf{A}^{SbD}\right)\right|_{T_{0}/K=10}^{T_{RL}^{\max}/I=0.2}\approx10^{-2}$)
and a significant time saving with respect to the \emph{StD} method
(i.e., $\left.\Delta t\right|_{T_{0}/K=10}^{T_{RL}^{\max}/I=0.2}\approx80$
\%). For completeness, the behavior of $\xi\left(\mathbf{A}^{SbD}\right)$
and $\Delta t$ versus $\frac{T_{0}}{K}$ and $\frac{T_{RL}^{\max}}{I}$
is shown in Fig. 6(\emph{a}) ($T_{RL}^{\max}/I=0.2$) and Fig. 6(\emph{b})
($T_{0}/K=10$), respectively. As expected, an increase of $T_{0}$
and/or $T_{RL}^{\max}$ yields a better approximation of the \emph{PF},
the values of $\xi\left(\mathbf{A}^{SbD}\right)$ being smaller because
of the higher accuracy of the \emph{SM} predictions thanks to the
larger set of \emph{I/O} training samples. However, the {}``price
to pay'' is that a greater number of \emph{FW} calls implies a lower
time saving (Fig. 6). To have a more immediate understanding of the
positive effects in choosing the selected $\left(T_{0},\, T_{RL}^{\max}\right)$-setup,
Figure 7(\emph{a}) shows that the \emph{SbD} converges to a \emph{PF},
$\mathbf{A}^{SbD}$, that carefully approximates the actual one, $\mathbf{A}^{act}$,
and it is very close to that found by the \emph{StD} method, $\mathbf{A}^{StD}$
{[}$\xi\left(\mathbf{A}^{StD}\right)=4.2\times10^{-4}$ - Fig. 7(\emph{a}){]}.
Interestingly, the solutions of the initial training set, $\mathbf{T}_{0}$,
as well as those of the initial \emph{PF}, $\mathbf{A}_{0}^{SbD}$,
are both very far from the optimal \emph{PF} {[}i.e., $\xi\left(\mathbf{A}_{0}^{SbD}\right)=3.1\times10^{1}$
- Fig. 7(\emph{a}){]}. This points out that there has been a remarkable
evolution of the \emph{PF} throughout the iterations. At the same
time, it is evident the key-role of the \emph{RL} strategy for updating
the \emph{SM} at each \emph{FW} evaluation carried out during the
\emph{MOP} solution. As a matter of fact, the \emph{PF} that would
be determined by the \emph{SbD}-based method without adaptively enhancing
the accuracy of the \emph{SM} is very poor as established by the corresponding
error value {[}i.e., $\left.\xi\left(\mathbf{A}^{SbD}\right)\right|_{\mathrm{{w/o\, RL}}}=2.9\times10^{1}$
- Fig. 7(\emph{a}){]}. As for the number of \emph{FW} calls during
the optimization, Figure 7(\emph{b}) gives the representative points
in the plane \{$\Phi_{1}\left(\underline{\Omega}\right)$, $\Phi_{2}\left(\underline{\Omega}\right)$\}
of all \emph{FW}-evaluated trial solutions by both the \emph{SbD}
and the \emph{StD} methods. As it can be inferred, the \emph{SbD}
enables a significant reduction of the computational cost, the total
number of evaluations being $C_{FW}^{SbD}=3015$ versus $C_{FW}^{StD}=15015$.

\noindent Similar outcomes arise when increasing the complexity of
the \emph{MOP} at hand, as well. For instance, the ability of the
\emph{SbD} to converge to the actual \emph{PF} is confirmed also for
a wider solution space with $K=7$ variables, $\xi\left(\mathbf{A}^{SbD}\right)=6.85\times10^{-3}$
($\approx$ $\xi\left(\mathbf{A}^{StD}\right)=3.61\times10^{-3}$)
being the value of the approximation error (Fig. 8). Moreover, since
$P=35$, $I=2.0\times10^{4}$ \cite{Deb 2003}, $T_{0}=70$, $T_{RL}=3584$
($T_{RL}^{\max}=4\times10^{3}$), the time saving amounts to $\Delta t\simeq82\%$.

\noindent The last test case of this Section refers to an even more
challenging scenario with $Q=3$ objectives. Once again, there is
a good agreement between $\mathbf{A}^{SbD}$ and $\mathbf{A}^{act}$
(Fig. 9) (i.e., $\xi\left(\mathbf{A}^{SbD}\right)=5.86\times10^{-3}$
close to $\xi\left(\mathbf{A}^{StD}\right)=5.53\times10^{-3}$) and
the time saving is kept close to $\Delta t\simeq80\%$.

\subsection{Assessment in Designing 77 GHz Automotive Radar Antennas\label{sub:4.2 - Design-of-77GHz-Automotive}}

\noindent The suitability of the proposed \emph{SbD} method to deal
with high-dimension \emph{MOP}s as well as the robustness of the setup
of its control parameters derived in Sect. \ref{sub:4.1 - Sensitivity-Analysis}
are then assessed by addressing the computationally-expensive design
of automotive radar antennas operating in the 77 {[}GHz{]} band.

\noindent The first numerical experiment is concerned with the synthesis
of a slotted substrate integrated waveguide (\emph{SIW}) antenna,
whose parametric model is sketched in Fig. 10. Such a radiating structure
has been modeled and simulated ($\Delta t_{FW}=580$ {[}sec{]}) with
the Ansys HFSS \emph{FW} solver \cite{HFSS 2021} by considering a
Rogers RT/duroid 6002 substrate (relative permittivity $\epsilon_{r}=2.94$
and loss tangent $\tan\delta=0.0012$) of thickness $h_{s}=1.27\times10^{-1}$
{[}mm{]}. It comprises $N=6$ slots etched over a metallic copper
film of thickness $h_{c}=35$ {[}$\mu$m{]} and roughness $\rho_{c}=1$
{[}$\mu$m{]}. The width $w_{1}$ of the input section of the tapered
microstrip line, which is used to feed the \emph{SIW}, has been chosen
to yield a characteristic impedance of $Z_{in}=50$ {[}$\Omega${]}.
Moreover, the diameter of the vias, their spacing, and the width of
the \emph{SIW} have been set by following the guidelines in the state-of-the-art
literature to excite the dominant mode $TE_{10}$ (i.e., $d_{via}=2.03\times10^{-1}$
{[}mm{]}, $s_{via}=4.06\times10^{-1}${[}mm{]}, and $w_{SIW}=1.85$
{[}mm{]} - Fig. 10) \cite{Deslandes 2006}.

\noindent The arising \emph{MOP} has been defined with the $Q=3$
objectives detailed in Sect. \ref{sub:Application-to-mm-Wave} by
assuming $f_{\min}=76$ {[}GHz{]}, $f_{\max}=79$ {[}GHz{]}, $B=7$
, $S_{11}^{th}=-15$ {[}dB{]}, $SLL^{th}=-20$ {[}dB{]}, and $BDD^{th}=0.25$
{[}deg{]}. As for the dimension of the solution space, $K=10$ geometrical
descriptors, $\underline{\Omega}=\left\{ \Omega_{k};\, k=1,...,K\right\} $,
have been optimized (Fig. 10) by considering the following assignment:
$\left(\Omega_{1},\,\Omega_{2}\right)=\left(l_{1},\, w_{2}\right)$,
$l_{1}$ and $w_{2}$ being the \emph{DoF}s that control the \emph{SIW}
input section; $\left(\Omega_{3},\,\Omega_{4}\right)=\left(s_{y,1},\, s_{y,2}\right)$,
$s_{y,1}$ and $s_{y,2}$ being the initial and the final offsets;
$\left(\Omega_{5},\,\Omega_{6},\,\Omega_{7}\right)=\left(s_{s},\, s_{l},\, s_{w}\right)$,
$s_{s}$, $s_{l}$, and $s_{w}$ being the spacing, the length, and
the width of the slot, respectively; $\left(\Omega_{8},\,\Omega_{9},\,\Omega_{10}\right)=\left(s_{x,1},\, s_{x,2},\, s_{x,3}\right)$,
$s_{x,1}$, $s_{x,2}$, and $s_{x,3}$ being the offset from the longitudinal
axis of the first $\frac{N}{2}$ slots (i.e., $n=1,...,\frac{N}{2}$),
while the other ones (i.e., $n=\left(\frac{N}{2}+1\right),...,N$)
are yielded with a mirroring operation to keep the geometric/electric
symmetry of the structure.

\noindent Figure 11(\emph{a}) shows the \emph{SbD}-optimized solutions
belonging to the final ($i=I$) \emph{PF} $\mathbf{A}^{SbD}$ {[}$P=50$,
$T_{0}=100$, $T_{RL}=2.05\times10^{3}$ ($T_{RL}^{\max}=4.0\times10^{3}$){]}
that comprises $A^{SbD}=8$ trade-off \emph{SIW} designs. The \emph{PF}
solutions that globally minimize each objective {[}i.e., $\underline{\Omega}_{q}^{SbD}$
($q=1,...,Q$) being $\underline{\Omega}_{q}^{SbD}=\arg$ $\left\{ \min_{\underline{\Omega}\in\mathbf{A}^{SbD}}\Phi_{q}\left(\underline{\Omega}\right)\right\} ${]}
are reported, as well. For comparison purposes, the same plot for
the \emph{StD} is shown in Fig. 11(\emph{b}), as well.

\noindent Figure 12 details the performance of $\underline{\Omega}_{q}^{SbD}$
($q=1,...,Q$) in terms of bandwidth {[}Fig. 12(\emph{a}){]}, \emph{SLL}
{[}Fig. 12(\emph{b}){]}, and \emph{BDD} {[}Fig. 12(\emph{c}){]}. As
expected, the solution $\underline{\Omega}_{S_{11}}^{SbD}$ is the
unique solution fully compliant with the $S_{11}$ requirement since
$\left|S_{11}\left(\left.f\right|\underline{\Omega}_{S_{11}}^{SbD}\right)\right|_{dB}\leq S_{11}^{th}$
whatever $f\in\left[f_{\min},\, f_{\max}\right]$ {[}Fig. 12(\emph{a}){]},
but it corresponds to a sub-optimal \emph{SLL} {[}i.e., $SLL\left(\left.f\right|\underline{\Omega}_{S_{11}}^{SbD}\right)\leq-18.9$
{[}dB{]} - Fig. 12(\emph{b}){]} and \emph{BDD} {[}i.e., $\left|BDD\left(\left.f\right|\underline{\Omega}_{S_{11}}^{SbD}\right)\right|\leq1$
{[}deg{]} - Fig. 12(\emph{c}){]} performance. The other two trade-off
solutions present a similar, but complementary, behavior (Fig. 12).
For illustrative purposes, Figure 13 plots the normalized power pattern
along the $\phi=90$ cut at the central frequency (i.e., $f=77.5$
{[}GHz{]}). It turns out that the best \emph{SLL} value is $SLL\left(\left.f\right|\underline{\Omega}_{SLL}^{SbD}\right)\leq-21.8$
{[}dB{]} {[}Fig. 12(\emph{b}){]}, while the beam pointing is very
stable in the whole working band when $\underline{\Omega}=\underline{\Omega}_{BDD}^{SbD}$
{[}i.e., $BDD\left(\left.f\right|\underline{\Omega}_{BDD}^{SbD}\right)=0$
{[}deg{]} - Fig. 12(\emph{c}){]}.

\noindent By comparing the \emph{SbD}-optimized \emph{PF} with that
found by the \emph{StD} method {[}Fig. 11(\emph{a}) vs. Fig. 11(\emph{b}){]},
it turns out that the interpolation surfaces look very similar as
well as the distribution of the representative points, while the computational
costs for determining those solutions are very different, the time
saving being equal to $\Delta t\simeq89$ \% when using the \emph{MO-SbD}
instead of the \emph{StD} method. Let us now consider the best trade-off
solution, according to the {}``Minimum Manhattan Distance'' (\emph{MMD})
criterion \cite{Rocca 2022} ($\underline{\Omega}^{*}$ - see Appendix
II), found by the two methods (i.e., $\underline{\Omega}^{SbD}$ {[}Fig.
11(\emph{a}){]} and $\underline{\Omega}^{StD}$ {[}Fig. 11(\emph{b}){]}).
The frequency behavior of the $S_{11}$ {[}Fig. 14(\emph{a}){]}, the
\emph{SLL} {[}Fig. 14(\emph{b}){]}, and the \emph{BDD} {[}Fig. 14(\emph{c}){]}
of these solutions is reported in Figure 14, while the corresponding
central frequency patterns are shown in Fig. 15(\emph{a}). As it can
be observed, the two solutions exhibit almost identical performance
thanks to the similarity of the antenna layouts {[}Fig. 15(\emph{b})
vs. Fig. 15(\emph{c}){]} also confirmed by the values of their geometric
descriptors (Tab. III).

\noindent To further assess the potentialities of the \emph{MO-SbD}
method in designing automotive radar antennas, the second experiment
deals with the synthesis of a perturbed travelling-wave open stubs
(\emph{PTOS}) comb-line array composed by $N=7$ series-fed radiating
elements {[}Fig. 16(\emph{a}){]} \cite{Zhang 2011}. The guideline
of the design is that of considering a different width for each microstrip
line connecting the stubs {[}i.e., $l_{w,1}\neq l_{w,2}\neq l_{w,3}\neq l_{w,4}$
- Fig. 16(\emph{a}){]} so that a better impedance matching can be
achieved along different sections of the antenna layout, which is
terminated on a matched load ($Z_{L}=50$ {[}$\Omega${]}). Moreover,
the stubs are allowed to have different widths {[}i.e., $o_{w,1}\neq o_{w,2}\neq o_{w,3}\neq o_{w,4}$
- Fig. 16(\emph{a}){]} to let them implement the desired current tapering
to lower the \emph{SLL} of a uniform structure \cite{Zhang 2011}
\footnote{\noindent Analogously to the \emph{SIW} design, the second half of
open stubs (i.e., $n=\left(\left\lceil \frac{N}{2}\right\rceil +1\right),...,N$)
are obtained from the first $\left\lceil \frac{N}{2}\right\rceil $
ones by means of a mirroring operation to enforce a better sidelobes
symmetry around the broadside {[}Fig. 16(\emph{a}){]}.%
}. Accordingly, the antenna layout has been parameterized with $K=11$
\emph{DoF}s, $\underline{\Omega}=\left\{ l_{l,1},\, l_{l,2},\, l_{w,1},\, l_{w,2},\, l_{w,3},\, l_{w,4},\, o_{w,1},\, o_{w,2},\, o_{w,3},\, o_{w,4},\, o_{l}\right\} $
{[}Fig. 16(\emph{a}){]}.

\noindent Figure 16(\emph{b}) shows the \emph{SbD}-synthesized \emph{PF}
solutions, while Figure 17 compares the performance indexes of $\underline{\Omega}_{S_{11}}^{SbD}$,
$\underline{\Omega}_{SLL}^{SbD}$, $\underline{\Omega}_{BDD}^{SbD}$,
and $\underline{\Omega}^{SbD}$ versus the frequency, $f\in\left[f_{\min},\, f_{\max}\right]$.
Generally speaking, the results confirm the effectiveness of the \emph{SbD}
method in efficiently (i.e., $\Delta t\simeq89\%$) synthesizing antenna
designs that fulfil contrasting requirements so that the designer
can select the most suitable layout for different specific applications.
The higher (with respect to the \emph{SIW}), but yet acceptable beam
squint {[}i.e., $\left|BDD\left(f\right)\right|\leq2$ {[}deg{]} -
Fig. 17(\emph{c}){]}, is expected and mainly due to the travelling-wave
structure \cite{Choi 2006} at hand. 

\noindent For completeness, representative radiation patterns at $f=77.5$
{[}GHz{]} are shown in Fig. 18(\emph{a}), while the \emph{HFSS} screen-shot
and the \emph{DoF}s of the optimal trade-off solution synthesized
with the \emph{SbD} are reported in Fig. 18(\emph{b}) and Tab. IV,
respectively.

\section{\noindent Conclusions and Final Remarks \label{sec:Conclusions-and-Final}}

\noindent A new computationally-efficient method for the \emph{MO}
design of complex \emph{EM} devices has been proposed by leveraging
on an effective integration of evolutionary optimization strategies
and \emph{AI} concepts. The arising \emph{MO-SbD} method allows one
an effective sampling of the solution space to provide the decision
maker a set of optimal trade-off solutions that fit conflicting requirements
(e.g., impedance matching and far-field radiation features in antenna
design).

\noindent The numerical validation, concerned with benchmark \emph{MOP}s
as well as real antennas for automotive radars, has outputted the
following main outcomes:

\begin{itemize}
\item the \emph{MO-SbD} method performs an effective sampling of high-dimensional
solution spaces being able to retrieve a \emph{PF} of compromise solutions
close to that found by the \emph{StD} approach and to the analytically-computed
one in the case of benchmark functions;
\item the huge time saving ($\Delta t\geq80\%$) of the \emph{MO-SbD} is
achieved thanks to the exploitation of \emph{AI-}based mechanisms
(i.e., \emph{SM} and \emph{RL}), while the \emph{StD} approach is
highly computationally-demanding since it is exclusively based on
\emph{FW} evaluations;
\item the \emph{MO-SbD} method proved to be a reliable, effective, and computationally
efficient tool for the \emph{MO} design of \emph{mm}-wave automotive
radar antennas working in the 77 {[}GHz{]} band either based on \emph{SIW}
or \emph{PTOS} architectures.
\end{itemize}
Future works, beyond the scope of this paper, will be aimed at exploiting
(\emph{i}) properly customized deep learning architectures to enable
an unprecedented prediction accuracy \cite{Massa 2019} as well as
(\emph{ii}) adaptive and/or Compressive sampling strategies \cite{Massa 2015}
to generate highly-informative and low-cardinality training sets.
Finally, the prototyping of the synthesized \emph{mm}-wave automotive
radiators as well as their experimental validation, besides the \emph{FW}
numerical assessment with \emph{HFSS} in this paper, are under investigation
by some industrial partners.

\section*{\noindent Acknowledgements}

\noindent This work has been partially supported by the Italian Ministry
of Education, University, and Research within the PRIN 2017 Program,
for the Project {}``Cloaking Metasurfaces for a New Generation of
Intelligent Antenna Systems (MANTLES)'' (Grant No. 2017BHFZKH - CUP:
E64I19000560001) and the Project \char`\"{}CYBER-PHYSICAL ELECTROMAGNETIC
VISION: Context-Aware Electromagnetic Sensing and Smart Reaction (EMvisioning)\char`\"{}
(Grant no. 2017HZJXSZ - CUP: E64I19002530001), within the Program
\char`\"{}Progetti di Ricerca Industriale e Sviluppo Sperimentale
nelle 12 aree di specializzazione individuate dal PNR 2015-2020\char`\"{},
Specialization Area \char`\"{}Smart Secure \& Inclusive Communities\char`\"{}
for the Project \char`\"{}Mitigazione dei rischi naturali per la sicurezza
e la mobilita' nelle aree montane del Mezzogiorno (MITIGO)\char`\"{}
(Grant no. ARS01\_00964), and within the Program \char`\"{}Smart cities
and communities and Social Innovation\char`\"{} for the Project \char`\"{}Piattaforma
Intelligente per il Turismo (SMARTOUR)\char`\"{} (Grant no. SCN\_00166
- CUP: E44G14000040008). Moreover, it benefited from the networking
activities carried out within the Project {}``SPEED'' (Grant No.
61721001) funded by National Science Foundation of China under the
Chang-Jiang Visiting Professorship Program. A. Massa wishes to thank
E. Vico for her never-ending inspiration, support, guidance, and help.

\section*{\noindent Appendix I}

\noindent In (\ref{eq:PF-Stationarity}), the maximum crowding distance
$\Gamma\left(\mathbf{A}_{i-w}\right)$ is defined as \cite{Roudenko 2004}\begin{equation}
\Gamma\left(\mathbf{A}_{i-w}\right)=\max_{a=2,...,\left(A_{i-w}-1\right)}\left\{ \sum_{q=1}^{Q}D_{q}\left(\underline{\Omega}_{i-w}^{\left(a\right)}\right)\right\} \label{eq:}\end{equation}
where the $q$-th ($q=1,...,Q$) distance is computed as\begin{equation}
D_{q}\left(\underline{\Omega}_{i-w}^{\left(a\right)}\right)=\Phi_{q}\left(\underline{\Omega}_{i-w}^{\left(a+1\right)}\right)-\Phi_{q}\left(\underline{\Omega}_{i-w}^{\left(a-1\right)}\right)\label{eq:}\end{equation}
by setting\begin{equation}
\begin{array}{c}
\left.\underline{\Omega}_{i-w}^{\left(a\right)}\right\rfloor _{a=1}=\arg\left\{ \min_{v=1,...,A}\left[\Phi_{q}\left(\underline{\Omega}_{i-w}^{\left(v\right)}\right)\right]\right\} \\
\left.\underline{\Omega}_{i-w}^{\left(a\right)}\right\rfloor _{a=A_{i-w}}=\arg\left\{ \max_{v=1,...,A}\left[\Phi_{q}\left(\underline{\Omega}_{i-w}^{\left(v\right)}\right)\right]\right\} \end{array}\label{eq:}\end{equation}
and ordering the remaining {[}$a=2,...,\left(A_{i-w}-1\right)${]}
\emph{PF} elements so that\begin{equation}
\Phi_{q}\left(\underline{\Omega}_{i-w}^{\left(a-1\right)}\right)\leq\Phi_{q}\left(\underline{\Omega}_{i-w}^{\left(a\right)}\right)\leq\Phi_{q}\left(\underline{\Omega}_{i-w}^{\left(a+1\right)}\right).\label{eq:}\end{equation}
Moreover, the average crowding distance over a window of $W$ iterations
is defined as \cite{Roudenko 2004}\begin{equation}
\overline{\Gamma}_{i}=\frac{1}{W}\sum_{w=0}^{W-1}\Gamma\left(\mathbf{A}_{i-w}\right).\label{eq:}\end{equation}

\section*{\noindent Appendix II}

\noindent According to the \emph{MMD} criterion, the best trade-off
solution, $\underline{\Omega}^{*}$, is given by \cite{Rocca 2022}\begin{equation}
\underline{\Omega}^{*}=\arg\left\{ \min_{a=1,...,A}\left\Vert \underline{\Phi}^{'}\left(\underline{\Omega}^{\left(a\right)}\right)-\underline{\Phi}^{ideal}\right\Vert _{1}\right\} \label{eq:}\end{equation}
where $\left\Vert \,.\,\right\Vert _{1}$ is the $\ell_{1}$-norm,
while the $q$-th ($q=1,...,Q$) entry of $\underline{\Phi}^{'}\left(\underline{\Omega}^{\left(a\right)}\right)=\left\{ \Phi_{q}^{'}\left(\underline{\Omega}^{\left(a\right)}\right);\, q=1,...,Q\right\} $
and $\underline{\Phi}^{ideal}=\left\{ \Phi_{q}^{ideal};\, q=1,...,Q\right\} $
is given by

\noindent \begin{equation}
\Phi_{q}^{'}\left(\underline{\Omega}^{\left(a\right)}\right)=\frac{\Phi_{q}\left(\underline{\Omega}^{\left(a\right)}\right)}{\max_{a=1,...,A}\left\{ \Phi_{q}\left(\underline{\Omega}^{\left(a\right)}\right)\right\} -\min_{a=1,...,A}\left\{ \Phi_{q}\left(\underline{\Omega}^{\left(a\right)}\right)\right\} }\label{eq:}\end{equation}
and\begin{equation}
\Phi_{q}^{ideal}=\min_{a=1,...,A}\left\{ \Phi_{q}^{'}\left(\underline{\Omega}^{\left(a\right)}\right)\right\} ,\label{eq:}\end{equation}
respectively.

\newpage
\section*{FIGURE CAPTIONS}

\begin{itemize}
\item \textbf{Figure 1.} \emph{Dominance}/$\varepsilon$\emph{-Dominance}
($Q=2$) - Illustrative example with the solutions $\underline{\Omega}^{\left(1\right)}$,
$\underline{\Omega}^{\left(2\right)}$, and $\underline{\Omega}^{\left(3\right)}$:
$\underline{\Omega}^{\left(1\right)}\prec\underline{\Omega}^{\left(3\right)}$,
$\underline{\Omega}^{\left(2\right)}\prec\underline{\Omega}^{\left(3\right)}$,
$\underline{\Omega}^{\left(1\right)}\nprec\underline{\Omega}^{\left(2\right)}$,
$\underline{\Omega}^{\left(2\right)}\nprec\underline{\Omega}^{\left(1\right)}$,
$\underline{\Omega}^{\left(1\right)}\prec_{\varepsilon}\underline{\Omega}^{\left(2\right)}$,
$\underline{\Omega}^{\left(1\right)}\prec_{\varepsilon}\underline{\Omega}^{\left(3\right)}$,
and $\underline{\Omega}^{\left(2\right)}\prec_{\varepsilon}\underline{\Omega}^{\left(3\right)}$.
\item \textbf{Figure 2.} Pictorial representation of the \textbf{}{}``confidence
hyper-volume'' of the prediction $\underline{\widetilde{\Phi}}\left(\underline{\Omega}\right)$.
\item \textbf{Figure 3.} \emph{SbD-Dominance} ($Q=2$) - Pictorial representation
of $\underline{\Omega}^{\left(1\right)}\prec_{sbD}\underline{\Omega}^{\left(2\right)}$
when (\emph{a}) $\underline{\widetilde{\Phi}}\left(\underline{\Omega}^{\left(1\right)}\right)$
and $\underline{\widetilde{\Phi}}\left(\underline{\Omega}^{\left(2\right)}\right)$,
(\emph{b}) $\underline{\Phi}\left(\underline{\Omega}^{\left(1\right)}\right)$
and $\underline{\widetilde{\Phi}}\left(\underline{\Omega}^{\left(2\right)}\right)$,
(\emph{c}) $\underline{\widetilde{\Phi}}\left(\underline{\Omega}^{\left(1\right)}\right)$
and $\underline{\Phi}\left(\underline{\Omega}^{\left(2\right)}\right)$,
(\emph{d}) $\underline{\Phi}\left(\underline{\Omega}^{\left(1\right)}\right)$
and $\underline{\Phi}\left(\underline{\Omega}^{\left(2\right)}\right)$.
\item \textbf{Figure 4.} \emph{MO-SbD Method} - Block diagram.
\item \textbf{Figure 5.} Sketch of the automotive radar scenario.
\item \textbf{Figure 6.} \emph{Sensitivity Analysis} (\emph{DTLZ1 Benchmark
Function}, $K=3$, $Q=2$ \cite{Deb 2001b}, $I=1.5\times10^{4}$,
$P=15$) - \textbf{}Behavior of the time saving, $\Delta t$, and
of the \emph{PF} approximation error, $\xi\left(\mathbf{A}^{SbD}\right)$,
versus (\emph{a}) $\frac{T_{0}}{K}$ ($\frac{T_{RL}^{\max}}{I}=0.2$)
and (\emph{b}) $\frac{T_{RL}^{\max}}{I}$ ($\frac{T_{0}}{K}=10$). 
\item \textbf{Figure 7.} \emph{Sensitivity Analysis} (\emph{DTLZ1 Benchmark
Function}, $K=3$, $Q=2$ \cite{Deb 2001b}, $I=1.5\times10^{4}$,
$P=15$, $T_{0}=30$, $T_{RL}^{\max}=3.0\times10^{3}$) - \textbf{}Representative
points in the ($\Phi_{1}$, $\Phi_{2}$)-plane of the \emph{MOP} solutions.
\item \textbf{Figure 8.} \emph{Sensitivity Analysis} (\emph{DTLZ1 Benchmark
Function}, $K=7$, $Q=2$ \cite{Deb 2001b}, $I=2.0\times10^{4}$,
$P=35$, $T_{0}=70$, $T_{RL}^{\max}=4.0\times10^{3}$) - \textbf{}Representative
points in the ($\Phi_{1}$, $\Phi_{2}$)-plane of the \emph{MOP} solutions.
\item \textbf{Figure 9.} \emph{Sensitivity Analysis} (\emph{DTLZ1 Benchmark
Function}, $K=7$, $Q=3$ \cite{Deb 2001b}, $I=2.0\times10^{4}$,
$P=35$, $T_{0}=70$, $T_{RL}^{\max}=4.0\times10^{3}$) - \textbf{}Representative
points in the ($\Phi_{1}$, $\Phi_{2}$, $\Phi_{3}$)-space of the
\emph{MOP} solutions.
\item \textbf{Figure 10.} \emph{Automotive Radar Antenna Design} (\emph{SIW
Radiator}, $K=10$, $Q=3$) - Sketch of the geometry of the antenna
model along with the \emph{DoF}s of the antenna synthesis problem.
\item \textbf{Figure 11.} \emph{Automotive Radar Antenna Design} (\emph{SIW
Radiator}, $K=10$, $Q=3$, $I=2.0\times10^{4}$, $P=50$, $T_{0}=100$,
$T_{RL}^{\max}=4.0\times10^{3}$) - Representative points in the ($\Phi_{1}$,
$\Phi_{2}$, $\Phi_{3}$)-space of (\emph{a}) the \emph{SbD} and (\emph{b})
the \emph{StD} \emph{PF}s solutions.
\item \textbf{Figure 12.} \emph{Automotive Radar Antenna Design} (\emph{SIW
Radiator}, $K=10$, $Q=3$, $I=2.0\times10^{4}$, $P=50$, $T_{0}=100$,
$T_{RL}^{\max}=4.0\times10^{3}$) - Behavior of (\emph{a}) the $S_{11}$,
(\emph{b}) the \emph{SLL}, and (\emph{c}) the \emph{BDD} versus the
frequency of \emph{MOP} solutions $\underline{\Omega}_{S_{11}}^{SbD}$,
$\underline{\Omega}_{SLL}^{SbD}$, and $\underline{\Omega}_{BDD}^{SbD}$.
\item \textbf{Figure 13.} \emph{Automotive Radar Antenna Design} (\emph{SIW
Radiator}, $K=10$, $Q=3$, $I=2.0\times10^{4}$, $P=50$, $T_{0}=100$,
$T_{RL}^{\max}=4.0\times10^{3}$) - Normalized elevation pattern,
$\mathcal{P}\left(\theta,\,\phi=90\right)$, radiated at $f=77.5$
{[}GHz{]} by the \emph{MOP} solutions $\underline{\Omega}_{S_{11}}^{SbD}$,
$\underline{\Omega}_{SLL}^{SbD}$, and $\underline{\Omega}_{BDD}^{SbD}$.
\item \textbf{Figure 14.} \emph{Automotive Radar Antenna Design} (\emph{SIW
Radiator}, $K=10$, $Q=3$, $I=2.0\times10^{4}$, $P=50$, $T_{0}=100$,
$T_{RL}^{\max}=4.0\times10^{3}$) - Behavior of the (\emph{a}) $S_{11}$,
(\emph{b}) \emph{SLL}, and (\emph{c}) \emph{BDD} versus frequency
for the compromise solutions found by the \emph{SbD}, $\underline{\Omega}^{SbD}$,
and the \emph{StD}, $\underline{\Omega}^{StD}$, methods.
\item \textbf{Figure 15.} \emph{Automotive Radar Antenna Design} (\emph{SIW
Radiator}, $K=10$, $Q=3$, $I=2.0\times10^{4}$, $P=50$, $T_{0}=100$,
$T_{RL}^{\max}=4.0\times10^{3}$) - Normalized elevation pattern,
$\mathcal{P}\left(\theta,\,\phi=90\right)$, radiated at $f=77.5$
{[}GHz{]} by (\emph{b}) the \emph{SbD} and (\emph{c}) the \emph{StD}
antenna layouts.
\item \textbf{Figure 16.} \emph{Automotive Radar Antenna Design} (\emph{PTOS
Radiator}, $K=11$, $Q=3$, $I=2.0\times10^{4}$, $P=55$, $T_{0}=110$,
$T_{RL}^{\max}=4.0\times10^{3}$) - (\emph{a}) Sketch of the geometry
of the antenna model along with the \emph{DoF}s of the antenna synthesis
problem and (\emph{b}) representative points in the ($\Phi_{1}$,
$\Phi_{2}$, $\Phi_{3}$)-space of the \emph{SbD} \emph{PF} solutions.
\item \textbf{Figure 17.} \emph{Automotive Radar Antenna Design} (\emph{PTOS
Radiator}, $K=11$, $Q=3$, $I=2.0\times10^{4}$, $P=55$, $T_{0}=110$,
$T_{RL}^{\max}=4.0\times10^{3}$) - Behavior of (\emph{a}) the $S_{11}$,
(\emph{b}) the \emph{SLL}, and (\emph{c}) the \emph{BDD} versus the
frequency of \emph{MO-SbD} representative solutions.
\item \textbf{Figure 18.} \emph{Automotive Radar Antenna Design} (\emph{PTOS
Radiator}, $K=11$, $Q=3$, $I=2.0\times10^{4}$, $P=55$, $T_{0}=110$,
$T_{RL}^{\max}=4.0\times10^{3}$) - (\emph{a}) Normalized elevation
pattern, $\mathcal{P}\left(\theta,\,\phi=90\right)$, radiated at
$f=77.5$ {[}GHz{]} by the \emph{MO-SbD} representative solutions
and (\emph{b}) layout of the \emph{SbD} antenna.
\end{itemize}

\section*{TABLE CAPTIONS}

\begin{itemize}
\item \textbf{Table I.} \emph{SbD-Dominance} ($Q=2$) - Rules for the condition
$\underline{\Omega}^{\left(1\right)}\prec_{sbD}\underline{\Omega}^{\left(2\right)}$.
\item \textbf{Table II.} \emph{MO-SbD} - Rules for the population update
($\mathbf{P}_{i-1}$ $\to$ $\mathbf{P}_{i}$) at the $i$-th ($i=1,...,I$)
iteration.
\item \textbf{Table III.} \emph{Automotive Radar Antenna Design} (\emph{SIW
Radiator}, $K=10$, $Q=3$, $I=2.0\times10^{4}$, $P=50$, $T_{0}=100$,
$T_{RL}^{\max}=4.0\times10^{3}$) - \emph{DoF}s values.
\item \textbf{Table IV.} \emph{Automotive Radar Antenna Design} (\emph{PTOS
Radiator}, $K=11$, $Q=3$, $I=2.0\times10^{4}$, $P=55$, $T_{0}=110$,
$T_{RL}^{\max}=4.0\times10^{3}$) - \emph{DoF}s values\emph{.}
\end{itemize}
\newpage
\begin{center}~\vfill\end{center}

\begin{center}\includegraphics[%
  width=0.90\columnwidth]{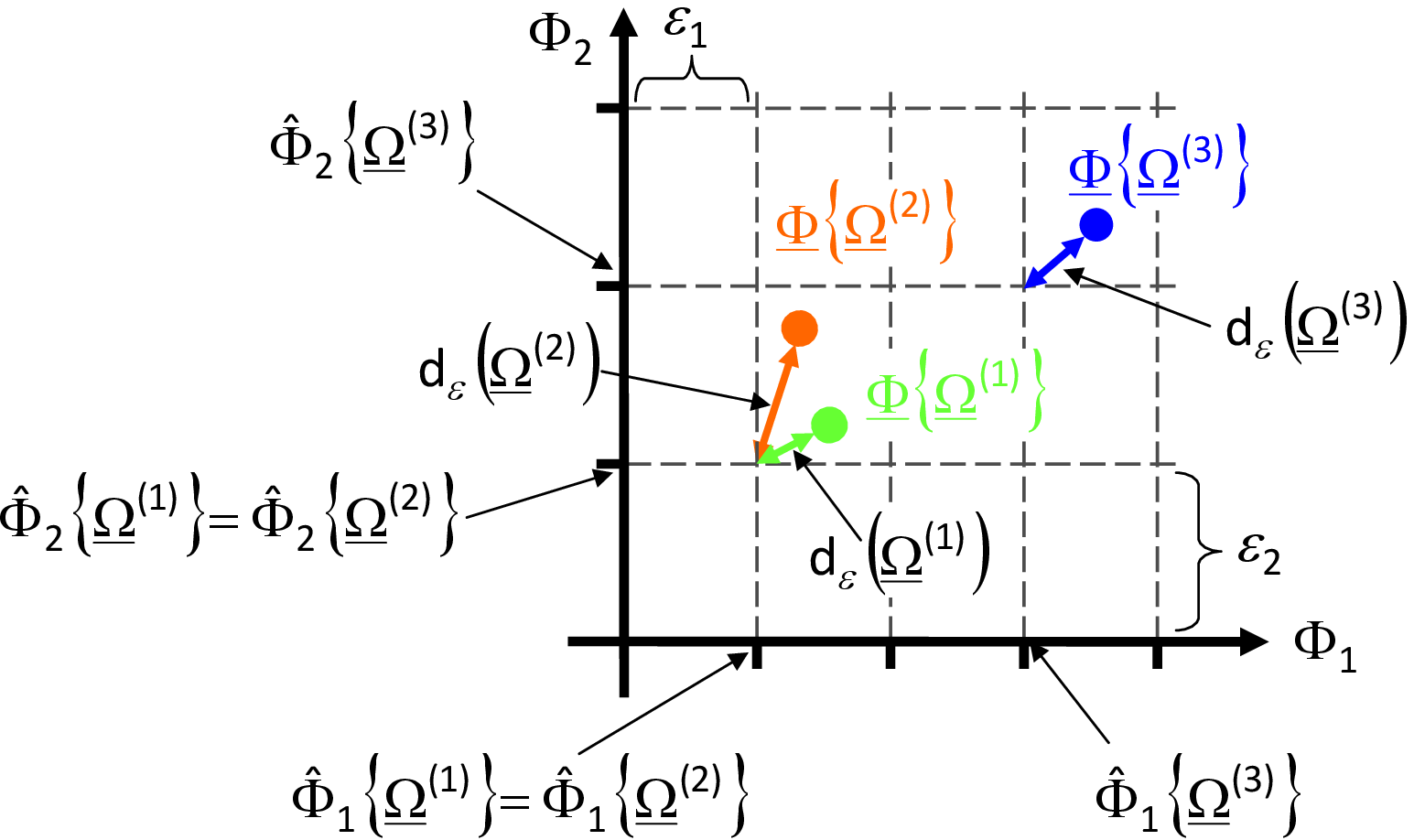}\end{center}

\begin{center}~\vfill\end{center}

\begin{center}\textbf{Fig. 1 - P. Rosatti et} \textbf{\emph{al.}}\textbf{,}
\textbf{\emph{{}``}}Multi-Objective System-by-Design for ...''\end{center}

\newpage
\begin{center}~\vfill\end{center}

\begin{center}\includegraphics[%
  width=0.75\columnwidth]{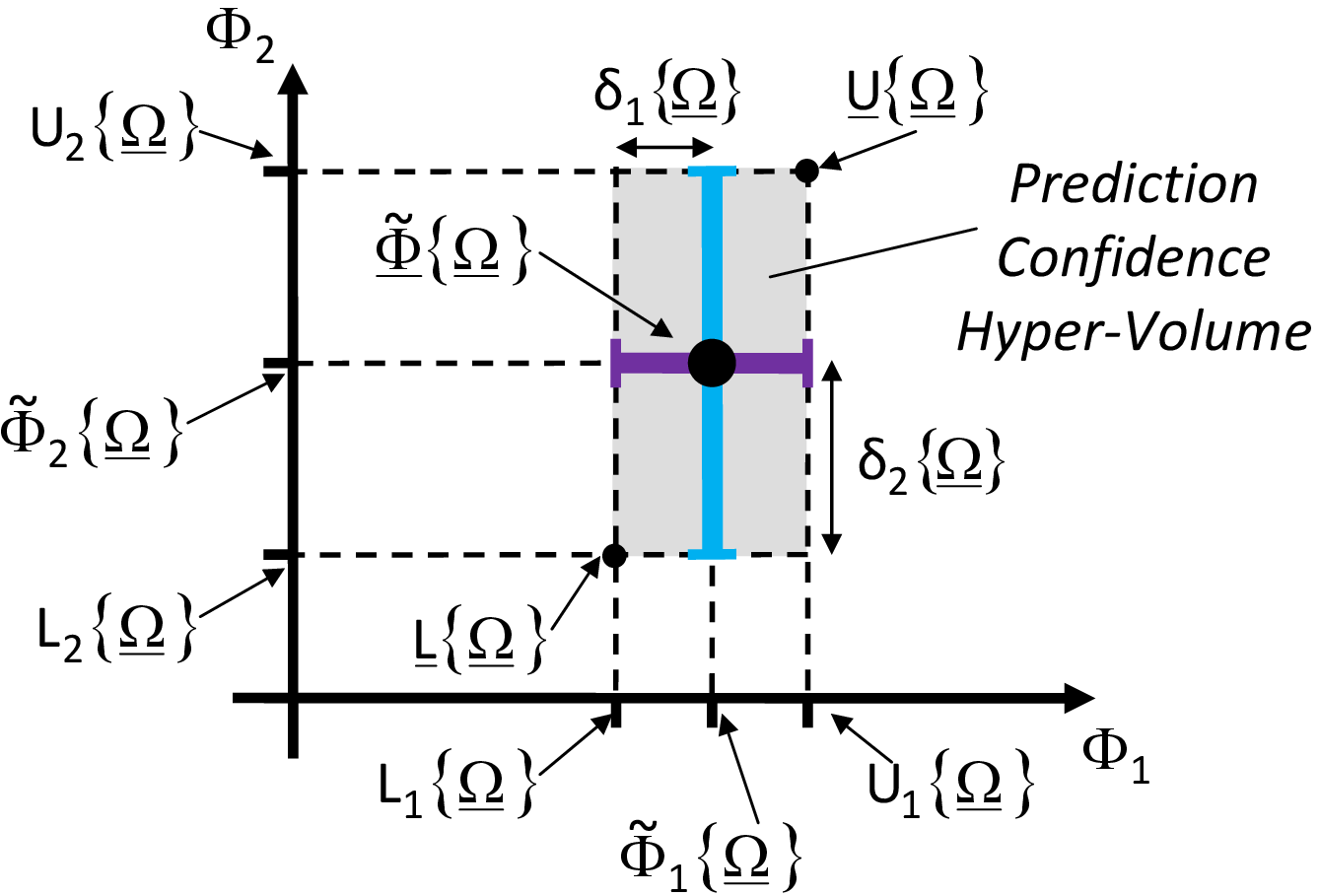}\end{center}

\begin{center}~\vfill\end{center}

\begin{center}\textbf{Fig. 2 - P. Rosatti et} \textbf{\emph{al.}}\textbf{,}
\textbf{\emph{{}``}}Multi-Objective System-by-Design for ...''\end{center}

\newpage
\begin{center}~\vfill\end{center}

\begin{center}\begin{tabular}{cc}
\includegraphics[%
  width=0.45\columnwidth]{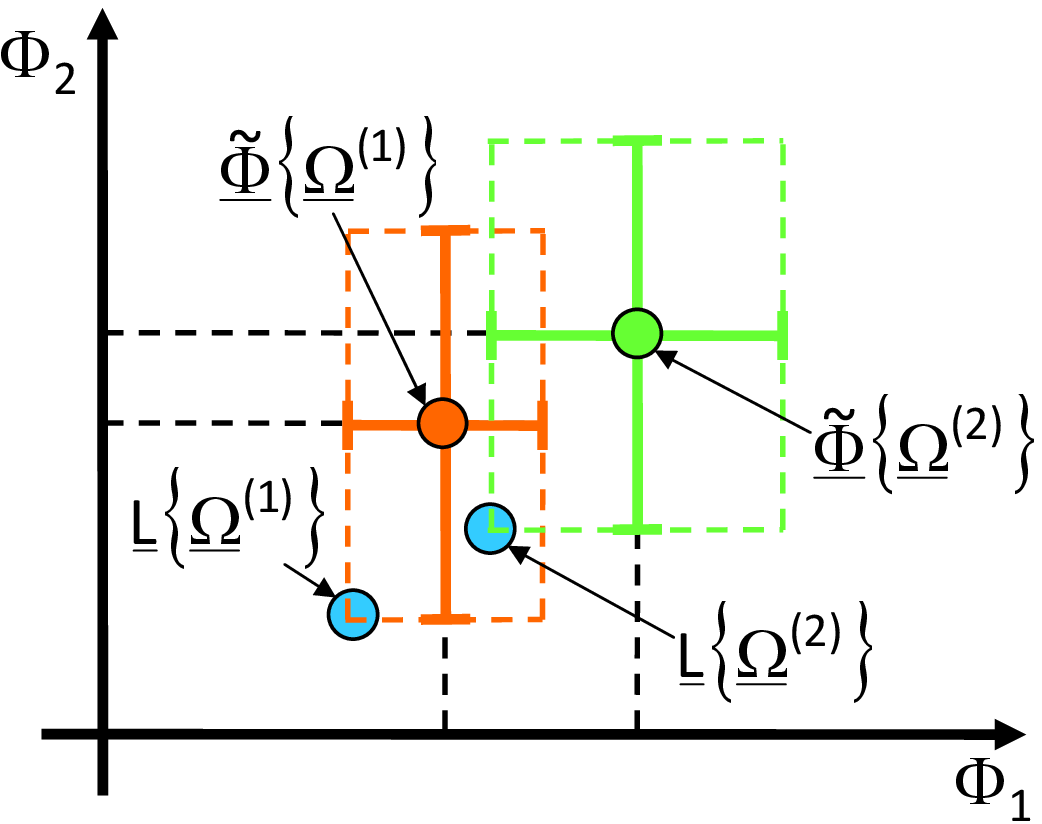}&
\includegraphics[%
  width=0.45\columnwidth]{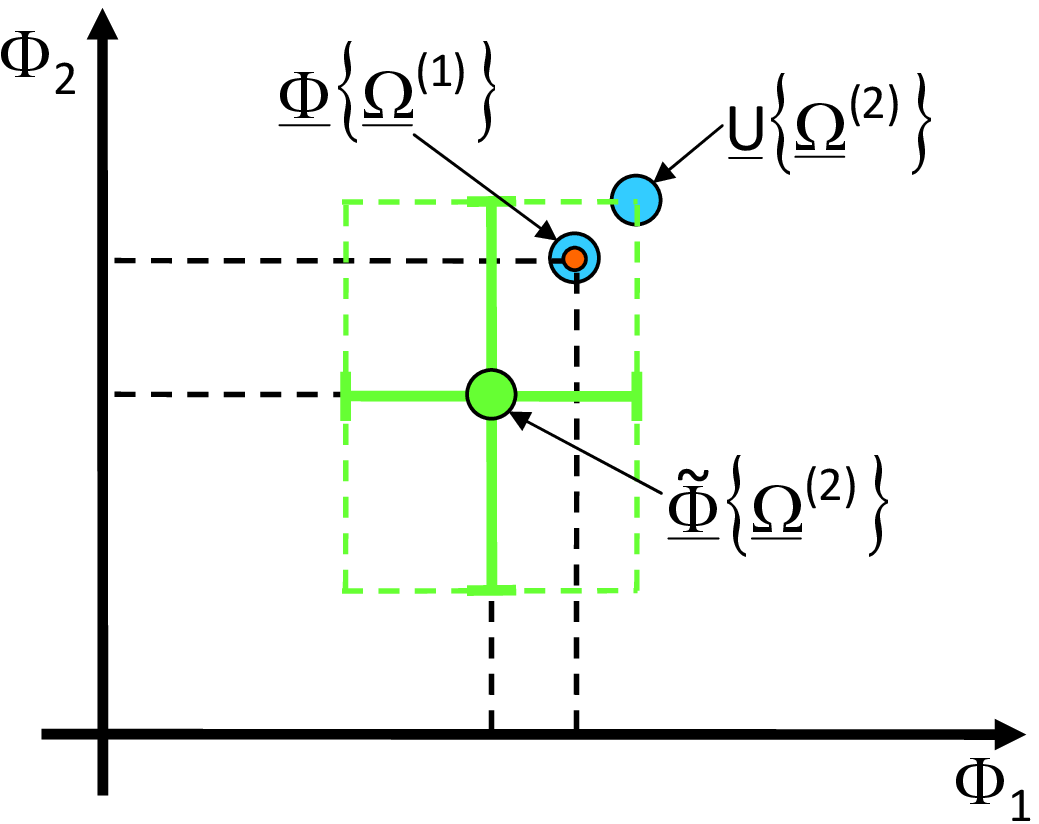}\tabularnewline
(\emph{a})&
(\emph{b})\tabularnewline
\includegraphics[%
  width=0.45\columnwidth]{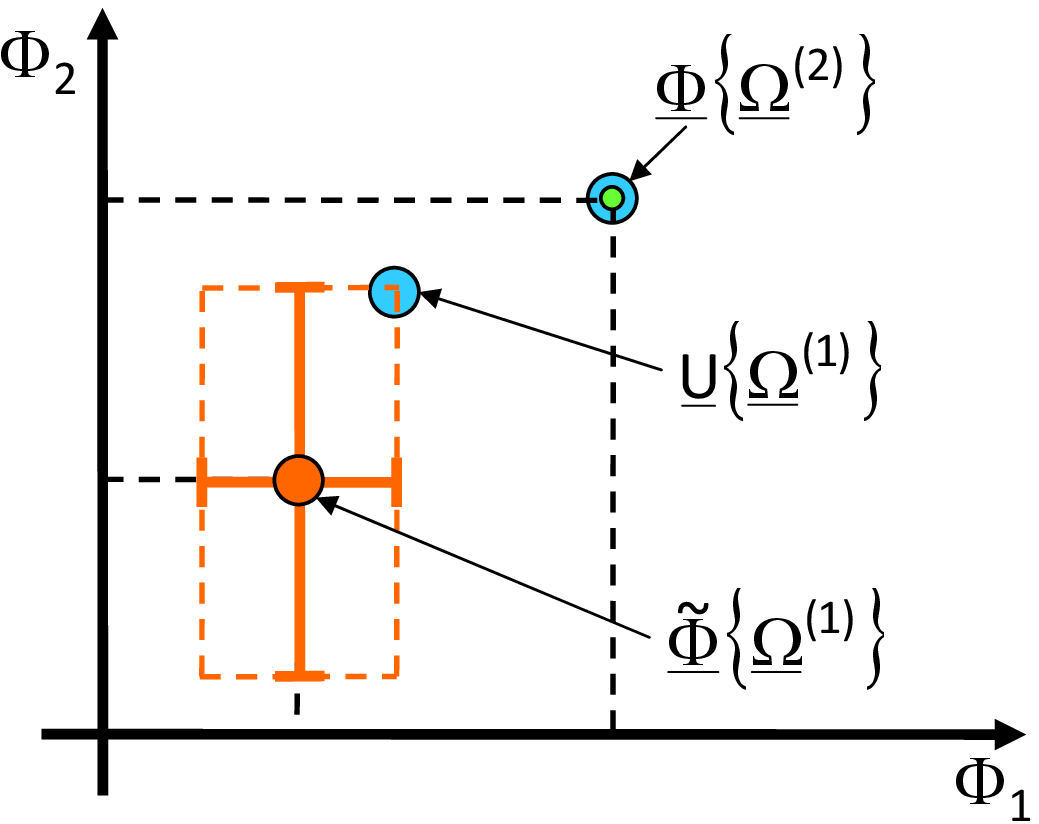}&
\includegraphics[%
  width=0.45\columnwidth]{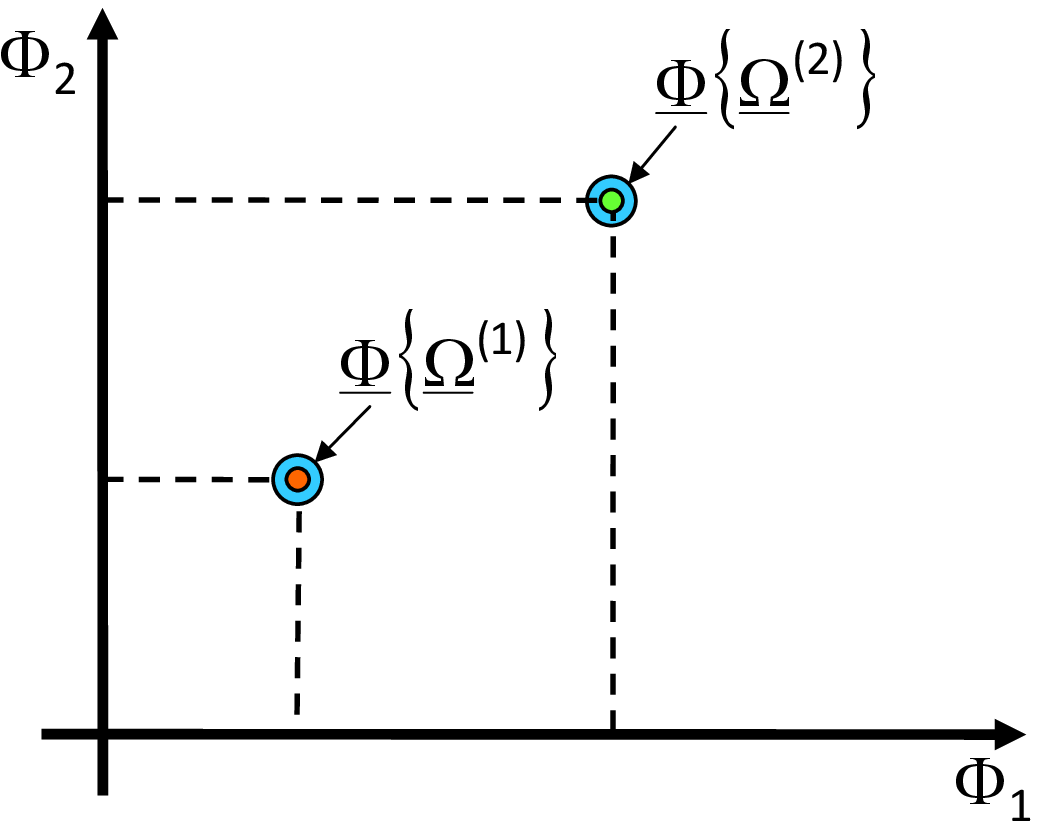}\tabularnewline
(\emph{c})&
(\emph{d})\tabularnewline
\end{tabular}\end{center}

\begin{center}~\vfill\end{center}

\begin{center}\textbf{Fig. 3 - P. Rosatti et} \textbf{\emph{al.}}\textbf{,}
\textbf{\emph{{}``}}Multi-Objective System-by-Design for ...''\end{center}

\newpage
\begin{center}~\vfill\end{center}

\begin{center}\includegraphics[%
  width=0.95\columnwidth]{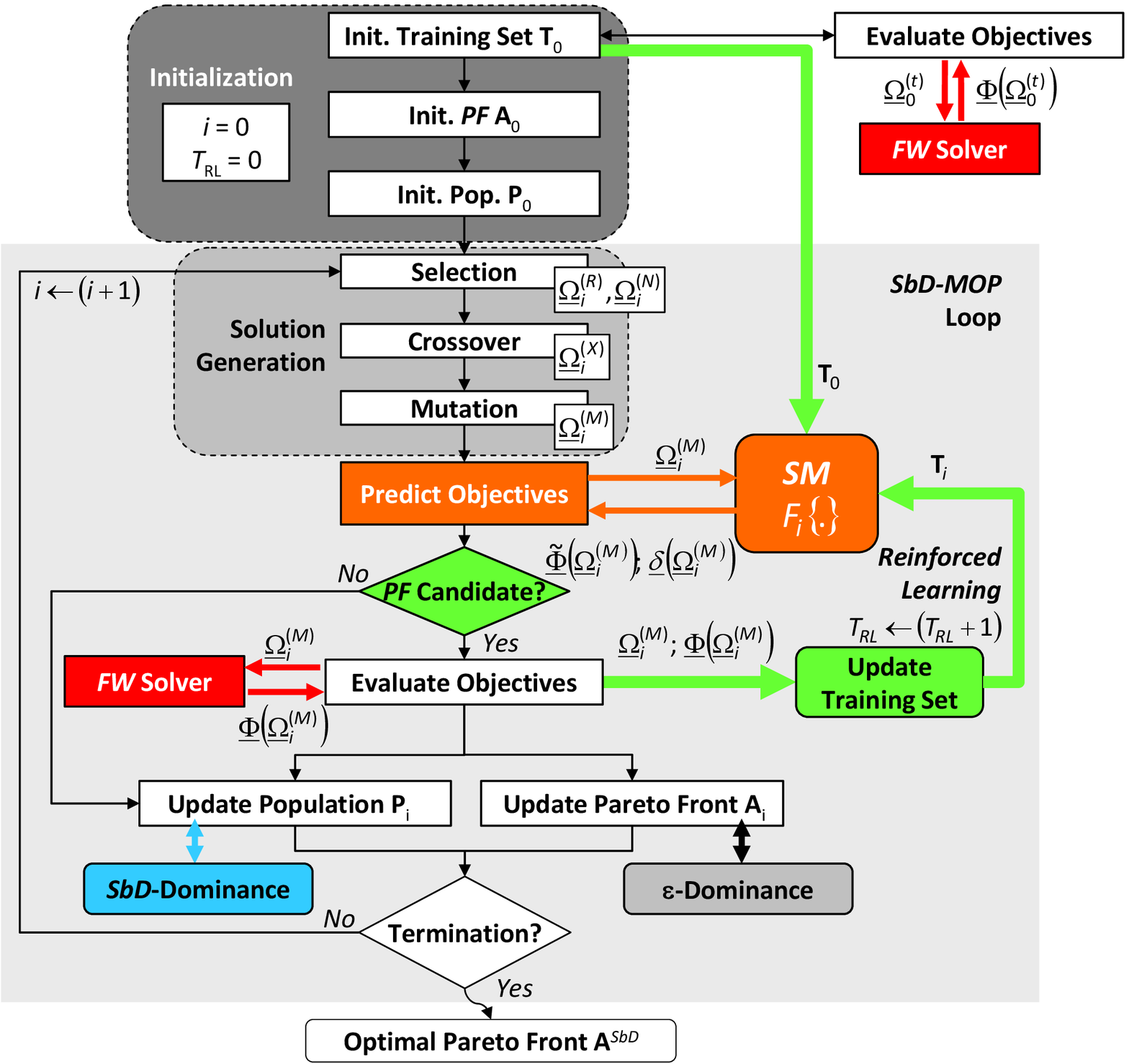}\end{center}

\begin{center}~\vfill\end{center}

\begin{center}\textbf{Fig. 4 - P. Rosatti et} \textbf{\emph{al.}}\textbf{,}
\textbf{\emph{{}``}}Multi-Objective System-by-Design for ...''\end{center}

\newpage
\begin{center}~\vfill\end{center}

\begin{center}\includegraphics[%
  width=0.75\columnwidth]{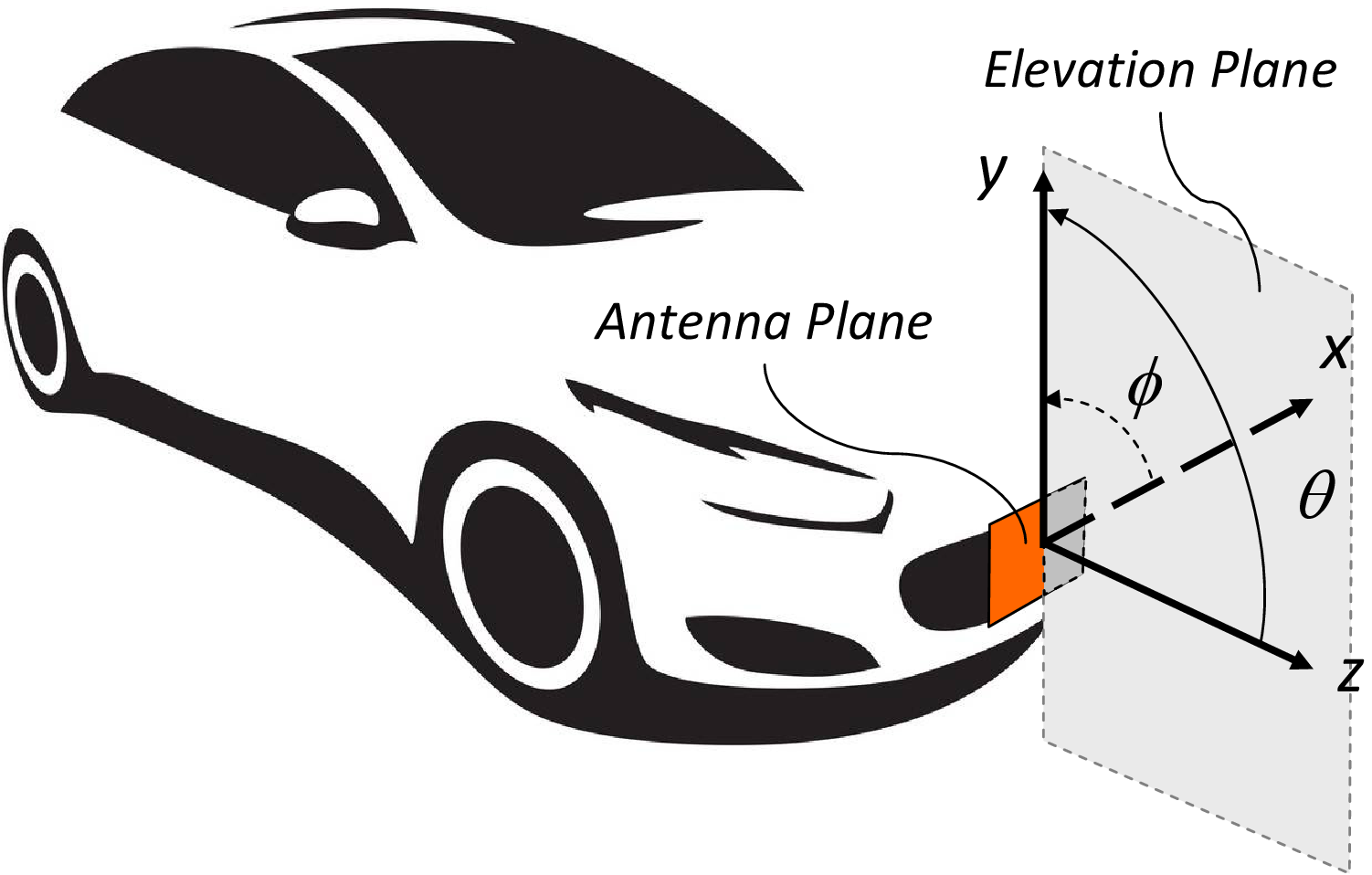}\end{center}

\begin{center}~\vfill\end{center}

\begin{center}\textbf{Fig. 5 - P. Rosatti et} \textbf{\emph{al.}}\textbf{,}
\textbf{\emph{{}``}}Multi-Objective System-by-Design for ...''\end{center}

\newpage
\begin{center}~\vfill\end{center}

\begin{center}\begin{tabular}{c}
\includegraphics[%
  width=0.75\columnwidth]{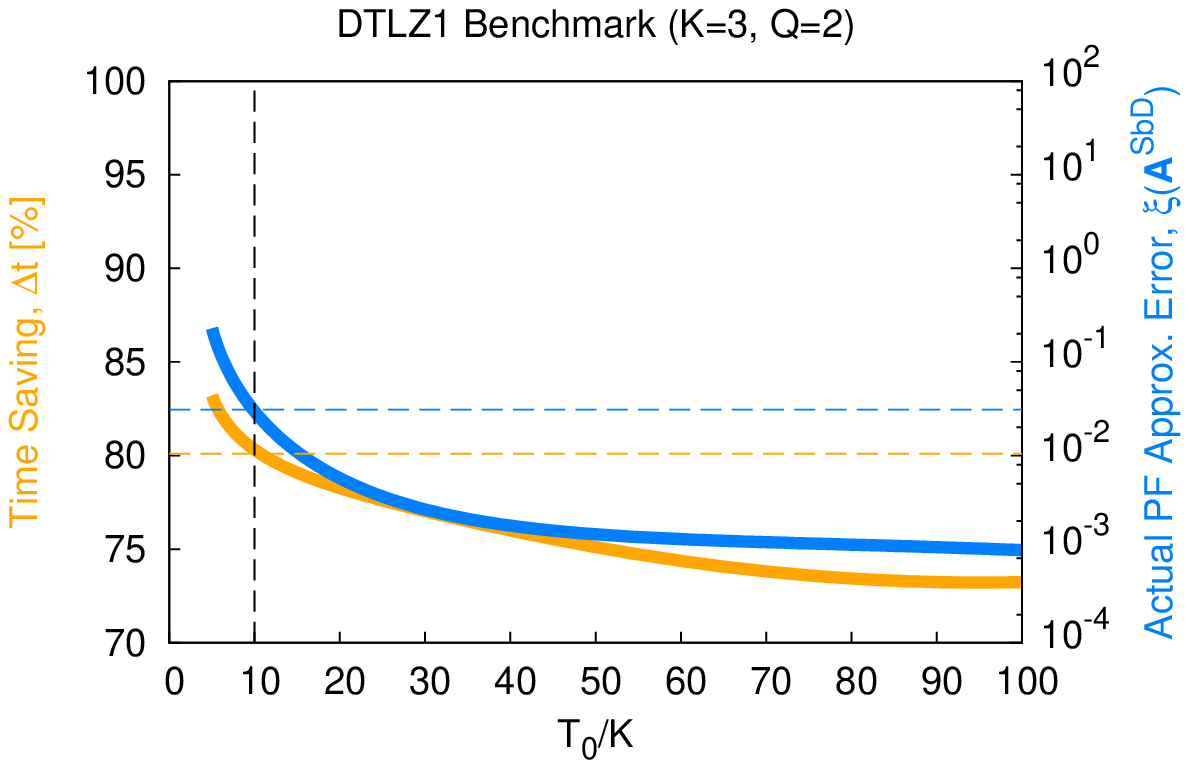}\tabularnewline
(\emph{a})\tabularnewline
\tabularnewline
\includegraphics[%
  width=0.75\columnwidth]{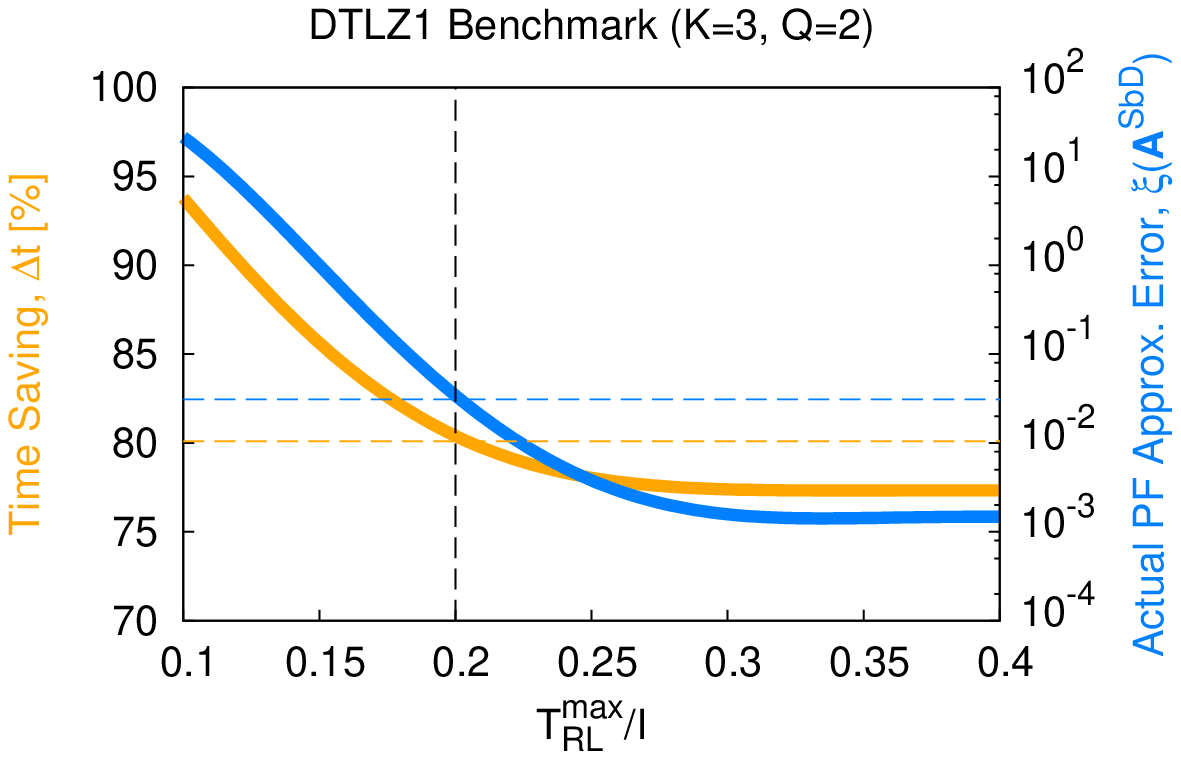}\tabularnewline
(\emph{b})\tabularnewline
\end{tabular}\end{center}

\begin{center}~\vfill\end{center}

\begin{center}\textbf{Fig. 6 - P. Rosatti et} \textbf{\emph{al.}}\textbf{,}
\textbf{\emph{{}``}}Multi-Objective System-by-Design for ...''\end{center}

\newpage
\begin{center}~\vfill\end{center}

\begin{center}\begin{tabular}{c}
\includegraphics[%
  width=0.85\columnwidth]{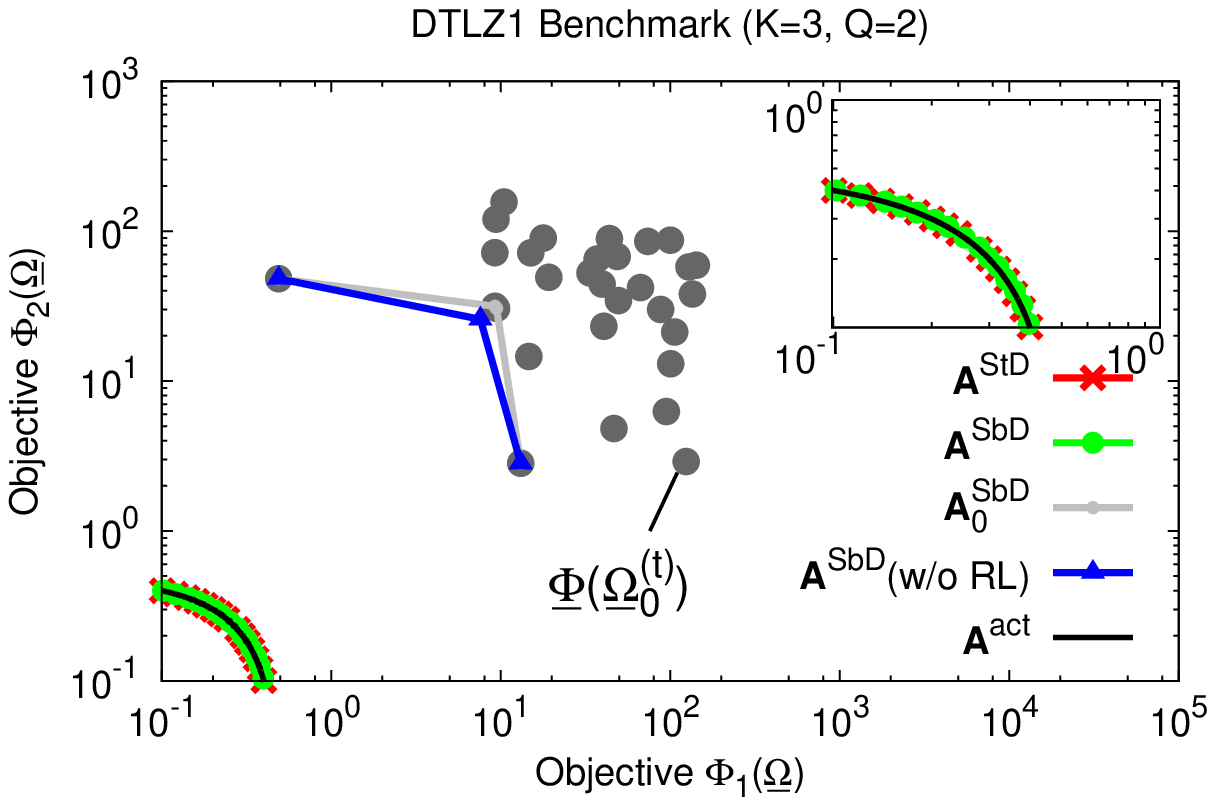}\tabularnewline
(\emph{a})\tabularnewline
\tabularnewline
\includegraphics[%
  width=0.85\columnwidth]{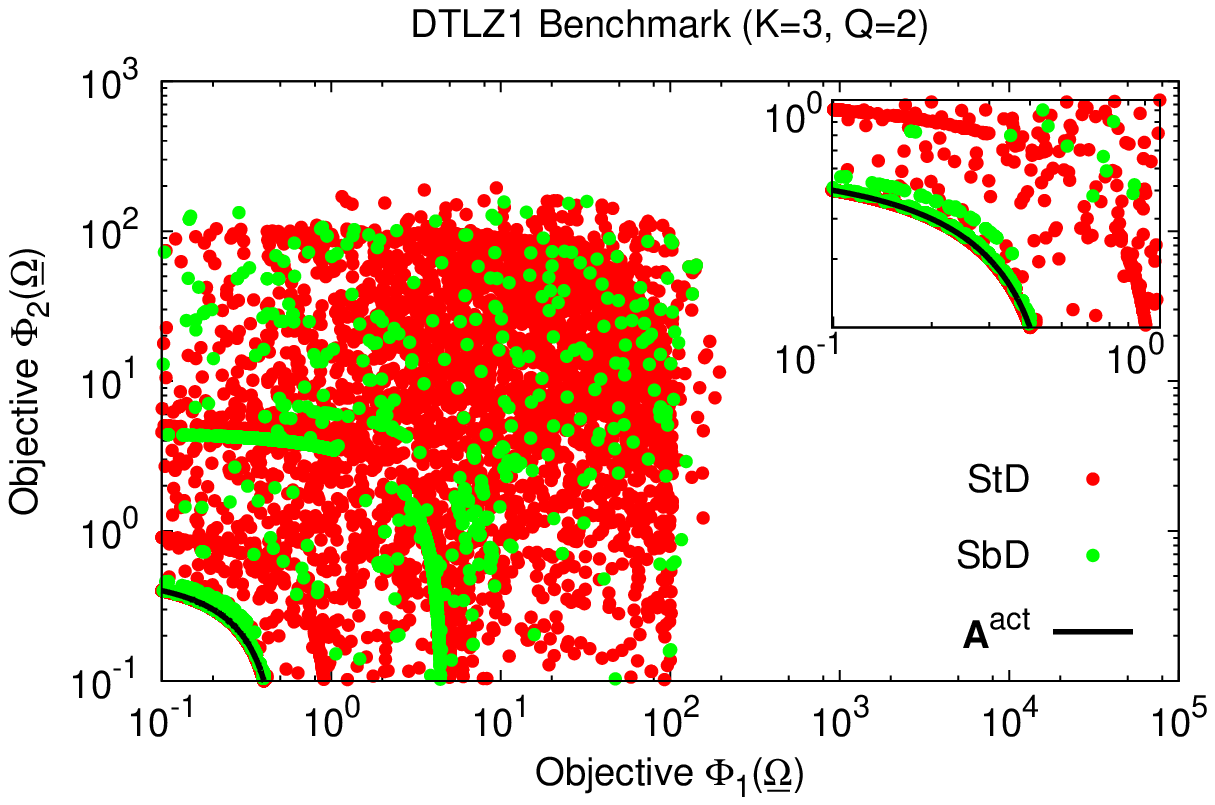}\tabularnewline
(\emph{b})\tabularnewline
\end{tabular}\end{center}

\begin{center}~\vfill\end{center}

\begin{center}\textbf{Fig. 7 - P. Rosatti et} \textbf{\emph{al.}}\textbf{,}
\textbf{\emph{{}``}}Multi-Objective System-by-Design for ...''\end{center}

\newpage
\begin{center}~\vfill\end{center}

\begin{center}\includegraphics[%
  width=0.90\columnwidth]{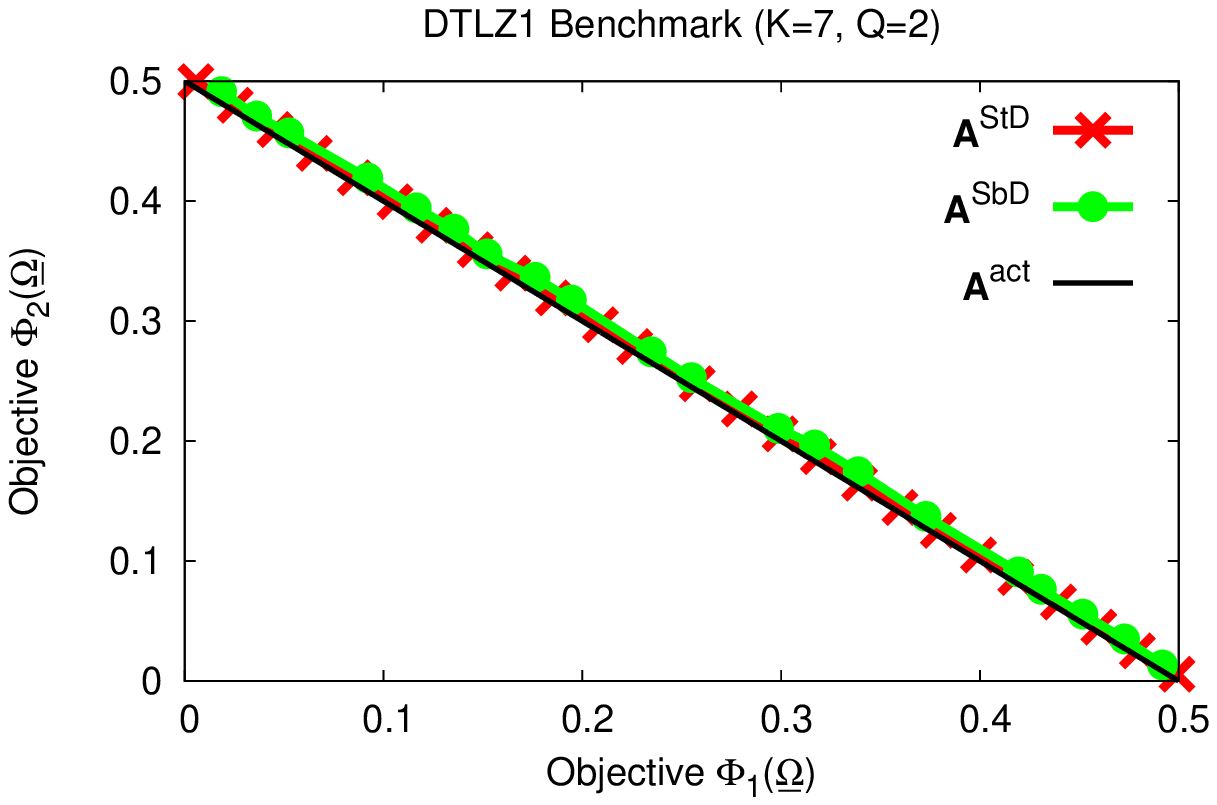}\end{center}

\begin{center}~\vfill\end{center}

\begin{center}\textbf{Fig. 8 - P. Rosatti et} \textbf{\emph{al.}}\textbf{,}
\textbf{\emph{{}``}}Multi-Objective System-by-Design for ...''\end{center}

\newpage
\begin{center}~\vfill\end{center}

\begin{center}\includegraphics[%
  width=0.90\columnwidth]{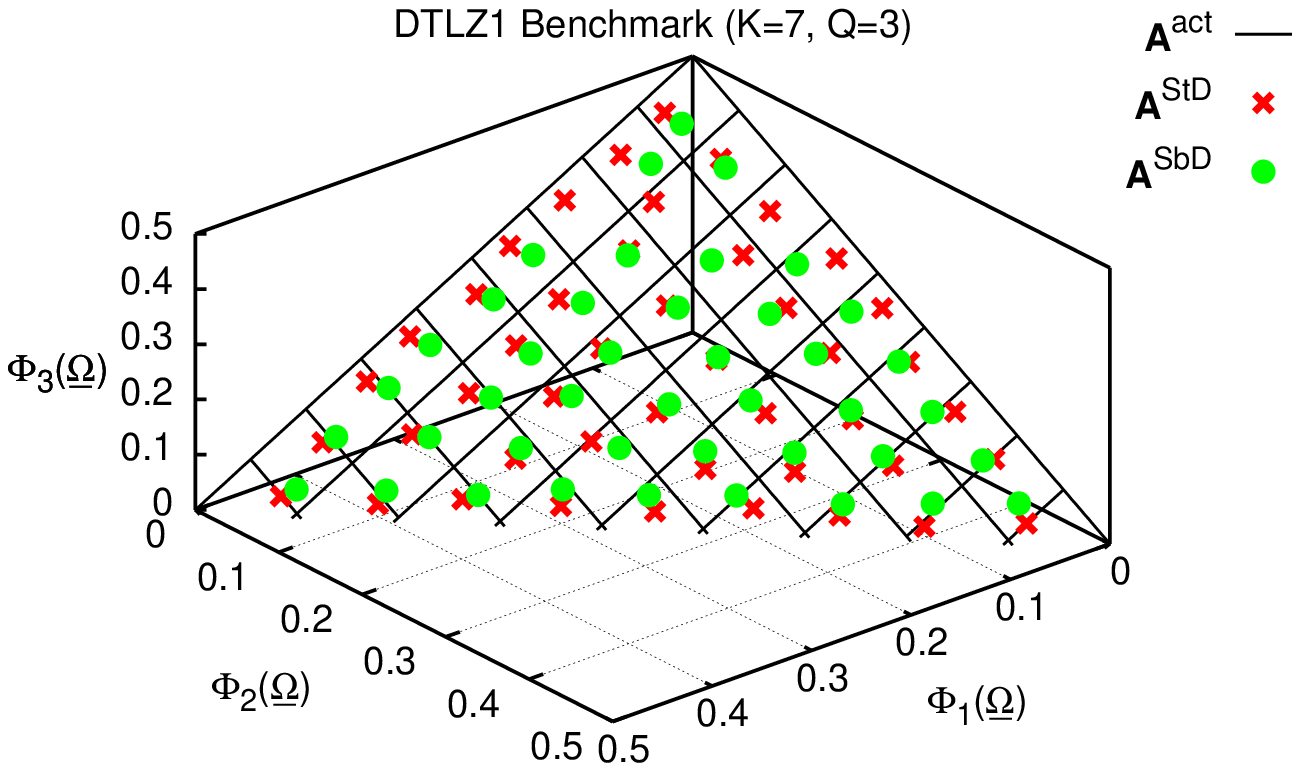}\end{center}

\begin{center}~\vfill\end{center}

\begin{center}\textbf{Fig. 9 - P. Rosatti et} \textbf{\emph{al.}}\textbf{,}
\textbf{\emph{{}``}}Multi-Objective System-by-Design for ...''\end{center}

\newpage
\begin{center}~\vfill\end{center}

\begin{center}\begin{tabular}{c}
\includegraphics[%
  width=1.0\columnwidth,
  keepaspectratio]{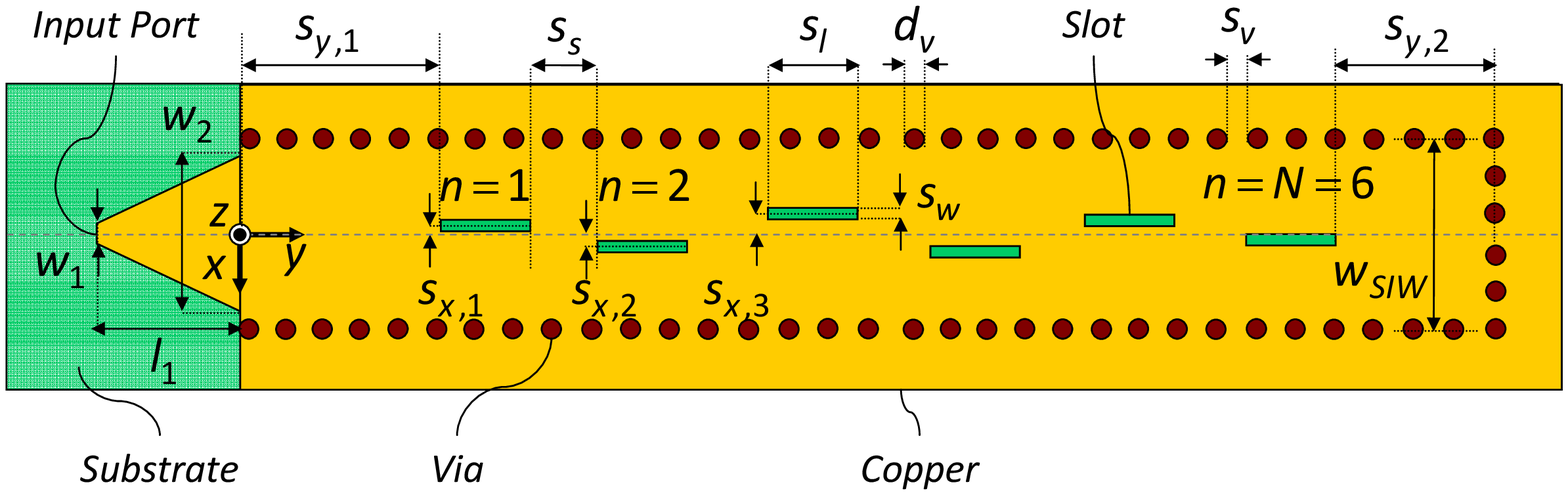}\tabularnewline
\end{tabular}\end{center}

\begin{center}~\vfill\end{center}

\begin{center}\textbf{Fig. 10 - P. Rosatti et} \textbf{\emph{al.}}\textbf{,}
\textbf{\emph{{}``}}Multi-Objective System-by-Design for ...''\end{center}

\newpage
\begin{center}~\vfill\end{center}

\begin{center}\begin{tabular}{c}
\includegraphics[%
  width=0.75\columnwidth,
  keepaspectratio]{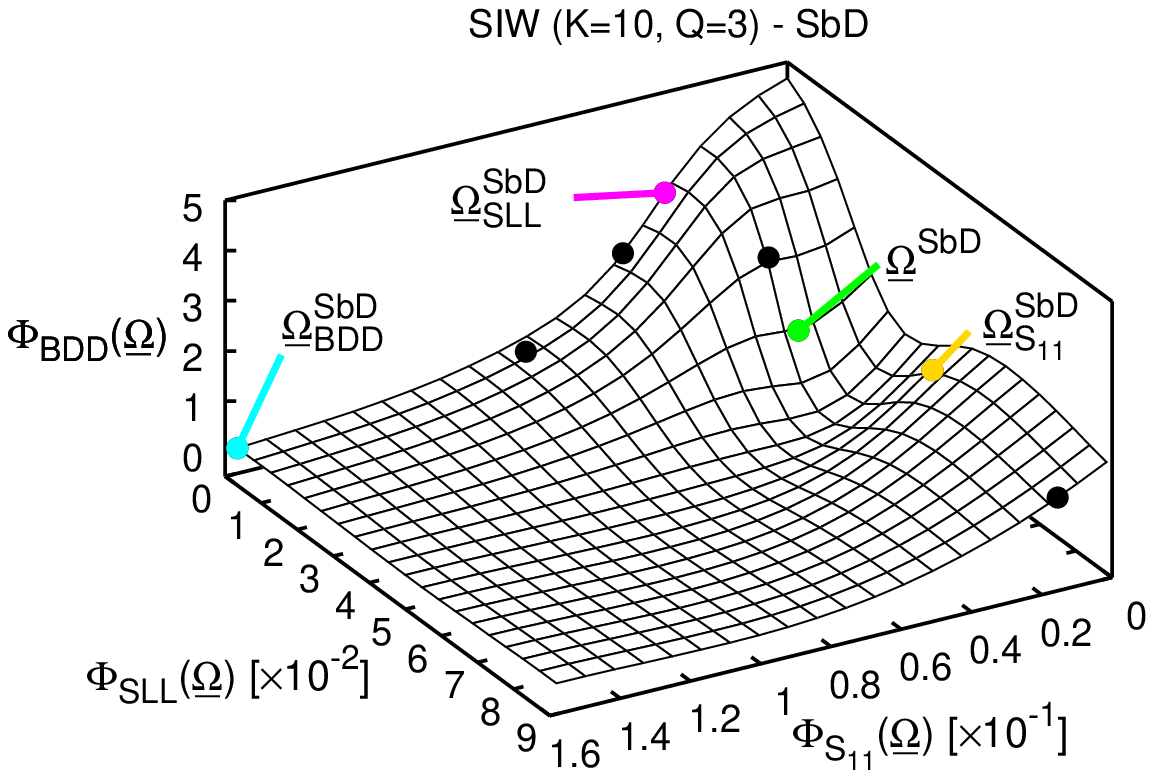}\tabularnewline
(\emph{a})\tabularnewline
\tabularnewline
\includegraphics[%
  width=0.75\columnwidth,
  keepaspectratio]{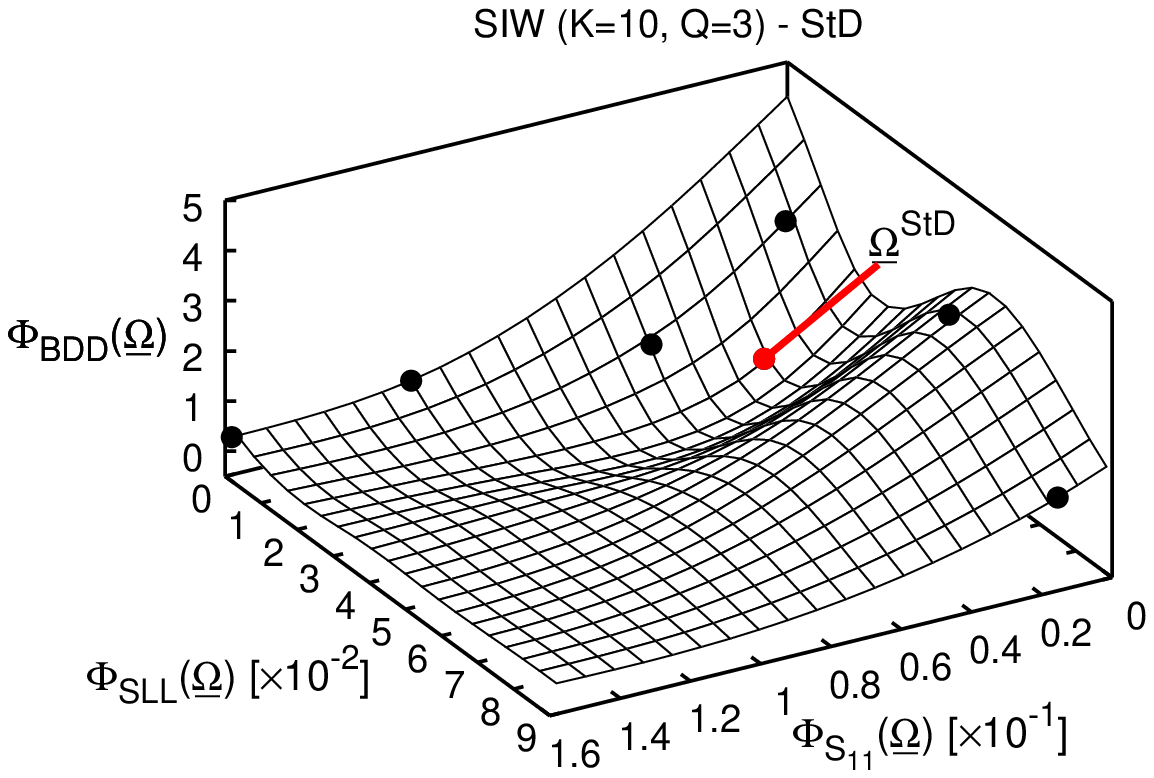}\tabularnewline
(\emph{b})\tabularnewline
\end{tabular}\end{center}

\begin{center}~\vfill\end{center}

\begin{center}\textbf{Fig. 11 - P. Rosatti et} \textbf{\emph{al.}}\textbf{,}
\textbf{\emph{{}``}}Multi-Objective System-by-Design for ...''\end{center}
\newpage

\begin{center}~\vfill\end{center}

\begin{center}\begin{tabular}{c}
\includegraphics[%
  width=0.50\columnwidth,
  keepaspectratio]{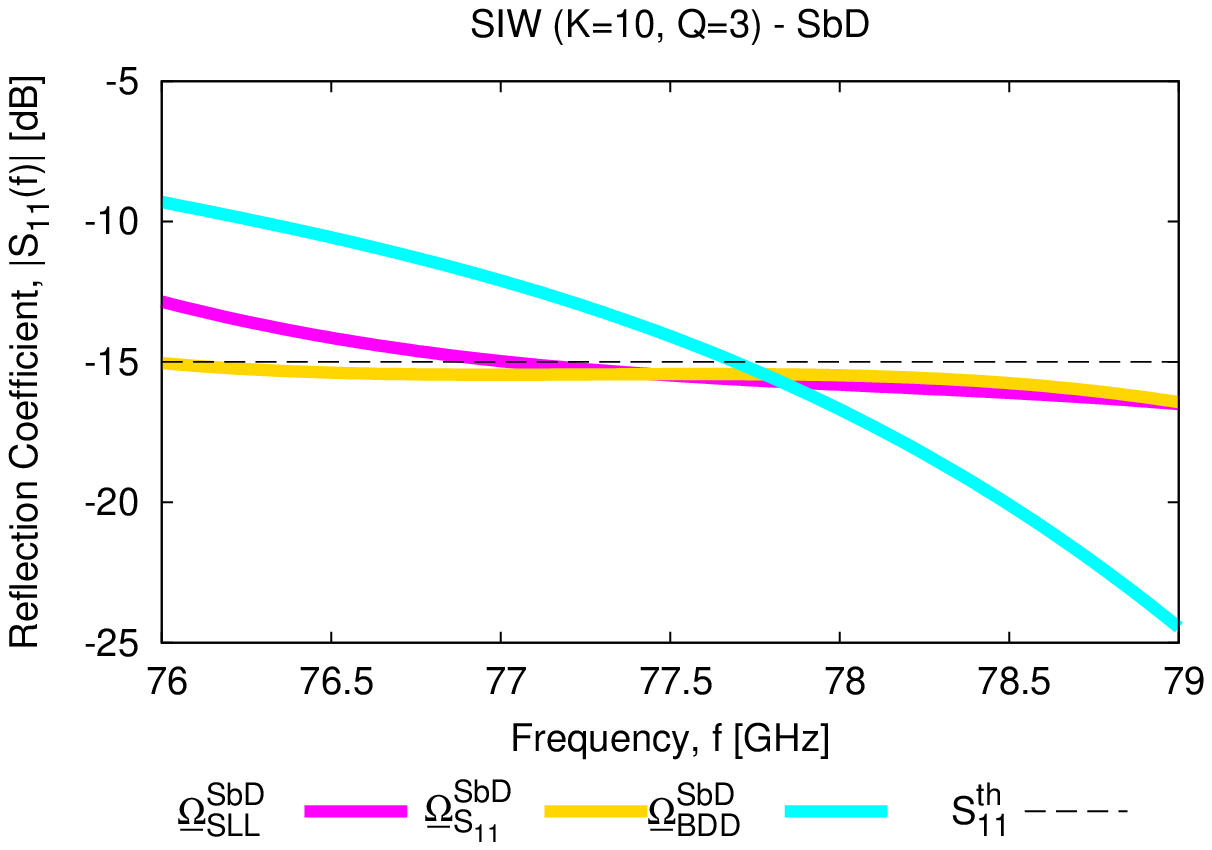}\tabularnewline
(\emph{a})\tabularnewline
\tabularnewline
\includegraphics[%
  width=0.50\columnwidth,
  keepaspectratio]{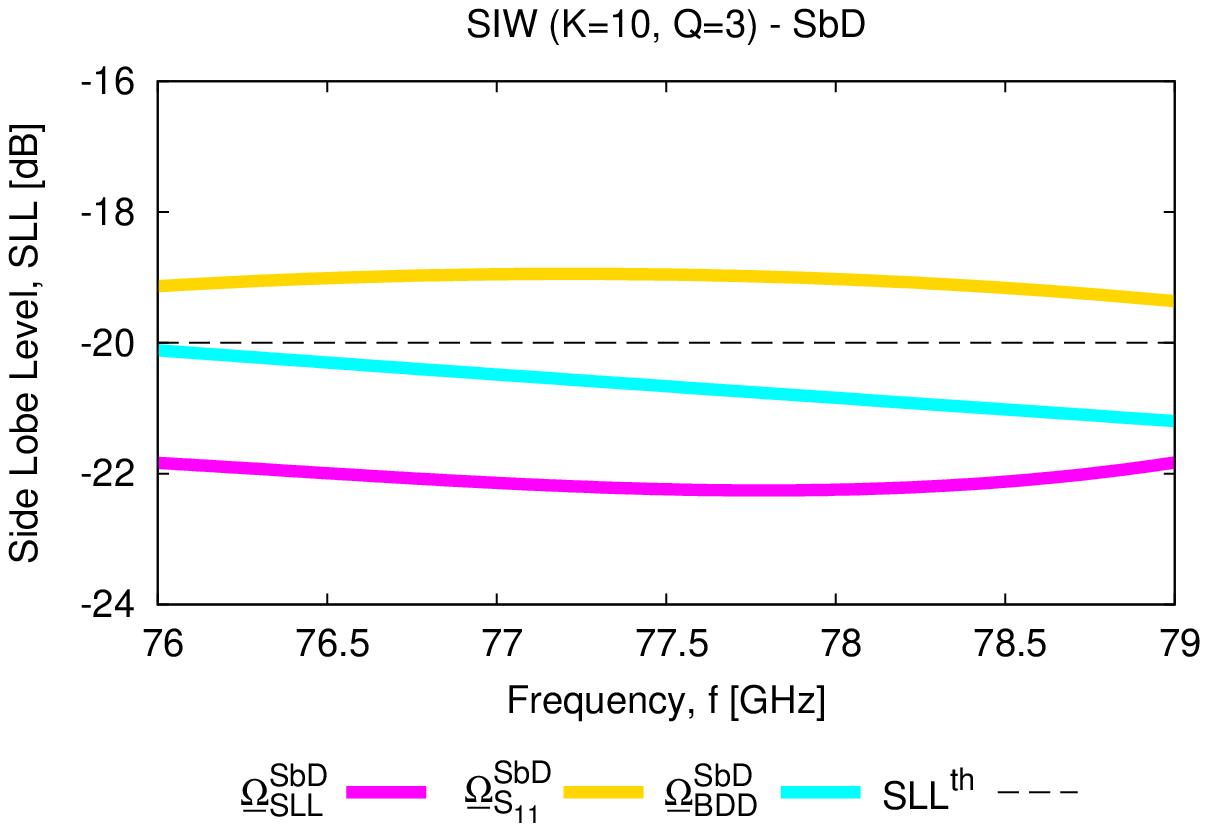}\tabularnewline
(\emph{b})\tabularnewline
\tabularnewline
\multicolumn{1}{c}{\includegraphics[%
  width=0.50\columnwidth,
  keepaspectratio]{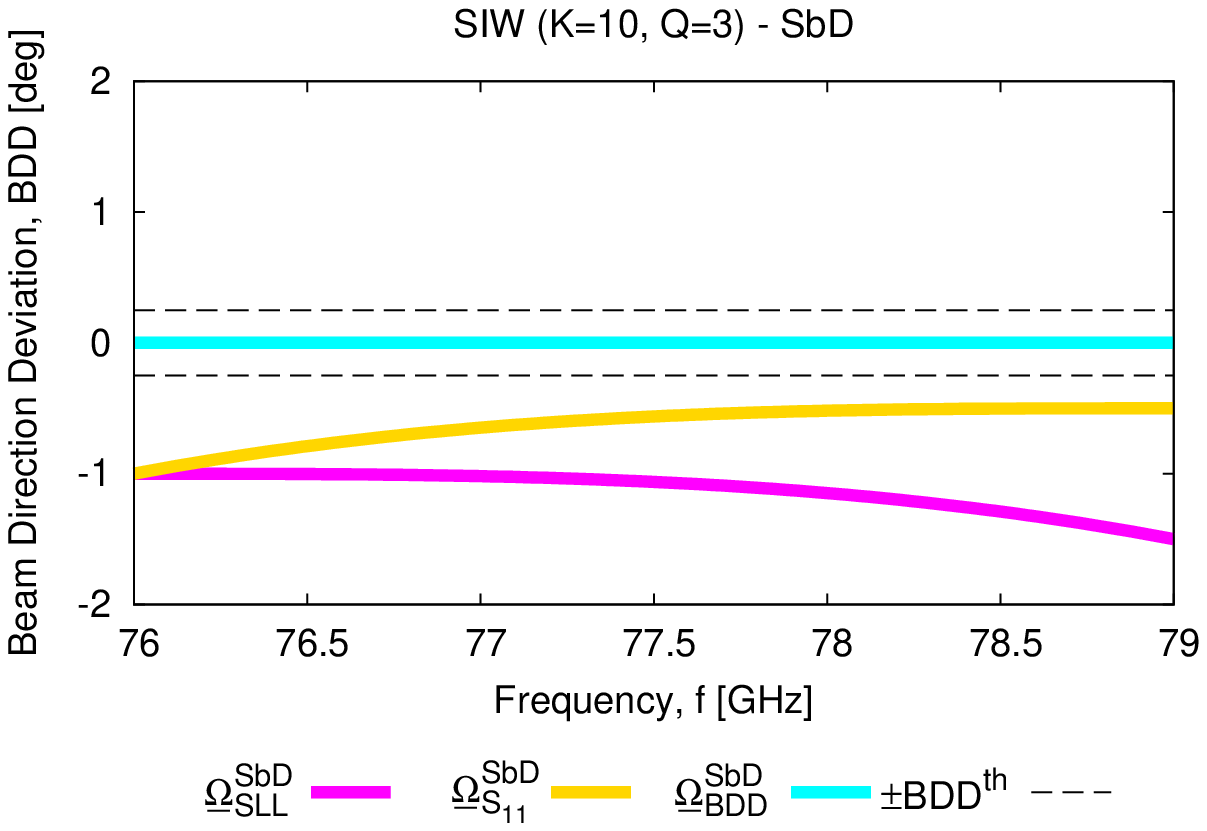}}\tabularnewline
\multicolumn{1}{c}{(\emph{c})}\tabularnewline
\end{tabular}\end{center}

\begin{center}~\vfill\end{center}

\begin{center}\textbf{Fig. 12 - P. Rosatti et} \textbf{\emph{al.}}\textbf{,}
\textbf{\emph{{}``}}Multi-Objective System-by-Design for ...''\end{center}
\newpage

\begin{center}~\vfill\end{center}

\begin{center}\includegraphics[%
  width=0.90\columnwidth,
  keepaspectratio]{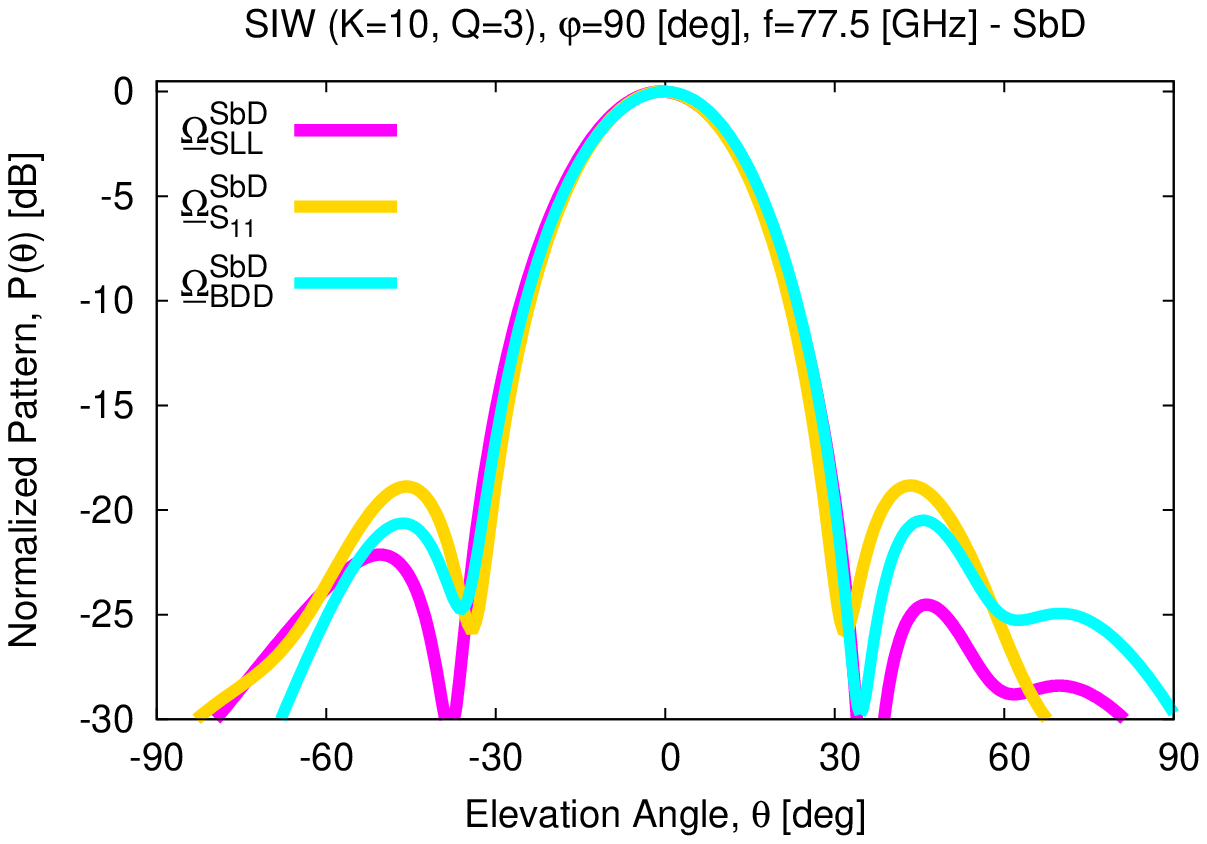}\end{center}

\begin{center}~\vfill\end{center}

\begin{center}\textbf{Fig. 13 - P. Rosatti et} \textbf{\emph{al.}}\textbf{,}
\textbf{\emph{{}``}}Multi-Objective System-by-Design for ...''\end{center}
\newpage

\begin{center}~\vfill\end{center}

\begin{center}\begin{tabular}{c}
\includegraphics[%
  width=0.50\columnwidth,
  keepaspectratio]{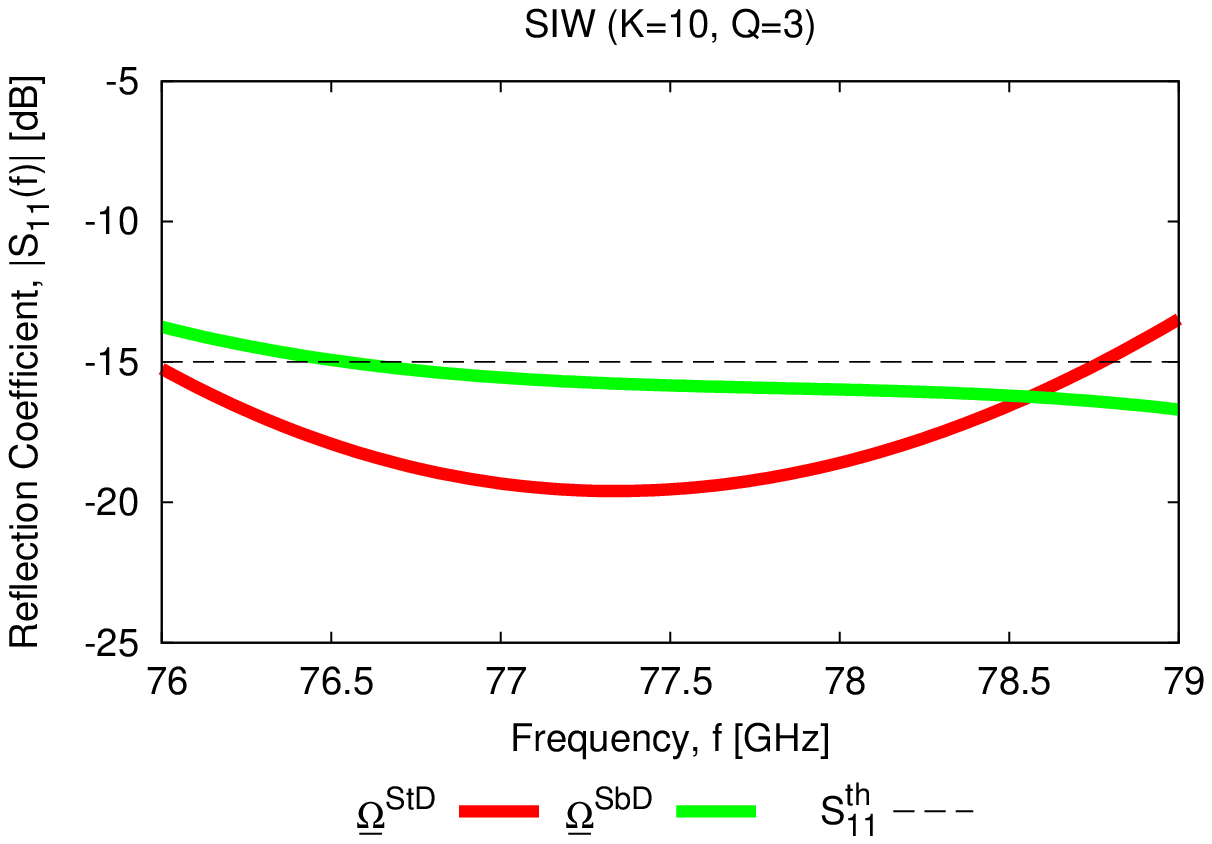}\tabularnewline
(\emph{a})\tabularnewline
\tabularnewline
\includegraphics[%
  width=0.50\columnwidth,
  keepaspectratio]{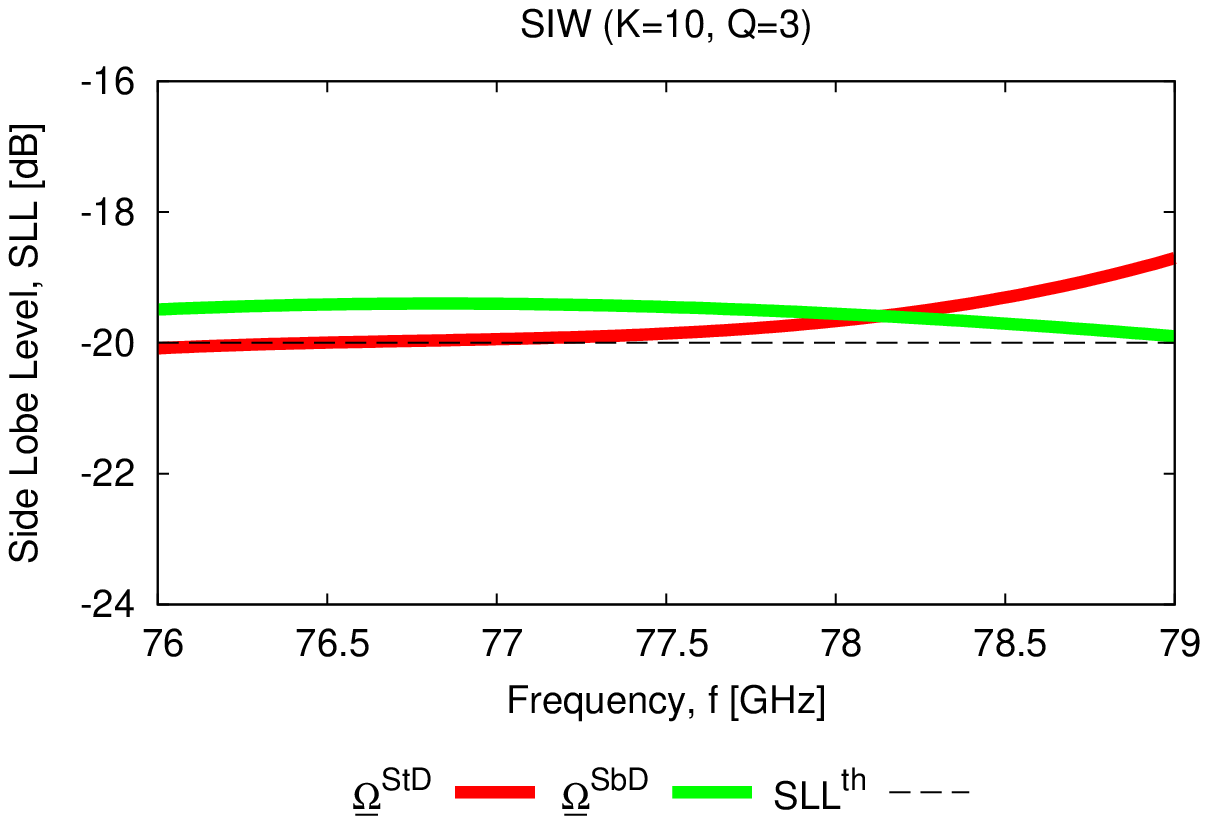}\tabularnewline
(\emph{b})\tabularnewline
\tabularnewline
\multicolumn{1}{c}{\includegraphics[%
  width=0.50\columnwidth,
  keepaspectratio]{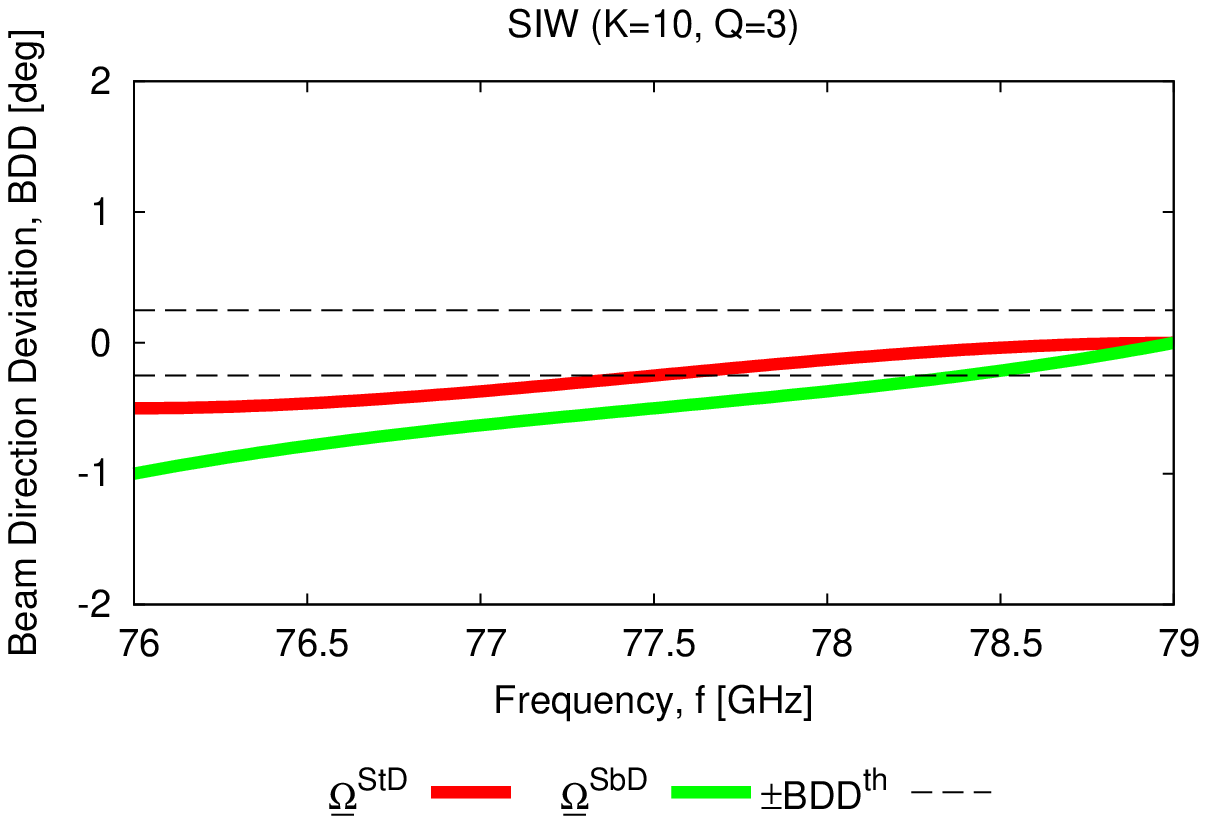}}\tabularnewline
\multicolumn{1}{c}{(\emph{c})}\tabularnewline
\end{tabular}\end{center}

\begin{center}~\vfill\end{center}

\begin{center}\textbf{Fig. 14 - P. Rosatti et} \textbf{\emph{al.}}\textbf{,}
\textbf{\emph{{}``}}Multi-Objective System-by-Design for ...''\end{center}
\newpage

\begin{center}~\vfill\end{center}

\begin{center}\begin{tabular}{c}
\includegraphics[%
  width=0.85\columnwidth,
  keepaspectratio]{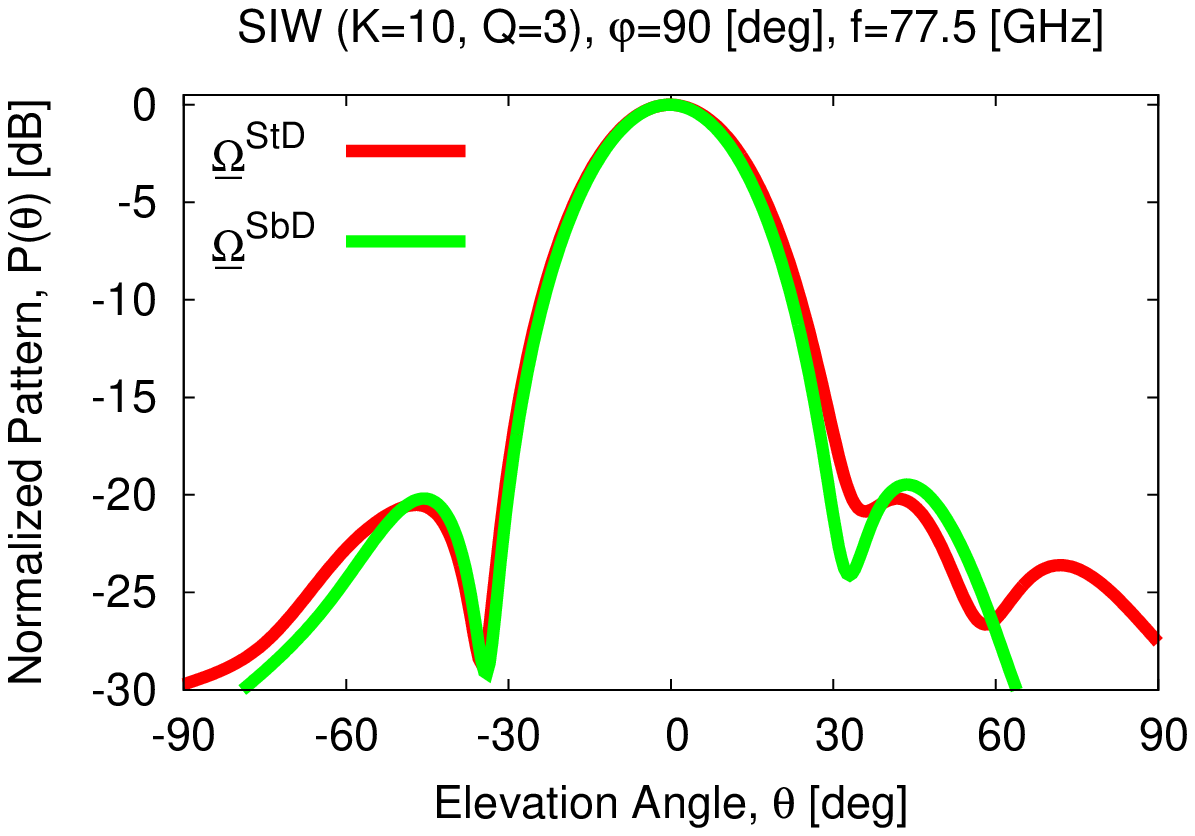}\tabularnewline
(\emph{a})\tabularnewline
\tabularnewline
\includegraphics[%
  width=0.85\columnwidth,
  keepaspectratio]{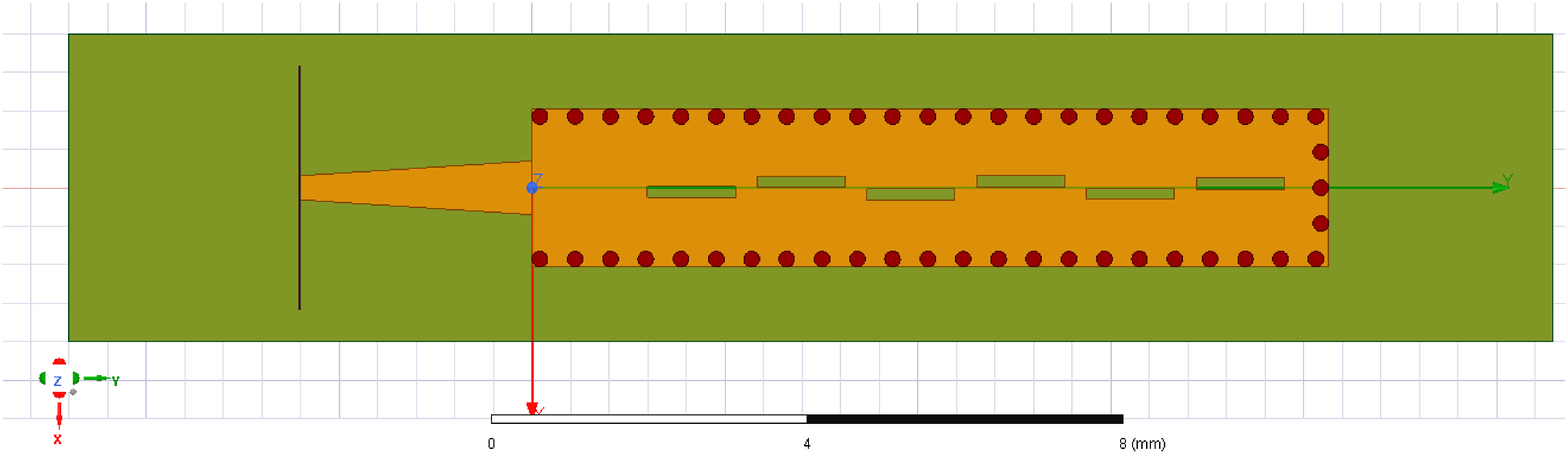}\tabularnewline
(\emph{b})\tabularnewline
\tabularnewline
\includegraphics[%
  width=0.85\columnwidth,
  keepaspectratio]{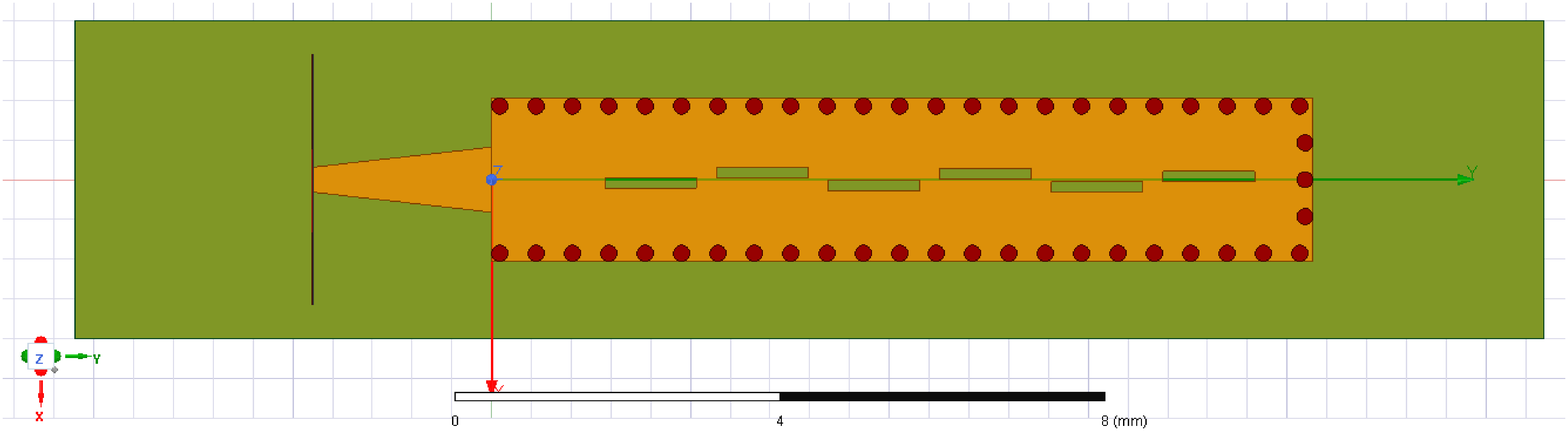}\tabularnewline
(\emph{c})\tabularnewline
\end{tabular}\end{center}

\begin{center}~\vfill\end{center}

\begin{center}\textbf{Fig. 15 - P. Rosatti et} \textbf{\emph{al.}}\textbf{,}
\textbf{\emph{{}``}}Multi-Objective System-by-Design for ...''\end{center}
\newpage

\begin{center}~\vfill\end{center}

\begin{center}\begin{tabular}{c}
\includegraphics[%
  width=1.0\columnwidth,
  keepaspectratio]{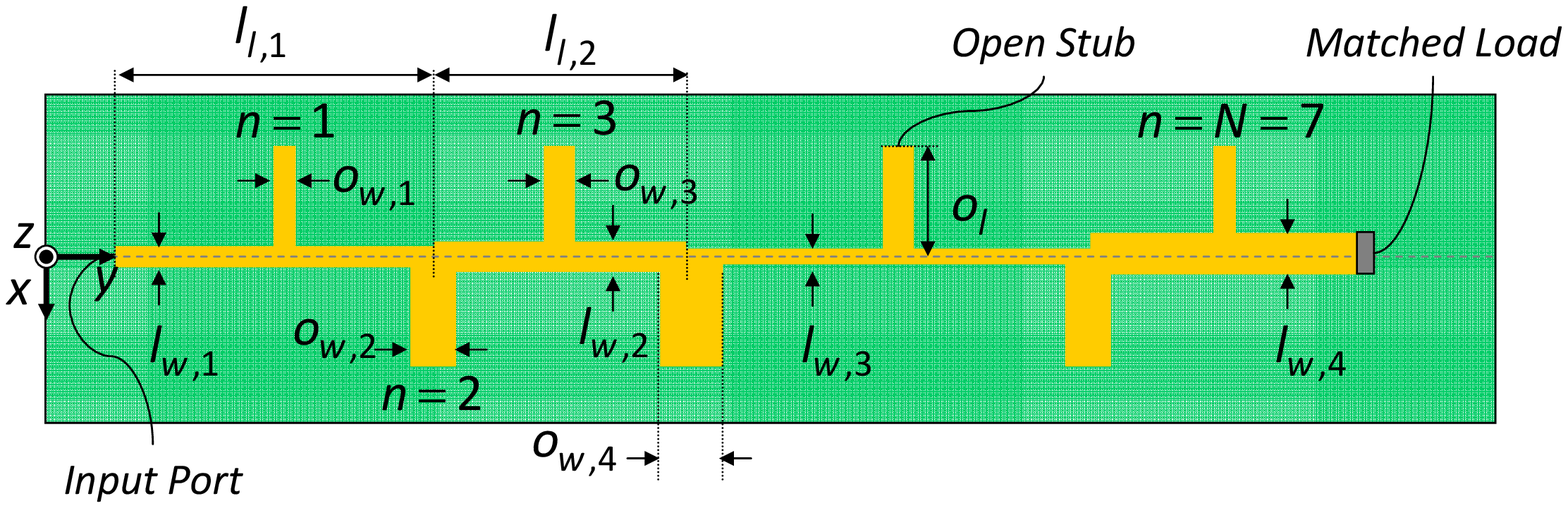}\tabularnewline
(\emph{a})\tabularnewline
\tabularnewline
\multicolumn{1}{c}{\includegraphics[%
  width=0.90\columnwidth,
  keepaspectratio]{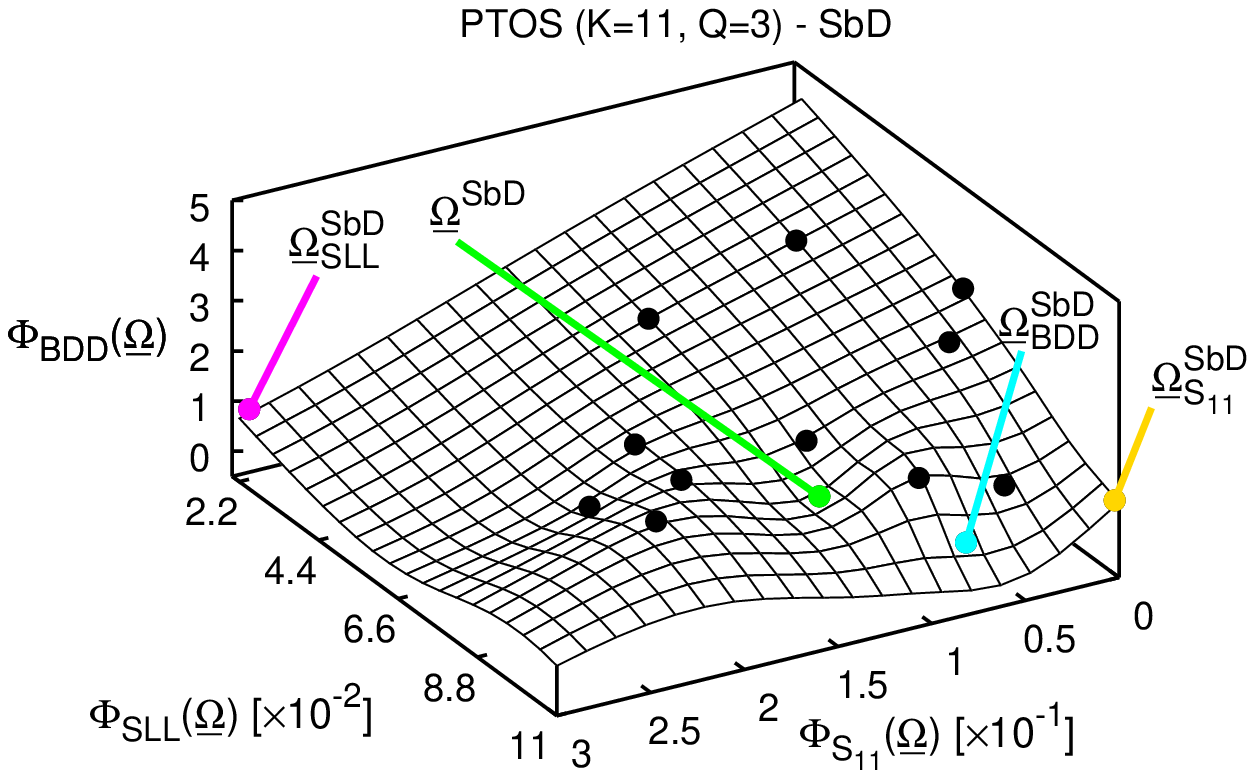}}\tabularnewline
\multicolumn{1}{c}{(\emph{b})}\tabularnewline
\end{tabular}\end{center}

\begin{center}~\vfill\end{center}

\begin{center}\textbf{Fig. 16 - P. Rosatti et} \textbf{\emph{al.}}\textbf{,}
\textbf{\emph{{}``}}Multi-Objective System-by-Design for ...''\end{center}
\newpage

\begin{center}~\vfill\end{center}

\begin{center}\begin{tabular}{c}
\includegraphics[%
  width=0.50\columnwidth,
  keepaspectratio]{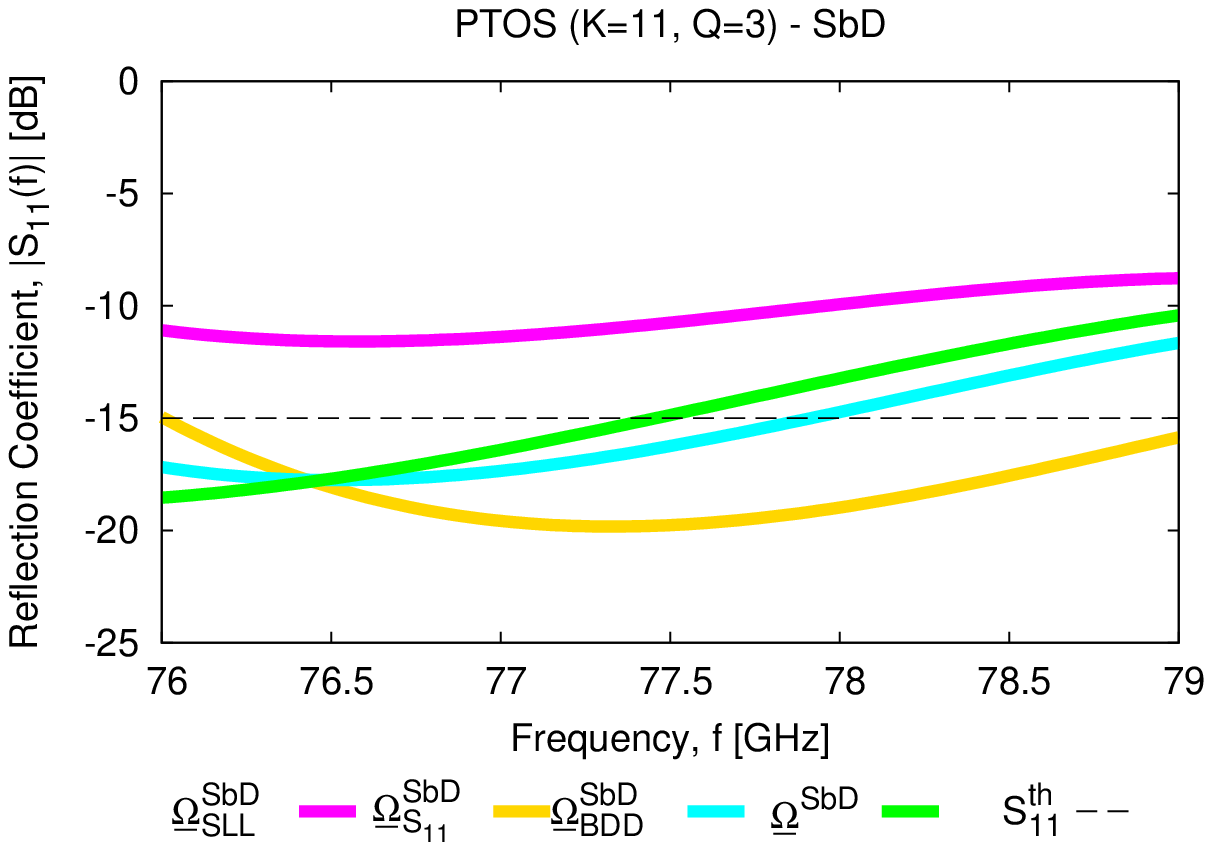}\tabularnewline
(\emph{a})\tabularnewline
\tabularnewline
\includegraphics[%
  width=0.50\columnwidth,
  keepaspectratio]{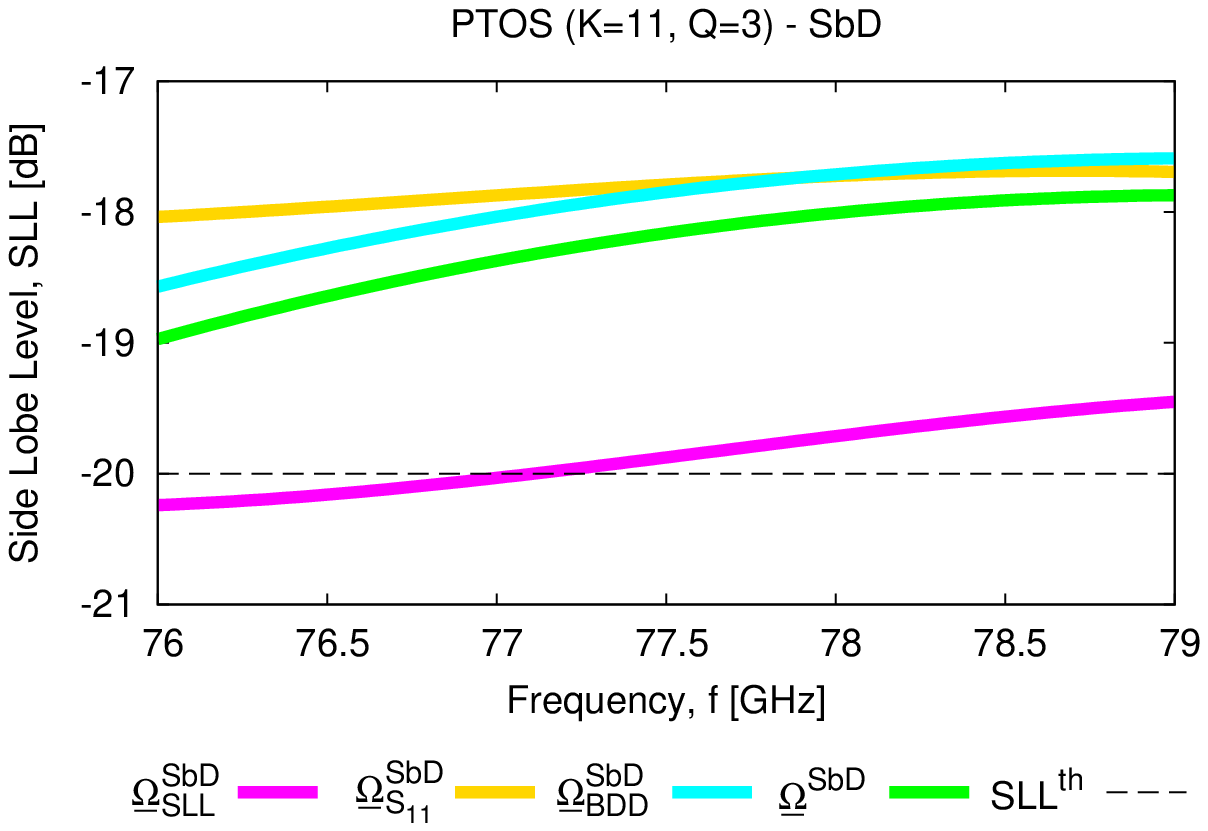}\tabularnewline
(\emph{b})\tabularnewline
\tabularnewline
\multicolumn{1}{c}{\includegraphics[%
  width=0.50\columnwidth,
  keepaspectratio]{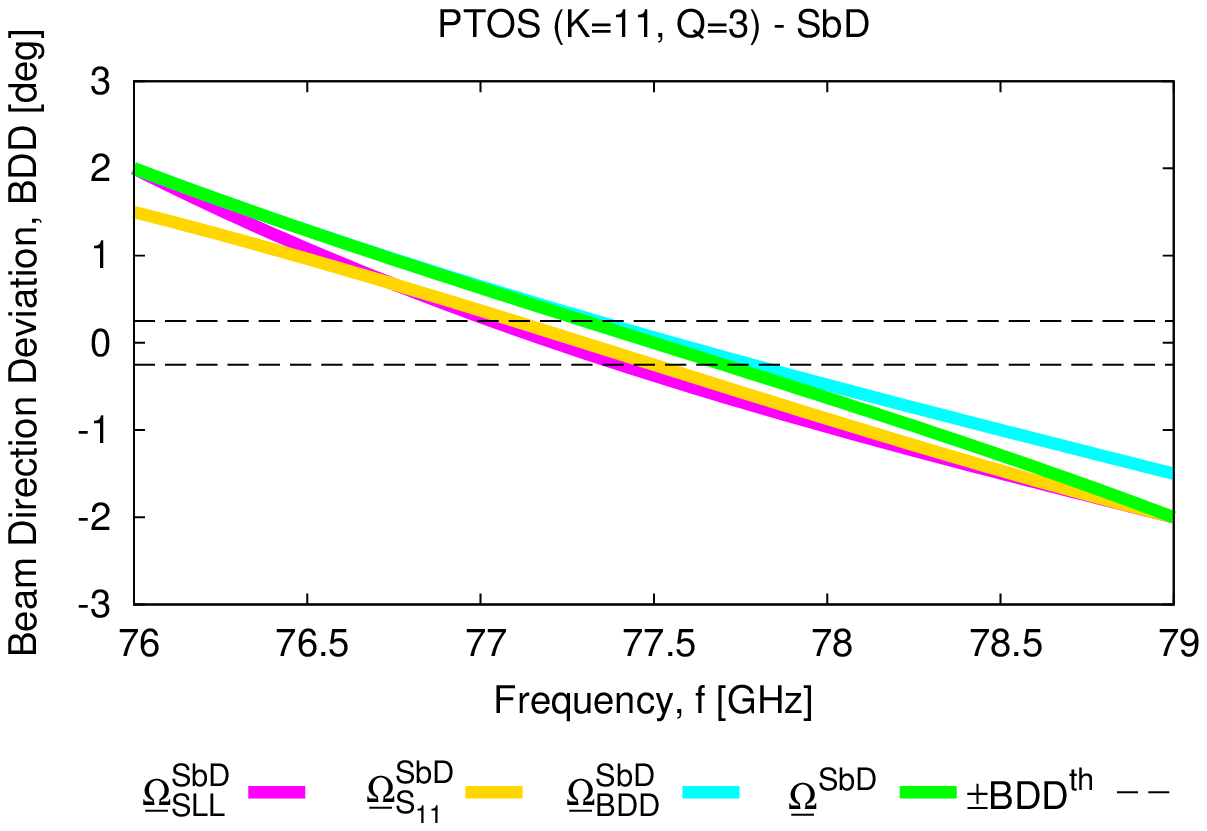}}\tabularnewline
\multicolumn{1}{c}{(\emph{c})}\tabularnewline
\end{tabular}\end{center}

\begin{center}~\vfill\end{center}

\begin{center}\textbf{Fig. 17 - P. Rosatti et} \textbf{\emph{al.}}\textbf{,}
\textbf{\emph{{}``}}Multi-Objective System-by-Design for ...''\end{center}
\newpage

\begin{center}~\vfill\end{center}

\begin{center}\begin{tabular}{c}
\includegraphics[%
  width=0.65\columnwidth,
  keepaspectratio]{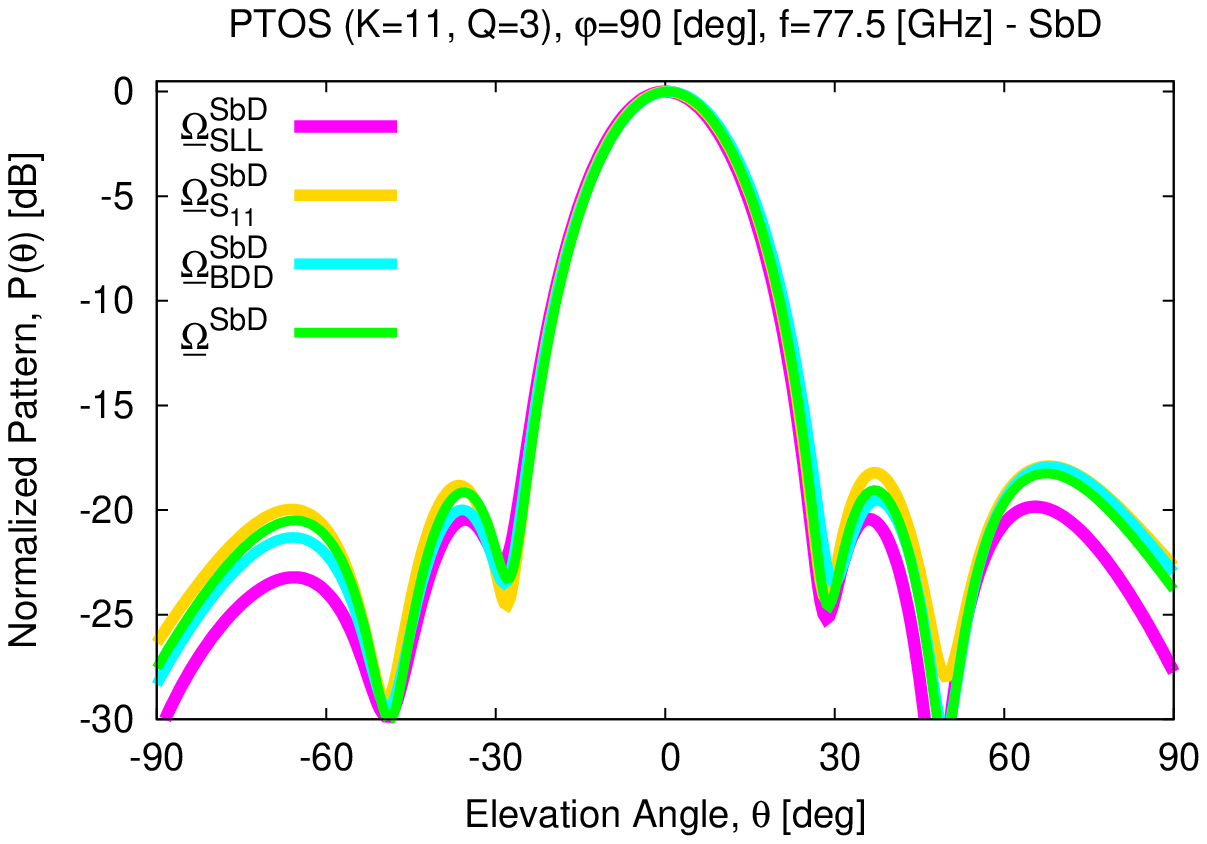}\tabularnewline
(\emph{a})\tabularnewline
\tabularnewline
\includegraphics[%
  width=0.90\columnwidth,
  keepaspectratio]{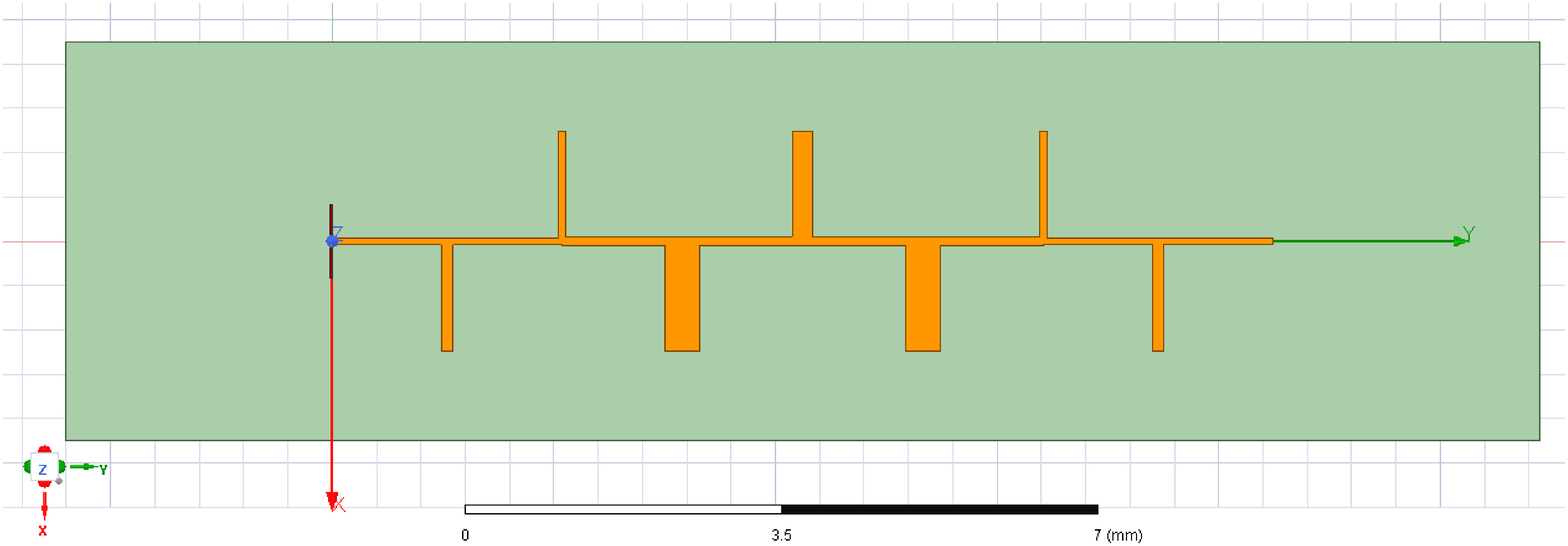}\tabularnewline
(\emph{b})\tabularnewline
\end{tabular}\end{center}

\begin{center}~\vfill\end{center}

\begin{center}\textbf{Fig. 18 - P. Rosatti et} \textbf{\emph{al.}}\textbf{,}
\textbf{\emph{{}``}}Multi-Objective System-by-Design for ...''\end{center}
\newpage

\begin{center}~\vfill\end{center}

\begin{center}\begin{tabular}{|c|c|c|}
\hline 
&
$\underline{\widetilde{\Phi}}\left(\underline{\Omega}^{\left(1\right)}\right)$&
$\underline{\Phi}\left(\underline{\Omega}^{\left(1\right)}\right)$\tabularnewline
\hline
\hline 
$\underline{\widetilde{\Phi}}\left(\underline{\Omega}^{\left(2\right)}\right)$&
$\left\{ \begin{array}{l}
\mathcal{L}_{q}\left(\underline{\Omega}^{\left(1\right)}\right)\leq\mathcal{L}_{q}\left(\underline{\Omega}^{\left(2\right)}\right)\\
\mathrm{for}\, q=1,...,Q\\
\exists q\in\left\{ 1,...,Q\right\} \\
\mathrm{for\, which}\,\mathcal{L}_{q}\left(\underline{\Omega}^{\left(1\right)}\right)<\mathcal{L}_{q}\left(\underline{\Omega}^{\left(2\right)}\right)\end{array}\right.$&
$\left\{ \begin{array}{l}
\Phi_{q}\left(\underline{\Omega}^{\left(1\right)}\right)\leq\mathcal{U}_{q}\left(\underline{\Omega}^{\left(2\right)}\right)\\
\mathrm{for}\, q=1,...,Q\\
\exists q\in\left\{ 1,...,Q\right\} \\
\mathrm{for\, which}\,\Phi_{q}\left(\underline{\Omega}^{\left(1\right)}\right)<\mathcal{U}_{q}\left(\underline{\Omega}^{\left(2\right)}\right)\end{array}\right.$\tabularnewline
&
{[}Fig. 3(\emph{a}){]}&
{[}Fig. 3(\emph{b}){]}\tabularnewline
\hline 
$\underline{\Phi}\left(\underline{\Omega}^{\left(2\right)}\right)$&
$\left\{ \begin{array}{l}
\mathcal{U}_{q}\left(\underline{\Omega}^{\left(1\right)}\right)\leq\Phi_{q}\left(\underline{\Omega}^{\left(2\right)}\right)\\
\mathrm{for}\, q=1,...,Q\\
\exists q\in\left\{ 1,...,Q\right\} \\
\mathrm{for\, which}\,\mathcal{U}_{q}\left(\underline{\Omega}^{\left(1\right)}\right)<\Phi_{q}\left(\underline{\Omega}^{\left(2\right)}\right)\end{array}\right.$&
$\left\{ \begin{array}{l}
\Phi_{q}\left(\underline{\Omega}^{\left(1\right)}\right)\leq\Phi_{q}\left(\underline{\Omega}^{\left(2\right)}\right)\\
\mathrm{for}\, q=1,...,Q\\
\exists q\in\left\{ 1,...,Q\right\} \\
\mathrm{for\, which}\,\Phi_{q}\left(\underline{\Omega}^{\left(1\right)}\right)<\Phi_{q}\left(\underline{\Omega}^{\left(2\right)}\right)\end{array}\right.$\tabularnewline
&
{[}Fig. 3(\emph{c}){]}&
{[}Fig. 3(\emph{d}){]}\tabularnewline
\hline
\end{tabular}\end{center}

\begin{center}~\vfill\end{center}

\begin{center}\textbf{Tab. I - P. Rosatti et} \textbf{\emph{al.}}\textbf{,}
\textbf{\emph{{}``}}Multi-Objective System-by-Design for ...''\end{center}

\newpage
\begin{center}~\vfill\end{center}

\begin{center}\begin{tabular}{|c||c|c|}
\hline 
Condition&
$\underline{\widetilde{\Phi}}\left(\underline{\Omega}_{i}^{\left(\mathcal{M}\right)}\right)$&
$\underline{\Phi}\left(\underline{\Omega}_{i}^{\left(\mathcal{M}\right)}\right)$\tabularnewline
\hline
\hline 
$\exists\underline{\Omega}\in\mathbf{P}_{i-1}:\,\underline{\Omega}\prec_{SbD}\underline{\Omega}_{i}^{\left(\mathcal{M}\right)}$&
$\mathbf{P}_{i}\leftarrow\mathbf{P}_{i-1}$&
$\mathbf{P}_{i}\leftarrow\mathbf{P}_{i-1}$\tabularnewline
\hline 
$\left\{ \begin{array}{l}
\exists\underline{\Omega}\in\mathbf{P}_{i-1}^{pred}:\,\underline{\Omega}_{i}^{\left(\mathcal{M}\right)}\prec_{SbD}\underline{\Omega}\\
\nexists\underline{\Omega}'\in\mathbf{P}_{i-1}:\,\underline{\Omega}'\prec_{SbD}\underline{\Omega}_{i}^{\left(\mathcal{M}\right)}\end{array}\right.$&
$\underline{\Omega}\leftarrow\underline{\Omega}_{i}^{\left(\mathcal{M}\right)}$&
$\underline{\Omega}\leftarrow\underline{\Omega}_{i}^{\left(\mathcal{M}\right)}$\tabularnewline
\hline 
$\left\{ \begin{array}{l}
\exists\underline{\Omega}\in\mathbf{P}_{i-1}^{sim}:\,\underline{\Omega}_{i}^{\left(\mathcal{M}\right)}\prec_{SbD}\underline{\Omega}\\
\nexists\underline{\Omega}'\in\mathbf{P}_{i-1}^{pred}:\,\underline{\Omega}'\prec_{SbD}\underline{\Omega}_{i}^{\left(\mathcal{M}\right)}\end{array}\right.$&
$\underline{\Omega}\leftarrow\underline{\Omega}_{i}^{\left(\mathcal{M}\right)}$&
$\underline{\Omega}\leftarrow\underline{\Omega}_{i}^{\left(\mathcal{M}\right)}$\tabularnewline
\hline 
$\left\{ \begin{array}{l}
\nexists\underline{\Omega}\in\mathbf{P}_{i-1}:\,\underline{\Omega}_{i}^{\left(\mathcal{M}\right)}\prec_{SbD}\underline{\Omega}\\
\nexists\underline{\Omega}'\in\mathbf{P}_{i-1}:\,\underline{\Omega}'\prec_{SbD}\underline{\Omega}_{i}^{\left(\mathcal{M}\right)}\\
\mathbf{P}_{i-1}^{pred}\neq\emptyset\end{array}\right.$&
$\mathcal{R}\left\{ \mathbf{P}_{i-1}^{pred}\right\} \leftarrow\underline{\Omega}_{i}^{\left(\mathcal{M}\right)}$&
$\mathcal{R}\left\{ \mathbf{P}_{i-1}^{pred}\right\} \leftarrow\underline{\Omega}_{i}^{\left(\mathcal{M}\right)}$\tabularnewline
\hline 
$\left\{ \begin{array}{l}
\nexists\underline{\Omega}\in\mathbf{P}_{i-1}:\,\underline{\Omega}_{i}^{\left(\mathcal{M}\right)}\prec_{SbD}\underline{\Omega}\\
\nexists\underline{\Omega}'\in\mathbf{P}_{i-1}:\,\underline{\Omega}'\prec_{SbD}\underline{\Omega}_{i}^{\left(\mathcal{M}\right)}\\
\mathbf{P}_{i-1}^{pred}=\emptyset\end{array}\right.$&
$\mathbf{P}_{i}\leftarrow\mathbf{P}_{i-1}$&
$\mathcal{R}\left\{ \mathbf{P}_{i-1}^{sim}\right\} \leftarrow\underline{\Omega}_{i}^{\left(\mathcal{M}\right)}$\tabularnewline
\hline
\end{tabular}\end{center}

\begin{center}~\vfill\end{center}

\begin{center}\textbf{Tab. II - P. Rosatti et} \textbf{\emph{al.}}\textbf{,}
\textbf{\emph{{}``}}Multi-Objective System-by-Design for ...''\end{center}

\newpage
\begin{center}~\vfill\end{center}

\begin{center}\begin{tabular}{|c|c|c|c|}
\hline 
$k$&
\emph{DoF}&
$\Omega_{k}^{SbD}$ {[}mm{]}&
$\Omega_{k}^{StD}$ {[}mm{]}\tabularnewline
&
&
{[}Fig. 15(\emph{b}){]}&
{[}Fig. 15(\emph{c}){]}\tabularnewline
\hline
\hline 
1&
$l_{1}$&
$2.99$&
$2.24$\tabularnewline
\hline
2&
$w_{2}$&
$0.70$&
$0.82$\tabularnewline
\hline
3&
$s_{y,1}$&
$1.50$&
$1.42$\tabularnewline
\hline
4&
$s_{y,2}$&
$0.47$&
$0.63$\tabularnewline
\hline
5&
$s_{s}$&
$0.28$&
$0.25$\tabularnewline
\hline
6&
$s_{l}$&
$1.14$&
$1.15$\tabularnewline
\hline
7&
$s_{w}$&
$0.15$&
$0.13$\tabularnewline
\hline
8&
$s_{x,1}$&
$0.06$&
$0.04$\tabularnewline
\hline
9&
$s_{x,2}$&
$0.08$&
$0.09$\tabularnewline
\hline
10&
$s_{x,3}$&
$0.09$&
$0.07$\tabularnewline
\hline
\end{tabular}\end{center}

\begin{center}~\vfill\end{center}

\begin{center}\textbf{Tab. III - P. Rosatti et} \textbf{\emph{al.}}\textbf{,}
\textbf{\emph{{}``}}Multi-Objective System-by-Design for ...''\end{center}

\newpage
\begin{center}~\vfill\end{center}

\begin{center}\begin{tabular}{|c|c|c|}
\hline 
$k$&
\emph{DoF}&
$\Omega_{k}^{SbD}$ {[}mm{]}\tabularnewline
&
&
{[}Fig. 18(\emph{b}){]}\tabularnewline
\hline
\hline 
1&
$l_{l,1}$&
$2.59$\tabularnewline
\hline
2&
$l_{l,2}$&
$2.71$\tabularnewline
\hline
3&
$l_{w,1}$&
$0.08$\tabularnewline
\hline
4&
$l_{w,2}$&
$0.11$\tabularnewline
\hline
5&
$l_{w,3}$&
$0.10$\tabularnewline
\hline
6&
$l_{w,4}$&
$0.07$\tabularnewline
\hline
7&
$o_{w,1}$&
$0.14$\tabularnewline
\hline
8&
$o_{w,2}$&
$0.08$\tabularnewline
\hline
9&
$o_{w,3}$&
$0.40$\tabularnewline
\hline
10&
$o_{w,4}$&
$0.24$\tabularnewline
\hline
11&
$o_{l}$&
$1.24$\tabularnewline
\hline
\end{tabular}\end{center}

\begin{center}~\vfill\end{center}

\begin{center}\textbf{Tab. IV - P. Rosatti et} \textbf{\emph{al.}}\textbf{,}
\textbf{\emph{{}``}}Multi-Objective System-by-Design for ...''\end{center}
\end{document}